\documentclass[12pt]{article}
\usepackage{amsmath,amssymb,amsfonts,graphicx,slashed,color,mathtools}
\usepackage{color}
\usepackage{tocloft}
\usepackage[colorlinks=true,linkcolor=black,citecolor=blue,linktoc=all]{hyperref}

\def\leigh{Robert G. Leigh}
\def\uiucaddress{\small Department of Physics, University of Illinois, 1110 W. Green St., 
Urbana IL 61801-3080, U.S.A. }
\def\title{\Large {
Torsion, Parity-odd Response and Anomalies in Topological States
}}

\parskip=\baselineskip

\newcommand{\cDsl}{{{\cal D}\kern-.65em /\,}}
\newcommand{\cDslsm}{{{\cal D}\kern-.5em /\,}}
\newcommand{\nabslsm}{\nabla\kern-.55em /}
\newcommand{\pasl}{\pa\kern-.55em /}
\newcommand{\psl}{p\kern-.45em /}
\newcommand{\Dsl}{D\kern-.65em /}
\newcommand{\Asl}{A\kern-.55em /}
\newcommand{\nabsl}{\nabla\kern-.65em /\kern+.2em}
\newcommand{\qsl}{q\kern-.5em /}
\newcommand{\ksl}{k\kern-.5em /}
\newcommand{\rsl}{r\kern-.5em /}
\newcommand{\lcw}{\mathring{\omega}}
\newcommand{\lcR}{\mathring{R}}
\newcommand{\torcplg}{{q_T}} 
\newcommand{\ue}{\underline{e}}

\newcommand{\cDslLCsq}{{\stackrel{\circ}{\cDsl^{\kern2pt 2}}}}

\newcommand{\tSigma}{\widetilde{\Sigma}}

\newcommand\cc[1]{#1^{^{\kern-6pt \circ}}\kern2pt}
 
\newcommand{\msD}{\mathcal{D}}

\font\mybb=msbm10 at 11pt
\def\bb#1{\hbox{\mybb#1}}

\def\bZ {\bb{Z}}

\newcommand{\re}{\mathbb{R}}

\newcommand{\pa}{\partial}

\newcommand{\beq}{\begin{equation}}
\newcommand{\eeq}{\end{equation}}
\newcommand{\beqn}{\begin{eqnarray}}
\newcommand{\eeqn}{\end{eqnarray}}
\newcommand{\w}{\wedge}
\newcommand{\mTr}{\mathrm{Tr}}
\newcommand{\mtr}{\mathrm{tr}}
\newcommand{\torcplq}{q_T}

\newcommand{\ble}{\widetilde{e}}
\newcommand{\blA}{\widetilde{A}}
\newcommand{\blw}{\widetilde{\omega}}
\newcommand{\blF}{\widetilde{F}}
\newcommand{\blD}{\widetilde{D}}
\newcommand{\blT}{\widetilde{T}}
\newcommand{\bllcw}{\mathring{\widetilde{\omega}}}
\newcommand{\blue}{\widetilde{\underline{e}}}

\newcommand{\bllcR}{\mathring{\widetilde{R}}}
\newcommand{\bllcD}{\mathring{\widetilde{D}}}
\newcommand{\blR}{\widetilde{R}}

\def\dalemb#1#2{{\vbox{\hrule height .#2pt
\hbox{\vrule width.#2pt height#1pt \kern#1pt
\vrule width.#2pt}
\hrule height.#2pt}}}

\begin{document}

\begin{center}
\title
\end{center}
\vskip 2 cm
\centerline{{\bf 
Onkar Parrikar, Taylor L. Hughes, \leigh}}

\vspace{.5cm}
\centerline{$^1$\it \uiucaddress}
\vspace{2cm}

\begin{abstract}
We study the response of a class of topological systems to electromagnetic and gravitational sources, including torsion and curvature. By using the technology of anomaly polynomials, we derive the parity-odd response of a massive Dirac fermion in $d=2+1$ and $d=4+1$, which provides a simple model for a topological insulator. We discuss the covariant anomalies of the corresponding edge states, from a Callan-Harvey anomaly-inflow, as well as a Hamiltonian spectral flow point of view. We also discuss the applicability of our results to other systems such as Weyl semi-metals. Finally, using dimensional reduction from $d=4+1$, we derive the effective action for a $d=3+1$ time-reversal invariant topological insulator in the presence of torsion and curvature, and discuss its various physical consequences.
\end{abstract}

\pagebreak
\tableofcontents
%

\section{Introduction}
Strong bonds between high-energy and condensed matter physics have been formed through the study of quantum field theory anomalies. Naively, anomalies simply represent the breaking of a classically preserved symmetry through quantum effects, but once one digs deeper one realizes the deep connections between anomalies, topological transport phenomena, bulk-boundary correspondence, and fermion representations that lie at the heart of some of the most interesting, and experimentally relevant, physical phenomena. With the discovery of topological insulators and topological phases of matter, anomalies have moved to the forefront of condensed matter physics\cite{HasanKane,Wenbook}. Many old ideas from high-energy physics, for instance \cite{Callan:1984sa, AlvarezGaume:1984nf} etc., have been repurposed and extended to explain properties of exotic materials that are being measured in experimental groups all over the world\cite{moore2009topological,xia2009observation,zhang2009experimental,HasanKane,chang2013,Kane2005A,bernevig2006c,konig2008}. One example is the connection between the bulk Hall conductivity in 2+1-d electron gasses in the quantum Hall state and the anomalous properties of the 1+1-d chiral fermion edge states at the boundaries of the samples\cite{Laughlin:1981jd,Haldane:1988zza,Kao:1996ey}. This type of bulk-boundary connection between bulk transport properties and anomalous transport of the gapless edge degrees of freedom underlies most of the interesting properties of topological phases of matter. In fact, each different field theory anomaly gives rise to a different type of transport phenomenon, for example, electrical or thermal transport. In recent years, there has also been a great deal of progress in understanding anomaly induced transport phenomena in hydrodynamics (see for instance \cite{Son:2009tf,Loganayagam:2012pz,Jensen:2012kj,Jensen:2013kka} and references therein).

By now there exists a mapping between most known quantum field theory anomalies (gauge and gravitational) and associated condensed matter phenomena in space-time dimensions $d\leq 4$ \cite{Qi:2008ew,Ryu:2010ah}. However, the role played by space-time torsion in anomaly physics is still poorly understood -- a notable example is the anomaly in the global chiral symmetry of $3+1-d$ fermions exposed to torsion \cite{Chandia:1997hu}. This anomaly implies the non-conservation of the chiral current when certain arrangements of dislocations and strain forces are applied to chiral fermions; it has also been the source of some controversy in high energy physics, the reason for which we will recount below. While dislocations and strain forces 
are not commonplace in our outward universe, they are ubiquitous in condensed matter systems. For example, effects of this anomaly should be seen if dislocations and strain are present in Weyl semi-metal materials, which have an electronic structure consisting of an even number of chiral fermions in 3+1-d\cite{nielsen1981,wan2011,turner2013,balentshalasz,haldane2014,matsuura2013,sid2013,grushin2013,jian2013,pavan2012,tewari2012,zyuzin2012,vazifeh2013,chen2013,haldane2014,chen2013weyl,rama2014}. Related effects will also appear in the response properties of time-reversal invariant topological insulators. Thus, while high-energy physicists may not ever have to worry about resolving the torsion anomaly puzzle in order to describe the fundamental properties of the Universe,\footnote{However, as we will see below, the role of torsion in anomaly inflow mechanisms suggests that it should play a role in some braneworld or holographic constructions \cite{Leigh:2008tt,Leigh:future}.}  condensed matter physicists should be concerned because it is something that can be measured. 

The goal of the present work is to resolve many of the uncertainties surrounding torsional anomalies by studying them in three explicit contexts analogous to the work done by Nielsen and Ninomiya for the Adler-Bell-Jackiw anomaly in crystals\cite{Nielsen:1983rb}, and also the work of Qi-Hughes-Zhang on the response properties of topological insulators\cite{Qi:2008ew}. The three systems that we will study are the boundary of a 4+1-d topological insulator which can harbor a single chiral fermion on its boundary\cite{Qi:2008ew}, Weyl semi-metals in which an even number of chiral fermions must be present so that the total chirality vanishes,  and 3+1-d time-reversal invariant topological insulators which contain no chiral fermions, but exhibit related response properties due to the dimensional reduction. In recent work \cite{Hughes:2012vg} we were  able to resolve a similar problem in 2+1-d fermionic insulators whereby torsional terms in the effective action of time-reversal breaking topological insulators were shown to correspond to Hall viscosity transport (see \cite{Avron:1995fg, Read:2008rn, Haldane:2009ke, PhysRevB.84.085316, Nicolis:2011ey,Hughes:2011hv, PhysRevLett.108.066805, Bradlyn:2012ea,Hughes:2012vg, PhysRevB.88.235124} and the review \cite{Hoyos:2014} for a detailed exposition to Hall viscosity in various systems) along with concomitant anomalies on the interface between topological phases, and we will now apply our techniques to the 3+1-d case. 

This  article is organized as follows: in Section \ref{sec:review} we will review the idea of torsion, how fermions couple to it, and its appearance in the $d=3+1$ chiral anomaly. Then in Section \ref{sec:anomalies} we will carefully derive the ``Chern-Simons-like'' parity-odd effective actions for massive Dirac fermions coupled to background curvature and torsion. In particular, we re-derive the response action for 2+1-d time-reversal breaking topological insulators and then present our main result which is the response action for the 4+1-d topological insulator. The 4+1-d case might seem irrelevant at first sight from a condensed matter perspective, but this is not so; it can be used to study torsion effects on chiral fermions by studying the boundary anomalies, and also the response of 3+1-d time-reversal invariant topological insulators by dimensional reduction. This is what we do next -- in Section \ref{sec:callanharvey}, we will carry out the 5-d bulk to 4-d boundary Callan-Harvey anomaly in-flow calculation\cite{Callan:1984sa} paying careful attention to the role of torsion, and in Section \ref{sec:spectralflow} we will give more microscopic Hamiltonian spectral flow arguments for the different anomaly types which illustrate the microscopic behavior of real material systems under the influence of torsion. From here we will discuss some consequences for Weyl semi-metals in Section \ref{sec:wsm}. Then in section \ref{sec:3dti}, we dimensionally reduce the $d=4+1$ parity-odd effective action to discuss some consequences for the 3+1-d time-reversal invariant topological insulator. We will end with some final discussion and conclusions.

\section{Review of Torsion and the Torsion Contribution to the Chiral Anomaly}\label{sec:review}
\subsection{Informal preliminaries}
In classical general relativity, torsion is simply taken to vanish, so that the geometric degrees of freedom can be captured solely by the metric tensor --- torsion can be regarded as a violation of the equivalence principle. In more general formulations of general relativity,\footnote{Here we refer to the first order formalism, in which the action of general relativity (the Palatini action for example) is regarded as depending on independent frame and connection variables. Details will be given below.} the types of matter usually considered provide no source for torsion, so even if it were allowed, one would find that it vanishes by equations of motion. If torsion is allowed, there is no natural choice for a (spin) connection, and both the metric (or more precisely, the frame) and the connection must be provided independently to specify a unique geometry.  

Condensed matter physics is not governed by general relativity. Nevertheless, it is often useful to formulate various concepts in geometric terms. Recently, in condensed matter, effects that are essentially connected to torsion have been brought to the forefront and include things like the Hall viscosity in Chern insulators\cite{Hughes:2011hv, Hughes:2012vg} and the properties of dislocations and disclinations in topological phases\cite{ran2009,ran2010weak,TeoKane,teohughes1,teohughes2,HughesYaoQi13,Benalcazar2013}. Torsion is most intuitively interpreted as the field strength tensor of the gauge potentials that encode translation invariance. A magnetic flux line of torsion is simply a dislocation, i.e., a particle encircling the torsional magnetic flux will be translated by an amount $b^{A}$ (where $A=0,1,2\ldots D$) which is the generalized Burgers' vector of the dislocation. The time component $b^0$ is the amount of translation in time,\footnote{One can envision a spatial Burgers' vector as a lattice dislocation. A temporal Burgers' vector arises, for example, in the presence of vorticity.} and the spatial components $b^{a}$ are the traditional Burgers' vector translation in space. Thus, to each torsional flux line we must associate a $d$-vector of fluxes $b^{A}$ instead of just a scalar flux for the $U(1)$ electromagnetic field. Since dislocations play a pivotal role in many aspects of the theory of crystalline solids, and in quantum-ordered crystals like charge density waves, the role of torsion must be carefully considered in condensed matter systems. 

If we only consider flat space without space-time curvature, we need only introduce geometric variables called co-frame fields $e^{A}$ to describe torsion. Each $e^{A}=e^{A}_{\mu}dx^{\mu}$ (where $\mu=t,x,y,z,\ldots$ and $A=0,1,...,d-1$) is a $1$-form vector potential with a label given by $A$. In flat space we can choose a gauge where the spin connection $1$-forms $\omega^{AB}_{\mu}dx^{\mu}\equiv 0$ so that the components of the torsion tensor are
\begin{equation}\label{torsioncomponentsnospinconn}
T^{A}_{\mu\nu}=\partial_\mu e^{A}_{\nu}-\partial_{\nu}e_{\mu}^A,
\end{equation}\noindent  
that is, $T^{A}$ is the field-strength $2$-form for the gauge potential $e^{A}.$ As an example,  if we have a dislocation-line localized at the origin in the $xy$-plane then $T^{A}_{xy}=b^{A}\delta(x)\delta(y).$  The generalized Burgers' vector $b^{A}$ of the localized dislocation is the torsion magnetic flux from each $e^{A}$ potential, or equivalently the circulation of $e^A$ around the dislocation in the $xy$ plane
\beq
b^{A}=\int d^2x\; \epsilon^{ij}T^{A}_{ij} = \oint e^A
\eeq
To make contact with more familiar condensed matter notation we note that the co-frame fields are simply a re-packaging of conventional elastic variables based on the displacement vector $u^{A}$ (where we allow for a time-displacement as well). In terms of the displacement vector the co-frame fields are (to linear order in displacements)
\begin{equation}
e^{A}_{\mu}=\delta^{A}_{\mu}-\frac{\partial u^{A}}{\partial x^{\mu}}
\end{equation}\noindent where the spatial components $w^{a}_{i}=\partial_i u^a$ are known conventionally as the \emph{distortion} tensor\cite{LandauElasticity}. The undeformed system is represented by the orthonormal frames $e^{A}_{\mu}=\delta^{A}_{\mu}$ which exist at every point in space-time. 

Similarly, lattice disclinations can be viewed as sources of curvature -- traversal around a disclination results in rotation. This effect can be encoded in link variables $\omega_i^{ab}$. Promoting this to space-time, we have the set of spin connection $1$-forms (valued in $\mathfrak{so}(d-1,1)$) ${\omega^{A}}_{B}={\omega^{A}}_{\mu;B}dx^\mu$ which are gauge potentials for local Lorentz invariance. The field strength ${R^A}_B$ for the spin connection is referred to as the \emph{curvature}. In fact, the spin connection can be grouped with the translation gauge potentials $e^{A}$ to form a kind of Poincar{\'e} gauge structure.\footnote{Formally, this can be seen by considering the coupling of a Dirac fermion (or any tensor) to a background frame and spin connection. The covariant derivative $\nabla_A$ generates translations, and the commutator of translations takes the form
\beqn
\left[\nabla_A,\nabla_B\right] &=& -T^C_{AB}\nabla_C+R_{CD;AB}J^{CD},
\eeqn
where $T$ is torsion, $R$ curvature and $J$ the generator of Lorentz transformations acting on the Dirac spinor. The commutator has an interpretation in terms of traversing a `closed' path, the result being a translation (if torsion is present) or a (Lorentz) rotation (if curvature is present). The standard relations between $e^A$, $\omega^A{}_B$ and $T^A$, $R^A{}_B$ will be given below in the following subsection.
}
 We refer the reader to Ref. \cite{Hughes:2012vg}, and references therein, for more discussion about the connection between the field-theory variables and conventional elasticity theory.

Now we will move on to discuss the well-known chiral anomaly. In 1+1-d, charged chiral fermions in the presence of an electric field will not conserve chiral charge.
This effect is captured by the anomalous Ward identity for the chiral (axial) current:
\begin{equation}
\partial_{\mu} j^{\mu}_{5}=\frac{q^2}{4\pi}\epsilon^{\mu\nu}F_{\mu\nu} \label{chiralanomaly}
\end{equation}\noindent where $q$ is the $U(1)$ charge.
This is problematic in the sense that it goes against all classical physical intuition about charge conservation. There are two common ways in which this problem is resolved: (i) if the chiral fermion appears as the low-energy description of a real 1+1-d material then it must always appear with its anti-chiral partner (a consequence of the Neilsen-Ninomiya no-go theorem (fermion doubling))\cite{Nielsen:1981hk} or (ii) the chiral fermion appears as the low-energy description on the boundary of a 2+1-d system, and the anti-chiral partner appears on the opposite boundary. In this case the total chiral charge of the two chiral fermions is passed back and forth through the 2+1-d bulk. One can show in case (ii) that when an electric field is applied parallel to the chiral edge state there is a bulk current  perpendicular to the applied electric field/edge, and the boundary chiral anomaly is attached to a bulk Hall effect; this is an example of the Callan-Harvey effect\cite{Callan:1984sa} and it appears in any 2D electron system exhibiting the integer quantum Hall effect. In case (i) the $U(1)$ axial charge is locally conserved but it can be converted between the low-energy left-handed (left-moving) and right-handed (right-moving) branches in the presence of an applied electric field. In this case there is no notion of a perpendicular Hall current since both chiral and anti-chiral fermions exist in the same local region of space. 

We note that because the frame field, and subsequently, the torsion $2$-form, carry an extra Lorentz index $A$, there is no Lorentz invariant contribution to the 1+1-d chiral anomaly from torsion. For a real crystalline material or a fluid at finite density, both of which naturally break Lorentz invariance, it is possible to generate a term of the form $\partial_{\mu} j^{\mu}_{5}\sim\theta_A \epsilon^{\mu\nu}T_{\mu\nu}^{A}$ for some field $\theta_A$ arising from the source of Lorentz violation. For example, this type of anomaly might be generated if we have left and right handed chiral fermions with different velocities, which is allowed in a condensed matter setting. For 1+1-d fermions different velocities means the density of states of the left and right movers are different, which can lead to a physically measurable consequence. We will not consider these effects in what follows, though they could appear in low-dimensional condensed matter materials and would be interesting to study in future work. 

In 3+1-d, the next dimension that supports chiral fermions, there is also a chiral anomaly in the presence of background electromagnetic fields, however it is only present when parallel electric and magnetic fields are applied. This is captured by the anomalous Ward identity
\begin{equation}
\partial_{\mu} j^{\mu}_{5}=\frac{q^3}{32\pi^2}\epsilon^{\mu\nu\rho\sigma}F_{\mu\nu}F_{\rho\sigma}=\frac{q^3}{4\pi^2}\vec{E}\cdot \vec{B}.
\end{equation}
 One can think of the anomaly as a two-step process in which one first turns on a uniform magnetic field and then a parallel electric field. The magnetic field will produce Landau levels in the low-energy chiral fermions, and there will be one Landau level that disperses chirally along the direction of the magnetic field. This dispersive Landau level is identical to a degenerate set of 1+1-d chiral fermions along the direction of the magnetic field, one chiral branch for each magnetic flux quantum. At this point the problem has been reduced back to decoupled copies of the 1+1-d case, and one can proceed by applying an electric field as the second step. The electric field will induce a non-conservation of charge for each 1+1-d chiral branch. The resolution of the non-conservation of chiral charge is solved using one of the two mechanisms presented earlier. Using the nomenclature from  recent condensed matter literature, one would say that chiral fermions occurring from case (i) appear in a Weyl semi-metal material\cite{nielsen1981,wan2011,turner2013,haldane2014,matsuura2013,sid2013,grushin2013,jian2013,pavan2012,tewari2012,zyuzin2012,vazifeh2013,chen2013} and from case (ii) one would state that the chiral fermions appear at the boundary of a 4+1-d topological insulator state\cite{Qi:2008ew}. 
 
 It is well-known that in addition to the electromagnetic contributions to the anomalous chiral conservation law, new terms are generated when the space-time in which the chiral fermion resides is curved or has torsion. As shown, for instance in \cite{AlvarezGaume:1983ig,Chandia:1997hu}, the Ward identity is modified in the presence of curvature and torsion to\footnote{These expressions should be taken to be schematic; the precise results will be presented later in the paper.}
 \begin{equation}
 \partial_{\mu} j^{\mu}_{5}=\frac{q^3}{32\pi^2}\epsilon^{\mu\nu\rho\sigma}F_{\mu\nu}F_{\rho\sigma}+\frac{q}{192\pi^2}\epsilon^{\mu\nu\rho\sigma}\frac{1}{4}R_{\mu\nu}^{\,\,ab}R_{\rho\sigma}^{\,\,cd}\eta_{ad}\eta_{bc}+C_{NY}.
 \end{equation}
 where $R_{\mu\nu}^{\,\,ab}$ is the Riemann curvature tensor and the Nieh-Yan term\cite{Nieh:1981ww} is given by
 \beq
 C_{NY} = \frac{q}{32\pi^2\ell^2}\epsilon^{\mu\nu\rho\sigma}\left(\eta_{ab}T^a_{\mu\nu}T^b_{\rho\sigma}-2R_{ab;\mu\nu}e^a_{\rho}e^b_{\sigma}\right)
 \eeq
with $\ell$ being a length scale. The consequences of the first term are well understood, and even the curvature dependent term has recently come under investigation in a condensed matter setting\cite{Ryu:2010ah,Stone:2012ud}, however the microscopic origin, and a clear condensed matter interpretation of the third term has not been considered.
 The coefficients of the first two terms are dimensionless and universal, while the Nieh-Yan term has a dimensionful coefficient, related to a UV scale\cite{Chandia:1997hu}. The reason the coefficients have different properties is  that the components of the co-frame $e^{A}_{\mu}$ are dimensionless and do not have the conventional natural units of $L^{-1}$ befitting the components of a  connection. Thus, the torsion field (\ref{torsioncomponentsnospinconn}) only has units of $L^{-1}$ and the anomalous Nieh-Yan  term needs a coefficient with units $\hbar/L^2$ so that the entire term has the  units of action when integrated over a space-time region. Usually, anomaly coefficients have a topological origin and are quantized as an integer multiplying fundamental constants. The Nieh-Yan term however has units, is sensitive to UV scales, and thus has no apparent universal interpretation. 
 
  In this article we have not set out to address the Nieh-Yan  term from a fundamental perspective, but instead we will provide a regularized derivation and a condensed matter interpretation of the consequences of this and other new torsional contributions to anomalies. Indeed, we do find that one can interpret the Nieh-Yan term as a contribution to the chiral anomaly, and its effects could possibly be observed, for example, in Weyl semi-metals.\footnote{In the context of topological insulators, the significance of UV scales is somewhat subtle. As we review below for example, the UV scale of an edge theory is related to a gap scale in the bulk. Thus, it is possible that anomalies depending on the UV scale in an edge theory have simple interpretations (by anomaly inflow) in terms of physics in the bulk. We expect that the same physics can arise in high energy theory, for example in brane-world scenarios, if either side of a brane corresponds to distinct topological phases. This possibility, as far as we are aware, has not been considered in the literature.}  A related effect also appears in the response of 3+1-d time-reversal invariant topological insulators to torsion where an axion-induced Nieh-Yan term gives rise to a surface Hall viscosity\cite{hidaka2013}. Before we get to these results, we will review the warm-up problem of the 2+1-d topological insulator that was covered in Refs. \cite{Hughes:2011hv, Hughes:2012vg} and then step up to the 4+1-d topological insulator. While considering 4+1-d may be a stretch for condensed matter minded readers, we can use two different properties of this system to study lower-dimensional systems that are relevant to experiments. We can first consider the gapless boundary modes of the 4+1-d topological insulator which will be standard 3+1-d chiral fermions as would be found in the \emph{bulk} of a Weyl semi-metal, and second, we can dimensionally reduce the 4+1-d insulator to obtain a time-reversal invariant strong topological insulator in 3+1-d using the framework set forth by Ref. \cite{Qi:2008ew}.  
  
\subsection{Formal preliminaries}
Before proceeding, we present here a brief introduction to the mathematical details of torsional gravity, fermions coupled to torsion, the corresponding symmetries, etc. (see \cite{Nieh:1981xk,Hughes:2012vg} for more details). As mentioned previously, conventionally, gravity is described in terms of the metric 2-tensor $g=g_{\mu\nu}dx^{\mu}\otimes dx^{\nu}$ on space-time. However, in order to couple fermions to gravity, it is essential that we use the first order formalism. In this language, we introduce the \emph{co-frame}, a local basis of 1-forms $e^A(x) = e^A_{\mu}(x)dx^{\mu}$ on space-time, such that
\beq
g = \eta_{AB}\;e^A\otimes e^B.
\eeq
The corresponding basis of tangent vector fields dual to the co-frame is called the \emph{frame} $\ue_A(x)$. In going from the metric to the co-frame, we have introduced a redundancy in our description, namely the \emph{local Lorentz} gauge symmetry 
\beq
e^A(x)\mapsto {\Lambda^A}_B(x)e^B(x)
\eeq
where $\Lambda$ is an $SO(1,d-1)$ matrix, $\Lambda^T\cdot \eta\cdot \Lambda=\eta$, with $\eta$ the constant Minkowski metric. Note that the local Lorentz transformation is {\it not} a space-time coordinate transformation, but a rotation/boost of the local orthonormal frame. In order to maintain covariance under this gauge symmetry, we must therefore introduce a connection 1-form ${\omega^A}_B$, which transforms under local Lorentz transformations as
\beq
{\omega^A}_B \mapsto {\left(\Lambda\cdot \omega\cdot \Lambda^{-1}- d\Lambda\cdot\Lambda^{-1}\right)^A}_B.
\eeq
The connection ${\omega^A}_B$ is often referred to as the \emph{spin connection}. Loosely speaking, we may think of $e^A$ and ${\omega^A}_B$ as gauge fields corresponding to local translations and local Lorentz rotations respectively. As has been mentioned above, the field strength 2-form corresponding to the co-frame 
\beq
T^A=de^A+{\omega^A}_B\wedge e^B
\eeq
is called \emph{Torsion}, while the field strength 2-form for the spin connection 
\beq
{R^A}_B = {d\omega^A}_B+{\omega^A}_C\wedge {\omega^C}_B
\eeq
is called \emph{Curvature}.\footnote{In a coordinate basis of 1-forms $dx^\mu$, the component forms of these expressions read
\beqn
T^A_{\mu\nu}&=& \pa_\mu e^A_\nu-\pa_\nu e^A_\mu+\omega_\mu^A{}_B e^B_\nu-\omega_\nu^A{}_B e^B_\mu\\
R^A{}_{B\mu\nu}&=& \pa_\mu \omega_\nu^A{}_B-\pa_\nu \omega_\mu^A{}_B+\omega_\mu^A{}_C\omega_\nu^C{}_B-\omega_\nu^A{}_C\omega_\mu^C{}_B.
\eeqn
The Riemann tensor $R^\lambda{}_{\rho\mu\nu}=e^A_\rho e_B^\lambda R^A{}_{B\mu\nu}$ can be expressed in terms of the Christoffel symbols in the usual way, but in the presence of torsion, the Christoffel symbol is not symmetric in its lower indices.
} Both torsion and curvature transform covariantly under local Lorentz transformations, $T^A\mapsto (\Lambda\cdot  T)^A$, $R^A{}_B\mapsto (\Lambda^T\cdot  R\cdot \Lambda)^A{}_B$. In standard discussions of general relativity, the torsion 2-form is set to zero. As a consequence, the spin connection is then uniquely determined in terms of the co-frame, and is called the \emph{Levi-Civita} connection, denoted herein by ${{\lcw}^A}_{\;\;B}$. However, the gravitational fields we will consider in this paper will be non-dynamical, and will be treated as background fields which determine the geometry in which the fermions propagate. As such, we will not set torsion to zero. There are two ways to view this: first, we might want to consider lattice systems, say, in which dislocations and disclinations are present. These are sources of torsion and curvature respectively, and so we would not want to set either to zero. Second, even in the absence of torsion or curvature in a given state of matter, we can regard $e^A$ and $\omega^A{}_B$ as {\it sources} for distinct operators. Thus, we can regard what we are doing in terms of a generating functional for correlation functions that determine transport properties, and as such we would have no reason to impose restrictions on sources (or their derivatives). 
This point is especially important in the present discussion, because Dirac/Weyl fermions carry spin, and as such the co-frame and the spin connection couple to independent fermion operators, namely the \emph{stress current} and the \emph{spin current} respectively. Thus, we will regard $e^A$ and ${\omega^A}_B$ as independent background fields, and treat them on an equal footing. However, we will find it notationally convenient to organize things in terms of the Levi-Civita connection occasionally. For future use, we also define the 3-form
\beq
H= \frac{1}{3!}H_{ABC}\;e^A\wedge e^B\wedge e^C \equiv \eta_{AB}\;e^A\wedge T^B.
\eeq
In fact, as will become clear in the following sections, the macroscopic properties of the fermionic models we consider, organize themselves in terms of an ``effective'' repackaged spin connection $\omega^{(c)}_{\mu;AB}=\left(\lcw_{\mu;AB}-\frac{c}{2}H_{\mu;AB}\right)$, for some constant $c$. Let us now move on to describe the coupling of fermions to the frame and the spin connection.  

The Dirac action in the presence of background gravity in $d=D+1$ space-time dimensions may be written as\footnote{We have written the action in this way, because it is this form for which the action is strictly real (not just up to a total derivative). This is crucial if we wish to study the system on a geometry with a boundary or other defects.} 
\beqn
S[\psi; e,\omega]
&=&
\frac{1}{D!}\int \epsilon_{A_1\ldots A_d}e^{A_1}\wedge \ldots\wedge e^{A_{D}}\wedge \left[ \frac{1}{2}\overline\psi \gamma^{A_d}\nabla\psi-\frac{1}{2}\overline{\nabla\psi}\gamma^{A_d}\psi- e^{A_d}m\overline\psi \psi\right]
\label{DiracActionInv}\\
&=&\int d^dx\det e \left[ \frac{1}{2}\overline\psi \gamma^A\nabla_{\underline{e}_A}\psi-\frac{1}{2}\overline{\nabla_{\underline{e}_A}\psi}\gamma^A\psi-m\overline\psi \psi
\right] \label{DiracAction}
\eeqn
where the Lorentz and gauge covariant derivative of the Dirac spinor is given by\footnote{$\gamma$'s with multiple indices correspond to anti-symmetrized quantities, e.g. $\gamma^{AB}=\frac12(\gamma^A\gamma^B-\gamma^B\gamma^A)$.} 
\beq
\nabla\psi=d\psi+\frac14 \omega_{AB}\gamma^{AB}\psi+qA\psi.
\eeq
Here we have also introduced a background electromagnetic (i.e. $U(1)$) connection $A$, with $q$ being the fermion charge. In odd space-time dimensions, the mass $m$ is real, and its sign will play a central role in determining the character of the resulting insulating state. The classical equation of motion for the spinor field involves the Dirac operator
\beq\label{DiracOp}
\cDsl=\gamma^A\ue_A^\mu\left(\pa_\mu+qA_{\mu}+\frac14 \omega_{\mu;BC}\gamma^{BC}+B_\mu\right)
\eeq
where $B\equiv \frac{1}{2}T^B(\underline{e}_A,\underline{e}_B)\;e^A$. 
The $B$ term arises upon integration by parts in deriving the equations of motion, and we note that it enters in such a way that it looks like it corresponds to an additional gauge field.\footnote{In fact, as explained in \cite{Nieh:1981xk}, the classical theory possesses a corresponding background scaling symmetry when $m=0$ under which the fields and background transform as $e^A(x)\mapsto e^{\Lambda(x)}e^A(x),\ \  {\omega^A}_B(x)\mapsto {\omega^A}_B(x),\ \  \psi(x)\mapsto e^{-(d-1)\Lambda(x)/2}\psi(x)$. We note from the definition of $B$ that under such a transformation, $B$ transforms like a gauge field $B \mapsto B+\frac{d-1}{2}d\Lambda$. However, this symmetry will not play much of a role in our discussion, so we leave it at that.} It is not of course independent of the spin connection, but does vanish with the torsion. Another way to write the Dirac operator is in terms of the Levi-Civita connection
 \beqn\label{DiracOpH}
\cDsl
&=&\gamma^A\ue_A^\mu\left(\pa_\mu+A_\mu+\frac14 \lcw_{\mu;BC}\gamma^{BC}\right)-\frac{1}{4}\frac{1}{3!}H_{ABC}\gamma^{ABC}.
\eeqn
The Dirac action shown above corresponds to `minimal coupling' of the frame and spin connection to the fermions. There is in fact another invariant term that we could add to the action
\beq
\int d^dx\;\mathrm{det}(e)\;H_{ABC}\;\bar\psi\gamma^{ABC}\psi.
\eeq
Although it is `non-minimal', it occurs at the same order in power counting as the other terms in the action.
Its inclusion has the effect of shifting the coefficient of the $H$ term in the Dirac operator, as in equation \eqref{DiracOpH}. Thus, there is a `torsional charge'  $\torcplq$, and we take the Dirac operator to be
\beq
\cDsl=\gamma^A\ue_A^\mu\left(\pa_\mu+A_\mu+\frac14 \lcw_{\mu;BC}\gamma^{BC}\right)-\frac{\torcplq}{4}\frac{1}{3!}H_{ABC}\gamma^{ABC}.
\eeq
Physically, $\torcplq$ can be thought of as the strength of the torsional coupling. While in the present case it is possible to absorb the torsion coupling into the definition of $H$, this is not true in general, because different species of fermions might have different coupling strengths. 

The Dirac theory has background diffeomorphism and local Lorentz gauge symmetry. In order to explore these, we begin by defining the following current 1-forms
\beq
J = q\;\overline\psi \gamma^A\ue_A\psi,\;\;J^A= \frac12(\overline\psi\gamma^A \nabla \psi-\overline{\nabla\psi}\gamma^{A}\psi),\;\;{J^A}_B
=\frac14 \ue_C\overline\psi {\gamma^{CAD}}\eta_{DB}\psi \label{spin current}
\eeq
which we will refer to as the \emph{charge} current, \emph{stress} current, and \emph{spin} current respectively. These couple respectively to the \(U(1)\) gauge field $A$, co-frame $e^A$, and spin connection ${\omega^A}_B$ in the classical action. The components of the current $J^A$ give a (not necessarily symmetric) notion of the ``stress-energy tensor''\footnote{For reasons that will become apparent below, we should resist the temptation to symmetrize the stress-energy tensor at this point.} via $T_{\mu\nu}= J^A_{\mu} e^B_{\nu} \eta_{AB}$. Also note that the spin current \(J^{AB}_{\mu}\) vanishes in \(d=2\). 

Invariance of the classical action under background diffeomorphisms follows immediately from writing it as the integral of a top form, as in \eqref{DiracActionInv}. We will take the action of local diffeomorphisms on fermions and background fields as 
\beq
\delta \psi = i_{\underline\xi}\nabla\psi,\qquad \delta e^A = D\xi^A+i_{\underline\xi}T^A,\qquad \delta \omega_{AB} = i_{\underline\xi}R_{AB},\qquad \delta A = i_{\underline\xi}F \label{covdiff}
\eeq
where $D$ is the Lorentz-covariant derivative\footnote{The Lorentz covariant derivative acting on a $p$-form with Lorentz indices ${K^{A_1\cdots A_M}}_{B_1\cdots B_N}$ reads $$D{K^{A_1\cdots A_M}}_{B_1\cdots B_N}=d{K^{A_1\cdots A_M}}_{B_1\cdots B_N}+{\omega^{A_1}}_{C_1}\wedge{K^{C_1\cdots A_M}}_{B_1\cdots B_N}+\cdots-(-1)^p{K^{A_1\cdots A_M}}_{C_1\cdots B_N}\wedge {\omega^{C_1}}_{B_1}+\cdots $$}, $\underline\xi$ is a vector field with compact support and $ i_{\underline\xi}$ is the interior product of $\underline\xi$ with a differential form.\footnote{For $\alpha=\frac{1}{p!}\alpha_{\mu_1\cdots\mu_p} dx^{\mu_1}\wedge\cdots dx^{\mu_p}$ a $p$-form, and $\underline{\xi}=\xi^\mu\pa_\mu$ a vector field, the interior product is defined as  $$i_{\underline{\xi}}\alpha=\frac{1}{(p-1)!}\xi^\nu\alpha_{\nu\mu_1\cdots\mu_{p-1}}dx^{\mu_1}\wedge \cdots dx^{\mu_{p-1}}$$.} These transformations differ from ordinary diffeomorphisms by local Lorentz and $U(1)$ gauge transformations, so we will refer to these as \textsl{covariant diffeomorphisms}. Using Noether's theorem, it is straightforward to obtain the conservation equation 
\beq 
D*J^A - i_{\underline{e}^A}T_B\wedge *J^B-i_{\underline{e}^A}R_{BC}\wedge *J^{BC}-i_{\underline{e}^A}F\wedge *J = 0. \label{ClassicalDiffWI}
\eeq 
Some readers might be more familiar with the component form of this equation, which reads
\beq
\frac{1}{\mathrm{det}(e)}D_{\mu}\left(\mathrm{det}(e)\; J^{A\mu}\right)-\ue^{A\mu} T_{B;\mu\nu} J^{B\nu}-\ue^{A\mu} R_{BC;\mu\nu} J^{BC;\nu}-\ue^{A\mu}F_{\mu\nu} J^\nu=0,
\eeq
or when written in terms of the stress-energy tensor, we have
\[\nabla^{(\Gamma)}_\mu {T^{\mu}}_{\rho}- R_{BC\rho\nu} J^{BC\nu}- F_{\rho\nu} J^\nu=0.
\]
where $\nabla^{(\Gamma)}$ is the coordinate covariant derivative, involving the (torsionful) Christoffel symbol. 

Next, under an infinitesimal Lorentz transformation, the spinors and background fields transform as
\beq
\delta\psi = \frac14\theta_{AB}\gamma^{AB}\psi,\qquad\delta e^A = -{\theta^A}_Be^B,\qquad\delta{\omega^A}_{B} = -{(D\theta)^A}_B,\qquad \delta A=0\label{loclor}
\eeq
where ${\theta^A}_B$ are the infinitesimal angles parametrizing the transformation. Invariance of the Dirac action under these transformations is automatic, by construction. 
The corresponding Ward identity is
\beq
D*J^{AB}-e^{[A}\wedge *J^{B]} = 0. \label{ClassicalLorWI}
\eeq
In components, this evaluates to
\beq
\ue_B^\lambda \ue_A^\rho D_\mu J^{AB\mu} +T^{[\rho\lambda]}=0,
\eeq
where the second term is the anti-symmetric part of the stress-energy tensor. The physical interpretation of this equation is that of conservation of net angular momentum. 
 This is the classical result; usually, it is interpreted to mean that the stress-tensor can be made symmetric, by adding the appropriate `improvement' terms involving the spin current. Note however, that if this Ward identity is anomalous in the quantum theory (as is indeed the case for Weyl fermions in even dimensions), then this interpretation is problematic, and the anomaly must correspond to an irremovable \emph{anti-symmetric} part of the stress-energy tensor (certainly this must be true in $1+1$-dimensions, since the fermionic Lorentz current vanishes). In such a case, the usual improvement of the stress-energy tensor to make it symmetric must fail, in the sense that it cannot correspond to the addition of local counterterms. We note that this conclusion also holds in theories which are not necessarily Lorentz invariant, but which have any type of  spin-orbit (or orbital-orbit) coupling where the momenta couple to matrices. This covers a large class of condensed matter systems where the electronic degrees of freedom couple to the geometry via the frame (or a frame-like object) and spin-connection instead of purely the metric. For example, a model of the form $H=(p_{x}^2-p_{y}^2)\sigma^x + 2p_xp_y\sigma^y+m\sigma^z,$ which is a continuum theory for a model with a Chern number equal to $2$, and not Lorentz invariant, will exhibit the qualitative features we have discussed above albeit with some important modifications that we leave to future work. 

Finally, we remark that in even dimensions, it is also possible to couple \emph{chiral} fermions to the frame and connection. The action is a straightforward modification of \eqref{DiracAction}. The chiral theory also has the same symmetries as the Dirac theory at the \emph{classical} level, and the above conservation laws carry over straightforwardly to the chiral case. However, all the symmetries are spoilt by perturbative anomalies upon quantization. Chiral fermions show up as edge states in topological insulators, and we will see that their anomalies are intimately related with the bulk transport properties. 


\section{Parity odd effective actions}\label{sec:anomalies}
All types of free-fermion topological insulator/superconductor phases can be represented by massive Dirac Hamiltonians with various symmetries, i.e.,
\begin{equation}
H=\sum_{a=1}^{D}p_{a}\Gamma^a + m\Gamma^0\label{eq:HamDirac}
\end{equation}\noindent where $\{\Gamma^{A},\Gamma^{B}\}=2\eta^{AB}$ for $A,B= 0,1, 2,\ldots D$ and $\eta^{AB}$ is the flat Lorentz metric. In odd space-time dimensions the Hamiltonians of insulators without additional symmetries (called the unitary A class) are classified by an integer topological invariant $\nu.$ Non-trivial insulators, i.e., insulators where $\nu\neq 0$ are said to exhibit the D-dimensional quantum Hall effect, or just the quantum Hall effect if $D=2.$ These systems are gapped in the bulk, but harbor $D-1$-dimensional chiral fermions on their boundaries ($D-1$ would give an even-dimensional boundary space-time). The bulk remains gapped, unless the mass vanishes, at which point there is a topological phase transition between insulating states where $\nu$ differs by one. The precise value of $\nu$ is not determined by Eq. (\ref{eq:HamDirac}) alone but requires information about the regularization scheme to uniquely define $\nu.$ Throughout this article we will use Pauli-Villars (spectator fermion) type regularization as it matches the structure of many simplified condensed matter lattice-Dirac models including lattice models with Wilson mass terms. Our convention is to choose the regularization such that $m<0$ is the topological phase with $\nu=1$ and $m>0$ is the trivial phase with $\nu=0.$ We note that such a regularization is required even in the absence of all gravitational/torsional effects, as noted in Ref. \cite{Haldane:1988zza}, since otherwise a 2+1-d free-fermion model would give rise to a non-integer Hall conductivity.

 The topological insulator phase with $\nu=1$ will possess chiral boundary states that will produce anomalous currents in the presence of background electromagnetic and gravitational fields. These anomalous currents are matched by a bulk response of the topological insulating state where all anomalous current flowing from the boundary simply flows through the bulk to another boundary. Even without boundaries, the bulk of the material can respond similarly when background fields are present. The bulk response is captured by topological terms that appear in the effective action when the gapped fermions are integrated out in the presence of background fields. For instance, the effective action for a massive Dirac fermion in $d=2+1$ flat space-time in the presence of background electromagnetic fields, contains the parity-odd Chern-Simons term
\beq
S_{odd}[A] = \frac{\sigma_H}{2}\int_{M_{3}}A\wedge dA
\eeq
where $\sigma_H = \frac{1}{2}(1-\mathrm{sign}(m))\frac{q^2}{2\pi}$. The flow of the corresponding Hall current $*J_{bulk} = \sigma_H dA$ into the boundary between a trivial $\sigma_H=0$ phase and a topological $\sigma_H=q^2/2\pi$ phase, precisely matches the $U(1)$ anomaly of the edge chiral fermion (Eq. \eqref{chiralanomaly}). In this section, we derive such topological response terms in the fermion effective action in odd-dimensional space-times with curvature and torsion, from an anomaly polynomial which is naturally defined in one higher dimension. The relevant terms are easily identified as they violate parity and can be easily extracted. In our discussion below, we will use the techniques presented in \cite{AlvarezGaume:1984nf}, albeit adapted to the case with non-zero torsion. Our main emphasis, as mentioned previously, will be on torsional terms and the corresponding transport physics. In particular, we will see that including torsion results in UV divergences in the effective action, which we will carefully regulate. Although such divergences represent non-universal effects, the {\it difference} of such quantities between distinct phases is finite and is captured by the boundary physics.

\subsection{The anomaly polynomial}
\begin{figure}[!t]\label{fig1}
\centering
\includegraphics[height=7cm]{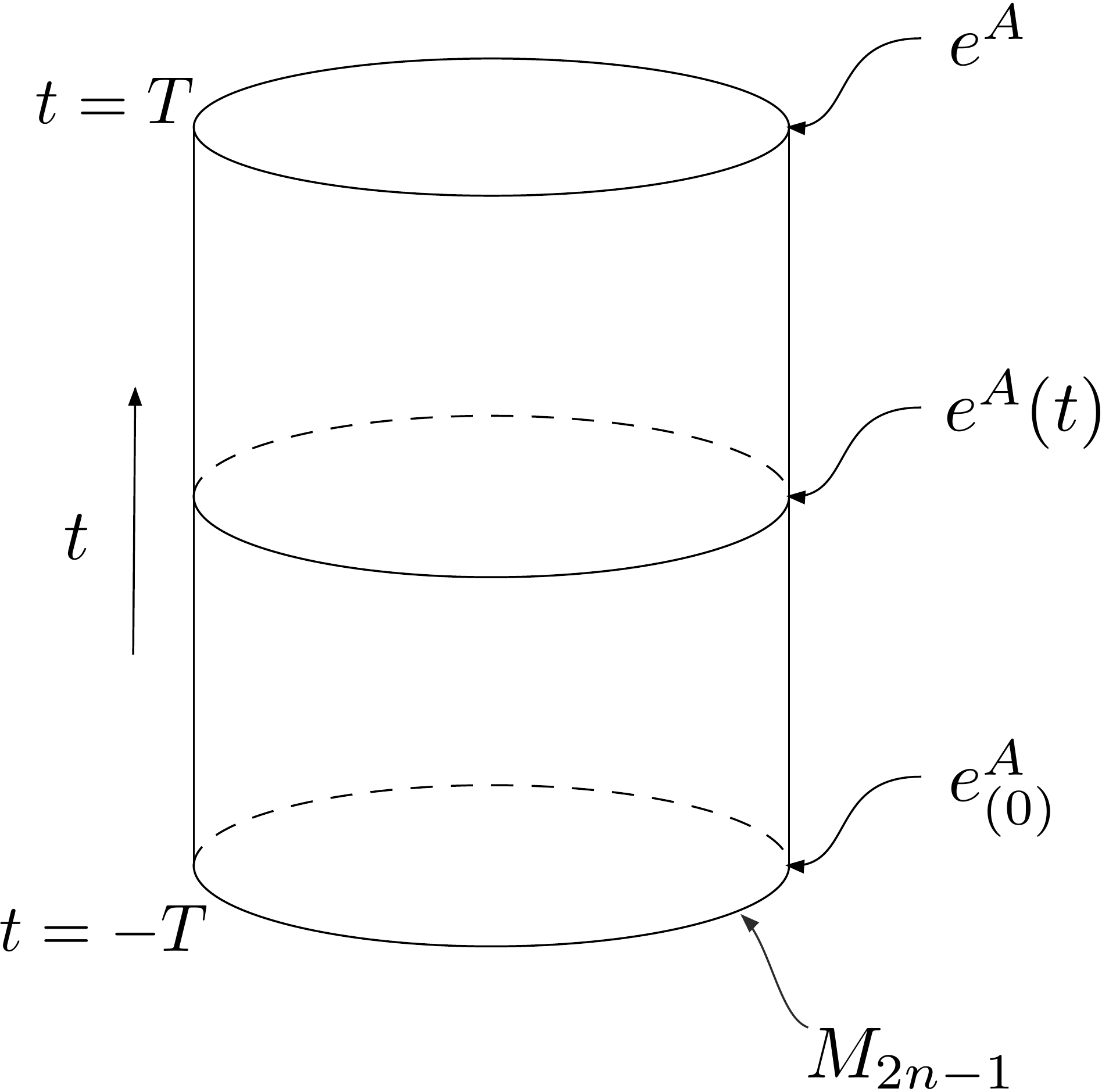}
\caption{An illustration of the one-parameter family of background co-frames, which interpolates between the fiducial co-frame $e^A_{(0)}$ and the co-frame in which we are interested $e^A$.}
\end{figure}

Let us consider massive Dirac fermions on a \(d=D+1=2n-1\)-dimensional manifold-without-boundary \(M_{2n-1}\), endowed (locally) with the co-frame \(e^A\), spin connection \(\omega_{AB}\), and a \(U(1)\) connection \(A\). In Euclidean signature, the fermionic quantum effective action is given by
\beq
S_{eff}[e,\omega,A] = -\mathrm{ln\;det}\left(i\slashed{\msD}_{2n-1}+im\right).
\eeq
Formally, we may rewrite the above as 
\beq
S_{eff}[e,\omega,A] = -\sum_{\lambda_k} \frac{1}{2}\mathrm{ln}\;(\lambda_k^2+m^2)-i\sum_{\lambda_k}\mathrm{tan}^{-1}\;\frac{m}{\lambda_k} 
\eeq
where \(\lambda_k\) are the eigenvalues of the Dirac operator: $i\slashed{\msD}_{2n-1}|\psi_k\rangle = \lambda_k|\psi_k\rangle$, $|\psi_k\rangle$ being the eigenstates. The parity violating piece must come with odd powers of $m$
\beq
S_{odd}[e,\omega,A] = -i\sum_{\lambda_k}\mathrm{tan}^{-1}\;\frac{m}{\lambda_k}. \label{formalpoddaction}
\eeq
In order to compute \eqref{formalpoddaction} as a functional of the background gauge and gravitational sources \((e^A,\omega_{AB},A)\), it is convenient to use the following strategy \cite{AlvarezGaume:1984nf}: imagine a one-parameter family of backgrounds \((e^A(t),\omega_{AB}(t),A(t))\) which \emph{adiabatically} interpolates between a fiducial background \((e^A_{(0)},\omega_{(0)\;AB},A_{(0)})\) and \((e^A,\omega_{AB},A)\) (see Fig. \ref{fig1}).\footnote{Note that this is merely a technique which facilitates the computation. Also, $t$ is an external parameter, and not to be confused with time.} For instance, we may choose the co-frame to be
\beqn
e^A(t) =\left\{\begin{matrix*}[l]
e^A_{(0)},&\qquad -\infty <t<-T\cr
\frac{1}{2}\left[1-\varphi(t)\right]\;e^A_{(0)}+\frac{1}{2}\left[1+\varphi(t)\right]\;e^A,&\qquad  -T \leq t \leq T\cr
e^A,&\qquad  T < t < \infty 
\end{matrix*}\right.\label{backfields}
\eeqn
where \(\varphi(t)\) is an arbitrary function which smoothly interpolates between \([-1,1]\) as \(t\) runs from \(-T\) to \(T\), for some large and positive \(T\). The other sources \(\omega_{AB}(t)\) and \(A(t)\) may be chosen similarly. 
This gives us a one-parameter family of Dirac operators \(\slashed{\msD}_{2n-1}(t)=\slashed{\msD}_{2n-1}[e^A(t),\omega_{AB}(t),A(t)]\) with eigenvalues \(\lambda_k(t)\). Taking a $t$-derivative of equation \eqref{formalpoddaction}, we obtain
\beq
\frac{dS_{odd}}{dt}(t) =i m\sum_{\lambda_k}\frac{1}{\lambda_k^2(t)+m^2}\;\frac{d\lambda_k}{dt}.
\eeq
Exponentiating the factor of $(\lambda_k^2+m^2)^{-1}$ and using $\frac{d\lambda_k}{dt} = \langle \psi_k(t)|i\frac{d\slashed{\msD}_{2n-1}}{dt}(t)|\psi_k(t)\rangle$, we therefore find
\beq
\int_{-\infty}^{\infty}dt\;\frac{d}{dt}S_{odd}(t) = -m\int_{-\infty}^{\infty}dt\int_{0}^{\infty}ds\;\mTr_{2n-1}\frac{d\slashed{\msD}_{2n-1}}{dt}e^{-s\left(m^2-\slashed{\msD}_{2n-1}^2(t)\right)}\label{effac1}
\eeq   
where \(\mTr_{2n-1}\) is the trace over the spectrum of \(\slashed{\msD}_{2n-1}(t)\). 

On the other hand, consider the \(2n\)-dimensional Dirac operator \(\slashed{\msD}_{2n}\) on the space \(M_{2n-1}\times \re\) given by\footnote{Here we take the Clifford matrices on \(M_{2n-1}\times \re\) to be \(\Gamma^0 = \sigma^1\otimes 1,\;\Gamma^A = \sigma^2\otimes \gamma^A\)}
\beq
\slashed{\msD}_{2n}=\sigma^1\otimes \frac{d}{dt}+\sigma^2\otimes \slashed{\msD}_{2n-1}(t).
\eeq
The square of \(\slashed{\msD}_{2n}\) is easily computed  
\beq
\slashed{\msD}_{2n}^2=\frac{d^2}{dt^2}+i\sigma^3\otimes\frac{d\slashed{\msD}_{2n-1}}{dt}+\slashed{\msD}_{2n-1}^2(t).
\eeq
Also note that the \(2n\)-dimensional chirality operator is given by \(\Gamma^{2n+1} = \sigma^3\otimes 1\). Now, define a \(2n\)-form \(\mathcal{P}^{(0)}(m)\) on \(M_{2n-1}\times \re\) by
\beq
\int_{M_{2n-1}\times \re} \mathcal{P}^{(0)}(m) = im\sqrt{\pi}\int_0^{\infty}ds\;s^{-1/2}\mTr_{2n}\Gamma^{2n+1}e^{-s\left(m^2-\slashed{\msD}_{2n}^2\right)}\label{unregpol}
\eeq
where \(\mTr_{2n}\) is trace over the spectrum of \(\slashed{\msD}_{2n}\) defined on \(M_{2n-1}\times \re\). Notice that \(\mTr_{2n}\Gamma^{2n+1}e^{s\slashed{\msD}_{2n}^2}\) is the integral over \(M_{2n-1}\times \re\) of the Atiyah-Singer index density, which is \emph{locally} exact. Since \(M_{2n-1}\) is taken to be without-boundary, \(\mathcal{P}^{(0)}(m)\) is a total derivative in \(t\). Using the assumption of adiabaticity we may carry out the trace in the \(t\)- direction to obtain 
\beq
\int_{M_{2n-1}\times \re} \mathcal{P}^{(0)}(m) =-m\int_{-\infty}^{\infty}dt \int_0^{\infty}ds\;\mTr_{2n-1}\frac{d\slashed{\msD}_{2n-1}}{dt}e^{-s\left(m^2-\slashed{\msD}_{2n-1}^2\right)}+\cdots
\eeq
where \(\cdots\) indicate terms with three or more $t$-derivatives. These terms drop out because the background fields are asymptotically \(t\)-independent (see Eq. \eqref{backfields}). Comparing with \eqref{effac1}, we conclude that
\beq
S_{odd}[e,\omega, A] - S_{odd}[e_{(0)},\omega_{(0)},A_{(0)}]=\int_{M_{2n-1}\times \re} \mathcal{P}^{(0)}(m).\label{podd3}
\eeq
Therefore, the parity odd fermion effective action \(S_{odd}[e,\omega,A]\) in \(d=2n-1\) may be interpreted as the ``Chern-Simons'' form correponding to the locally exact index polynomial \(\mathcal{P}^{(0)}(m)\) defined in \(2n\) dimensions. We will refer to \(\mathcal{P}^{(0)}(m)\) as the \emph{anomaly polynomial}. 

We will mainly focus on computing \(S_{odd}[e,\omega,A]\) in the limit where the mass scale \(|m|\) is taken to be much larger than all background curvature and torsion scales. Our general strategy to compute \(\mathcal{P}^{(0)}(m)\) in this limit will be as follows: in the limit \(s\mapsto 0\), there exists an \emph{asymptotic expansion}
\beq
\mTr_{2n}\Gamma^{2n+1}e^{s\slashed{\msD}_{2n}^2}\simeq \sum_{k=0}^{\infty}b_k s^{-n/2+k}
\eeq 
where the \(b_k\) are integrals over \(M_{2n}\) of polynomials in curvature, torsion, and their covariant derivatives. The important point is that it suffices to use this asymptotic expansion in order to extract terms in \eqref{unregpol} which survive in the limit where \(|m|\) is taken to be much larger than all background curvature and torsion scales. Unfortunately, as will become clear soon, the anomaly polynomial as defined above is divergent if the background spin connection is torsional. These are the same divergences that one would encounter in a direct computation of the \(2n-1\) dimensional parity odd effective action (for instance, by using Feynman diagrams) in the presence of background torsion. In order to remedy the situation, we introduce \(N\) Pauli-Villar's regulator fermions with coefficients \(C_i\) and masses \(M_i\), with \(i=1,2 \cdots N\). For convenience, we label \(C_0 =1\) and \(M_0 = m\). We then define the regularized anomaly polynomial
\beq
\mathcal{P}(m) = \sum_{i=0}^NC_i\mathcal{P}^{(0)}(M_i). \label{reganompol}
\eeq
The \(C_i\)'s and \(M_i\)'s may be determined by requiring UV finiteness. In a condensed matter context this type of regulator is natural in simple lattice Dirac models which are often used to describe topological insulators. These models contain massive spectator Dirac fermions at locations in the Brillouin zone far away from the region which contains the low-energy fermion(s). Indeed, upon including the spectator fermions of the lattice Dirac model (interpreted as Pauli-Villar's regulator fermions), the anomaly polynomial $\mathcal{P}(m)$ becomes finite in arbitrary even dimension; we postpone the proof to appendix B.

Since the anomaly polynomial is the (exterior) derivative of the parity odd effective action in \(2n-1\) dimensions, it encodes the \(2n-1\) dimensional transport coefficients for the two gapped phases. Furthermore, as has been explained in \cite{AlvarezGaume:1984nf, Hughes:2012vg}, covariant anomalies of the \(2n-2\)-dimensional edge theory can be extracted out of the fermion effective action in \(d=2n-1\) by computing Hall-type currents passing between the edges through the bulk. In this way, \(\mathcal{P}(m)\) encodes all the anomalies of the \(2n-2\) dimensional edge theory. Let us now apply the above formalism to explicitly compute the parity odd terms in the fermion effective actions in \(d=2+1\) and \(d=4+1\). 
\subsection{$d=2+1$}
We first begin with the asymptotic expansion (see Appendix \ref{app:asymptotic})
\beq
\mTr_4\;\Gamma^5e^{s\slashed{\msD}^2}\simeq\int_{M_3\times \re}\left(\frac{\torcplq}{16\pi^2 s}dH+\frac{1}{192\pi^2}\mtr\;R^{(-\torcplq)}\w R^{(-\torcplq)}+\frac{1}{8\pi^2}F\w F+\frac{\torcplq}{96\pi^2}d*d*dH+O(s)\right) \label{ind4}
\eeq
where we recall that \(H=e^A\w T_A\), and we have defined $R_{AB}^{(-\torcplq)}$ to be the curvature 2-form for the connection
\beq
\omega_{AB}^{(-\torcplq)} = \lcw_{AB}+\frac{\torcplq}{2}H_{ABC}e^C.
\eeq
The terms higher order in \(s\) may be ignored as they give rise to negative powers of \(m\). We may also drop the last term in \eqref{ind4} as it necessarily contains three or more \(t\)-derivatives, and does not pull back to the boundary for asymptotically $t$-independent backgrounds, as explained in the previous section. The unregulated polynomial \eqref{unregpol} is then given by
\beq
\mathcal{P}^{(0)}(m) =\frac{i\zeta_H^{(0)}}{2}dH+\frac{i\kappa_H^{(0)}}{2}\mtr\;R^{(-\torcplq)}\w R^{(-\torcplq)}+\frac{i\sigma_H^{(0)}}{2}F\w F.
\eeq
The unregulated transport coefficients may be computed from \eqref{unregpol} and \eqref{ind4} 
\newcommand{\sgn}{\mathrm{sign}}
\beqn
\zeta_H^{(0)}(m) &=& -\frac{\torcplq}{4\pi}\left[-\frac{m}{\sqrt{\pi\epsilon}}+\sigma_0m^2\right]\nonumber\\
\kappa_H^{(0)}(m) &=& \frac{1}{96\pi}\sigma_0\nonumber\\
\sigma_H^{(0)}(m) &=& \frac{q^2}{4\pi}\sigma_0
\eeqn
where \(\sigma_0 = \sgn(m)\), and \(\frac{1}{\sqrt{\epsilon}} \sim \Lambda\) is the UV cutoff. Introducing the Pauli-Villar's regulator fermions, and requiring finiteness in the limit \(\epsilon \mapsto 0\), we are led to the constraints 
\beq
\sum_{i=0}^NC_i = 0,\;\;\sum_{i=0}^NC_iM_i = 0.
\eeq
Even without the UV divergent term this action would need to be regularized  due to the fact that the Hall conductivity $\sigma_H^{(0)}(m)$ is not an integer multiple of $\tfrac{q^2}{2\pi}$ as it must be for a non-interacting system\cite{Haldane:1988zza}.  One possible choice for  \(\{C_i\}\) and \(\{M_i\}\) that solves the constraints can be inferred from the spectator fermion structure of the 2+1-d lattice Dirac model\cite{creutz} where 
\begin{center}
\begin{tabular}{|c|c|}
\hline
$M_i$ & $C_i$ \\\hline
$m$ & +\\
$m+2\Delta$ & -\\
$m+2\Delta$ & -\\
$m+4\Delta$ & + \\\hline
\end{tabular}
\end{center}
where the energy scale $\Delta$ is a large energy scale with \(|m| <<\Delta << \Lambda\). The regulated anomaly polynomial is then given by\footnote{We have also cancelled out a \(\sigma_0\)-independent (and hence independent of whether or not the system is in the topological or trivial phase) divergence proportional to \(dH\) by adding a counterterm. Such a counterterm is required only in \(d=2+1\), and not in higher dimensions.}
\beq
\mathcal{P}(m) =\frac{i\zeta_H}{2}dH+\frac{i\kappa_H}{2}\mtr\;R^{(-\torcplq)}\w R^{(-\torcplq)}+\frac{i\sigma_H}{2}F\w F
\eeq
with the regulated transport coefficients 
\beqn
\zeta_H &=& \frac{\torcplq m^2}{2\pi}\frac{1-\sigma_0}{2}\nonumber\\
\kappa_H &=& \frac{1}{48\pi}\frac{1-\sigma_0}{2}\nonumber\\
\sigma_H &=& \frac{q^2}{2\pi}\frac{1-\sigma_0}{2}.
\eeqn
Since the anomaly polynomial is  a total derivative, we may read off the parity odd effective action from the above as the corresponding Chern-Simons form
\beqn
S_{odd}[e,\omega,A]&=&\frac{i}{2}\int_{M_3}\left(\zeta_H\;e^A\wedge T_A+\sigma_H A\wedge dA\right.\nonumber\\
&+&\left.\kappa_H\;\mtr(\omega^{(-\torcplq)}\wedge d\omega^{(-\torcplq)}+\frac{2}{3}\omega^{(-\torcplq)}\wedge \omega^{(-\torcplq)}\wedge \omega^{(-\torcplq)})\right)
\eeqn
Expanding \(S_{odd}\) to linear order in torsion, we find
\beqn
S_{odd}[e,\omega,A]&=&\frac{i}{2}\int_{M_3}\left(\sigma_H A\wedge dA+\kappa_H\mtr(\lcw\wedge d\lcw+\frac{2}{3}\lcw\wedge\lcw\wedge\lcw)\right.\nonumber\\
&+&\left.\zeta_H\;e^A\wedge T_A-\torcplq\kappa_H\;\lcR e^A\wedge T_A+\cdots\right). \label{podd3lin}
\eeqn
which is the same action that was derived in \cite{Hughes:2012vg} by a more direct computation. It might seem odd that the coefficient of the $e^A\wedge T_A$ term is a dimensionful parameter, as opposed to the other coefficients, which are universal and quantized. We note that this is not an obstacle to gauge invariance: the quantization of both $\sigma_H$ and $\kappa_H$ is forced upon us by the requirement of gauge invariance under large gauge transformations. The $e^A\wedge T_A$ term on the other hand, is \emph{globally} well-defined (i.e., gauge, Lorentz, and diffeomorphism invariant), and hence requires no such quantization of it's coefficient.  

We now focus on the physics of the torsional terms. The \(\zeta_H\; e^A\wedge T_A\) term has the interpretation of \emph{Hall viscosity}, as has been explained in \cite{Hughes:2011hv,Hughes:2012vg}. Here we wish to delve a bit into the curvature correction \(\lcR\;e^A\wedge T_A\) since similar terms will appear in higher dimensions. We may loosely interpret this term as a local-curvature dependent Hall viscosity. On a space-time of the form $\mathbb{R}\times \Sigma$, with $\Sigma$ a constant curvature Riemann surface of Euler characteristic $\chi_\Sigma$ and area $A$, terms linear in torsion in \eqref{podd3lin} become
\beq\label{Sodd}
S_{odd}[e,\omega,A]= \frac{i}{2}\left(\zeta_H-\frac{4\pi\torcplq\kappa_H\chi_{\Sigma}}{A}\right) \int  e^A\wedge T_A.  
\eeq
For curvature and area preserving deformations of the co-frame, we thus find a \emph{shift} in the effective Hall viscosity $\mbox{\boldmath$\zeta_{H}$}$ relative to its flat space value 
\beq
\mbox{\boldmath$\zeta_{H}$} = \zeta_H -\frac{4\pi\kappa_H\chi_{\Sigma}}{A}.
\eeq
This effect is reminiscent of the Wen-Zee shift of the number density in a quantum Hall fluid in the presence of curvature. In fact, let us define the \emph{spin density} \(\mathfrak{s}\) of the Chern insulator as
\beq
\mathfrak{s} = \frac{1}{A}\int_{\Sigma} *J^{12}
\eeq
where \(J^{12}\) is the spatial component of the spin current \(J^{AB}\). To lowest order in torsion, this may be computed from the action\footnote{In particular, $J^{AB}$ is obtained by varying with respect to $\omega_{AB}$, holding $e^A$ fixed.} (\ref{Sodd}), and we see that the local spin density is also affected by the local curvature, and in fact satisfies 
\beq
\mbox{\boldmath$\zeta_{H}$} = -\mathfrak{s}. \label{readresult}
\eeq
Thus, the shift due to curvature may be interpreted as a shift in the spin density relative to its flat space value. Equation \eqref{readresult} is similar to the relation between Hall viscosity and spin presented in \cite{Read:2008rn,PhysRevB.84.085316}.

Although we will not consider them in this paper, we note that for $d=2+1$, the parity-even terms can similarly be computed with careful regularization. The complete effective action then arranges into chiral gravity, namely an $SL(2,\re)$ Chern-Simons term \cite{Hughes:2012vg}.
\subsection{$d=4+1$}
Since the primary goal of this article is to discuss $3+1$-d systems, let us now repeat the above analysis for \(d=4+1\), which we will subsequently use to determine the properties of $3+1$-d chiral fermions, and $3+1$-d time-reversal invariant topological insulators. We begin with the corresponding \(6\)-dimensional asymptotic expansion 
\beqn
\mTr_6\;\Gamma^7e^{s\slashed{\msD}_6^2} &\simeq& \int_{\mathbb{R}\times M_5}\left(-\frac{\torcplq}{32\pi^3 s}F\w dH-\frac{1}{384\pi^3}F\w \mtr\;R^{(-\torcplq)}\wedge R^{(-\torcplq)}-\frac{1}{48\pi^3}F\w F\w F\right.\nonumber\\
&-&\left. \frac{\torcplq}{192\pi^3}d\left(F\wedge *d*dH\right)+\frac{\torcplq}{384\pi^3 }d*d*(F\w dH)+O(s)\right).\label{ind6}
\eeqn
We do not consider \(O(s)\) terms as they lead to inverse powers of \(m\), and are generally of higher order in the curvature/torsion expansion. As before, we may also drop the last term in \eqref{ind6}, as it does not pull back to the boundary effective action. The unregulated anomaly polynomial is then easily obtained
\beq
\mathcal{P}^{(0)}(m)= \frac{i\zeta_H^{(0)}}{2}F\w dH + \frac{i\kappa_H^{(0)}}{2}F\wedge \mtr\;R^{(-\torcplq)}\wedge R^{(-\torcplq)}+\frac{i\sigma_H^{(0)}}{3}F\wedge F\wedge F+\frac{i\lambda^{(0)}}{2}d\left(F\w *d*dH\right)
\eeq
with the unregulated transport coefficients
\beqn
\zeta_H^{(0)}(m) &=& -\frac{q\torcplq}{8\pi^2}\left[-\frac{m}{\sqrt{\pi\epsilon}}+\sigma_0m^2\right]\nonumber\\
\kappa_H^{(0)}(m) &=& \frac{q}{192\pi^2}\sigma_0\nonumber\\
\sigma_H^{(0)}(m) &=& \frac{q^3}{16\pi^2}\sigma_0\nonumber\\
\lambda^{(0)}(m) &=& \frac{q\torcplq}{96\pi^2}\sigma_0.
\eeqn
The structure of divergences is the same as previously encountered in \(2+1\) dimensions - namely a linear divergence. In fact, more generally the structure of divergences (i.e. linear, quadratic etc.) of the parity-odd effective action is identical in \(d= 4n-1\) and \(d=4n+1\) (see Appendix \ref{app:expansion} for more details). Therefore, it suffices to use the Pauli-Villar's regulators we used in \(d=2+1\), which gives the regulated anomaly polynomial
\beq
\mathcal{P}(m)= \frac{i\zeta_H}{2}F\wedge dH  + \frac{i\kappa_H}{2}F\wedge \mtr\;R^{(-\torcplq)}\wedge R^{(-\torcplq)}+\frac{i\sigma_H}{3}F\wedge F\wedge F+\frac{i\lambda}{2}d\left(F\w *d*dH\right)
\eeq
with the regulated transport coefficients 
\beqn
\zeta_H &=& \frac{q\torcplq m^2}{4\pi^2}\frac{1-\sigma_0}{2}\nonumber\\
\kappa_H &=& \frac{q}{96\pi^2}\frac{1-\sigma_0}{2}\nonumber\\
\sigma_H &=& \frac{q^3}{8\pi^2}\frac{1-\sigma_0}{2}\nonumber\\
\lambda &=& \frac{q\torcplq}{48\pi^2}\frac{1-\sigma_0}{2}.
\eeqn
The parity odd effective action in \(d=4+1\) is then given by
\beqn
S_{odd}[e,\omega,A]&=&\frac{i}{2}\int_{M_5}\left(\zeta_H\;F\w e^A\w T_A+\frac{2\sigma_H}{3}\;A\w F\w F\right.\label{podd51}\\
&+&\left.\kappa_H\;F\w \mtr\;(\omega^{(-\torcplq)}\wedge d\omega^{(-\torcplq)}+\frac{2}{3}\omega^{(-\torcplq)}\w\omega^{(-\torcplq)}\w\omega^{(-\torcplq)})+\lambda\;F\w *d*dH\right).\nonumber 
\eeqn
As before, we stress that this should be regarded as giving rise to the leading (in powers of $|m|$) parity-violating terms in correlation functions of the charge, stress, and spin currents.
Once again, we may expand this to linear order in torsion to obtain
\beqn
&=&\frac{i}{2}\int_{M_5}\left(\frac{2\sigma_H}{3}\;A\w F\w F+\kappa_H\;F\wedge\mtr(\lcw\wedge d\lcw+\frac{2}{3}\lcw\wedge\lcw\wedge\lcw)\right. \\
&+&\left.\zeta_H\;F\w e^A\w T_A-\torcplq\kappa_H\;(\lcR\;F+2F_C\wedge \lcR^C+F^{CD}\lcR_{CD})\wedge e^A\w T_A+\lambda\;F\w*d*dH+\cdots \right)\nonumber
\eeqn
where we have introduced the notation \(F_A = F(\ue_A),\;F_{AB} = F(\ue_A,\ue_B), \;\lcR_B= \lcR_{AB}(\ue^A)\) and so on. 

Let us focus on the second line above. The term proportional to \(\zeta_H\) now represents a \emph{magneto-Hall viscosity}, which is to say a dissipationless viscosity in the presence of a magnetic flux through perpendicular spatial dimensions. To be more explicit, let us consider a simple example where we take the space-time manifold to be of the form \(M_5 = \mathbb{R}\times \Sigma\times \widetilde{\Sigma}\), with \(\Sigma\) and \(\widetilde{\Sigma}\) being two constant curvature Riemann surfaces with areas \(A\) and \(\widetilde A\). 
\begin{figure}[!t]
\centering
\includegraphics[width=13cm]{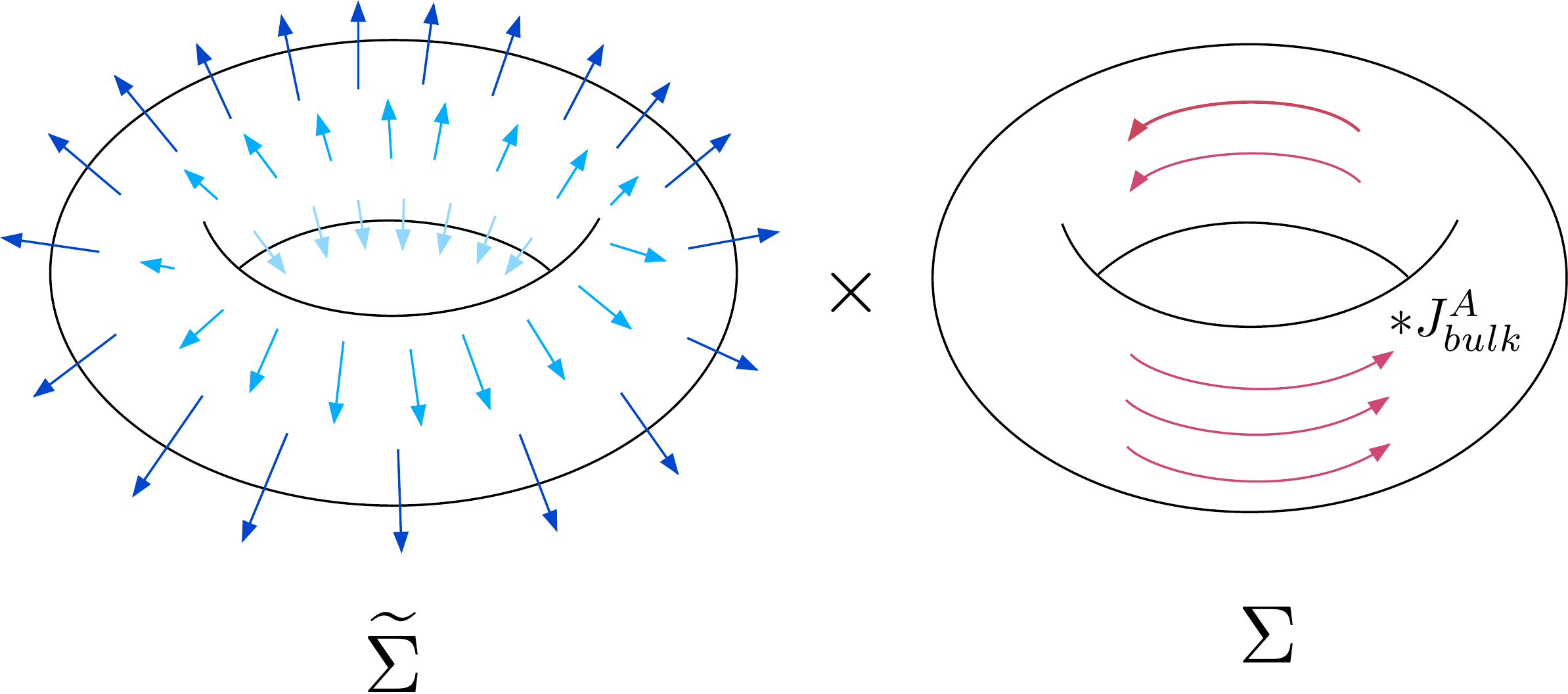}
\caption{An illustration describing the field setup for a magneto-Hall viscosity response: turning on a $U(1)$ flux through $\widetilde{\Sigma}$ gives rise to a Hall viscosity response on $\Sigma$.}
\end{figure}
If we turn on a \(U(1)\) magnetic flux of \(F= \frac{2\pi n}{q\widetilde{A}}vol_{\widetilde{\Sigma}}\) through \(\tSigma\) (for \(n\in \mathbb{Z}\)), then the effective dissipationless viscosity for co-frame deformations in the orthogonal surface \(\Sigma\) is given by 
\beq
\boldsymbol{\zeta}_H=n\frac{\torcplq m^2}{2\pi}\frac{1-\sigma_0}{2}.
\eeq
Just as in $2+1$-d, we also have curvature dependent corrections to the effective magneto-Hall viscosity. For the choice of \(M_5\) and \(F\) we are working with, the terms linear in torsion simplify to give us the following effective action on the subspace \(\Sigma\)
\beq
S_{odd}(\Sigma) = \frac{i}{2}\int_{\mathbb{R}\times \Sigma}\left\{\boldsymbol{\zeta}_H-\frac{\torcplq}{q}\kappa_H\left(2\pi n\lcR + \frac{32\pi^2 n\chi_{\tSigma}}{\widetilde{A}}\right)\right\}e^A\wedge T_A.
\eeq 
As before, if we restrict ourselves to curvature and area-preserving co-frame deformations on \(\Sigma\), we find that the effective magneto-Hall viscosity gets shifted from its flat space value to
\beq
\boldsymbol{\zeta}_H\mapsto \boldsymbol{\zeta}_H-\frac{\torcplq}{q} \kappa_H\left(\frac{32\pi^2n\chi_{\tSigma}}{\widetilde{A}}+ \frac{8\pi^2n\chi_{\Sigma}}{A}\right).
\eeq
Once again, the shift in the magneto-Hall viscosity may be interpreted as a shift in the spin density on \(\Sigma\) relative to the flat space value.

With the completed derivation of the $4+1$-d parity-violating terms in the effective action we are now ready to explore measurable consequences in real condensed matter systems. In the next two sections we will first consider the properties of isolated 3+1-d boundary chiral fermions and then discuss some aspects of the response properties of Weyl semi-metals that result from these effects. Finally we will discuss the dimensional reduction of the $4+1$-d action to $3+1$-d that will determine the response properties of the $3+1$-d time-reversal invariant topological insulator.

\section{Callan-Harvey Anomaly Inflow and Boundary Chiral Anomalies}\label{sec:callanharvey}

To study the properties of isolated chiral fermions, or pairs of chiral fermions in a Weyl semi-metal, we must consider their anomaly structure. One nice way to organize the anomalous currents is to consider the low-energy  chiral modes which are localized on an interface between topological and trivial phases in odd space-time dimensions. The case of 1+1 dimensional edge modes on the surface (interface between the non-trivial topological phase and the trivial vacuum) of a 2+1 dimensional topological insulator was discussed in detail in \cite{Hughes:2012vg}. Here we will deal with the case of 3+1 dimensional boundary modes, and their relationship with the 4+1 dimensional parity-odd transport coefficients described in the previous section. 

Consider then the non-trivial phase labelled by transport coefficients \((\sigma_H, \zeta_H, \kappa_H, \lambda)\) on a 4+1 dimensional manifold \(M_5\), separated from the trivial phase by a 3+1 dimensional interface \(\Sigma_4=\partial M_5\). One model for this system is a 4+1 dimensional Dirac fermion with mass \(m < 0\) on \(M_5\), and \(m > 0\) outside, with some interpolation region, the interface \(\Sigma_4\), which we refer to as the domain wall. In general, there could be multiple fermions with mass domain walls along \(\Sigma_4\), and their number decides \((\sigma_H,\zeta_H,\kappa_H,\lambda)\). The domain wall hosts 3+1-d chiral fermions, whose anomalies will encode the differences in \((\sigma_H,\zeta_H,\kappa_H,\lambda)\) between opposite sides of the domain wall. 

In order to avoid complicating our discussion, we will first explain the anomaly inflow only focusing on the first two terms in \eqref{podd51}, and later present the more general result. We start with the 4+1-d bulk effective action
\begin{equation}
S_{bulk} = i\frac{\sigma_H}{3}\int_{M_5}A\wedge F\wedge F+i\frac{\zeta_H}{2}\int_{M_5} F\wedge H \label{podd2}
\end{equation}
where we recall the notation \(H = e^A\wedge T_A\). The first term is the second (Abelian) Chern-Simons form and is diffeomorphism and Lorentz invariant, but not \(U(1)\) invariant. This gauge non-invariance must be compensated by the \emph{consistent} anomaly of the boundary/interface theory. This means that the boundary effective action \(S_{bdry}\) cannot be gauge invariant either. In fact, under a \(U(1)\) gauge transformation \(\delta A = d\alpha\), we must have
\beq
\delta_{\alpha} S_{bdry} = -\frac{i\sigma_H}{3}\int_{\Sigma_4}\alpha F\wedge F. 
\eeq
in order to cancel the gauge variation of the bulk Chern-Simons term. Interestingly, the second term in \eqref{podd2}  is gauge, diffeomorphism, and Lorentz invariant despite its similarity to the first term, and hence we do not expect it to contribute to consistent anomalies in the boundary. This is an important distinction between the two terms. Using these constraints, the consistent Ward identities on the boundary are\footnote{Note that the right hand side of equation \eqref{consdiff} originates from the fact that this Ward identity corresponds to a \emph{covariant} diffeomorphism, which involves an ordinary diffeomorphism plus a $U(1)$ and local Lorentz gauge transformation.}
\beq
d*J_{cons} = \frac{\sigma_H}{3}F\wedge F
\eeq
\beq
D*J^a_{cons} -i_{\ue^a}T_b\wedge *J^b_{cons}-i_{\ue^a}R_{bc}\wedge *J^{bc}_{cons}-i_{\ue^a}F\wedge *J_{cons}=-\frac{\sigma_H}{3}i_{\ue^a}A\wedge F\wedge F\label{consdiff}
\eeq
\beq
D*J^{ab}_{cons}+e^{[b}\wedge *J^{a]}_{cons}=0
\eeq
where lower-case Latin indices are local Lorentz indices on the boundary manifold \(\Sigma_4\). The Ward identities written in terms of consistent currents are clearly not gauge covariant since they depend on gauge-variant fields like the vector-potential $A.$ To remedy the situation, we must write these in terms of \emph{covariant currents}. Consider then, the variation of the bulk response action\footnote{Here we will assume that the boundary values of the variations $\delta e^A$ and $\delta\omega_{AB}$ are non-zero only when the Lorentz indices are those of the boundary. In other words, we are ignoring extrinsic effects here.}  
\beq
\delta S_{bulk} = \int_{M_5}\left(\delta A\wedge *J_{bulk}+\delta e_A\wedge *J^A_{bulk}+\delta \omega_{AB}\wedge *J_{bulk}^{AB}\right) +\int_{\Sigma_4}\left(\delta A\wedge *j+\delta e^a\wedge *j_a+\delta \omega_{ab}\wedge *j^{ab}\right)
\eeq
The conserved Hall currents in the bulk are given by 
\begin{subequations}\label{Hallcurrents}
\begin{align}
*J_{bulk} &= \sigma_HF\wedge F+\frac{\zeta_H}{2}dH\label{Hallcurrents1}\\
*J^A_{bulk} & = \zeta_H F\wedge T^A \label{Hallcurrents2'}\\
*J^{AB}_{bulk} & = -\frac{\zeta_H}{2} F\wedge e^A\wedge e^B \label{Hallcurrents3}
\end{align}
\end{subequations}
while the induced currents in the boundary are
\begin{subequations}\label{boundarycurrents}
\begin{align}
*j &=\frac{2}{3}\sigma_HA\w F+\frac{\zeta_H}{2} H\\
*j^a &= \frac{\zeta_H}{2} F\wedge e^a\\
*j^{ab} &= 0.
\end{align}
\end{subequations}
Define the covariant boundary currents \(J_{cov} = J_{cons} + j\), \(J_{cov}^a = J^a_{cons} + j^a\), and \(J^{ab}_{cov} = J^{ab}_{cons}+j^{ab}\). Then the Ward identities written in terms of these are
\begin{equation}
d*J_{cov} = \sigma_HF\wedge F + \frac{\zeta_H}{2}dH
\end{equation}
\begin{equation}
D*J_{cov}^a -i_{\ue^a}T_b\wedge *J_{cov}^b-i_{\ue^a}R_{bc}\wedge *J_{cov}^{bc}-i_{\ue^a}F\wedge *J_{cov}=\zeta_H F\wedge T^a
\end{equation}
\begin{equation}
D*J_{cov}^{ab}+e^{[b}\wedge *J_{cov}^{a]}=-\frac{\zeta_H}{2}F\wedge e^a\wedge e^b.
\end{equation}
These are referred to as the \emph{covariant anomalies} in the boundary theory. Notice that these precisely match the fluxes of bulk Hall currents \eqref{Hallcurrents} into \(\Sigma_4\)
\begin{subequations}
\begin{align}
\Delta Q &= \sigma_H\int_{\Sigma_4}F\wedge F+\frac{\zeta_H}{2}\int_{\Sigma} dH\\
\Delta Q^a & = \zeta_H\int_{\Sigma_4} F\wedge T^a\\
\Delta Q^{ab} & = -\zeta_H\int_{\Sigma_4} F\wedge e^a\wedge e^b.
\end{align}
\end{subequations}
Thus, the charge, momentum, and spin injected into the edge from the bulk are carried by the \emph{covariant} currents $J_{cov},  J^{a}_{cov},$ and  $J^{ab}_{cov}$ respectively. 

Having described the general idea of anomaly inflow in a simpler setting, we now give the full result for edge anomalies. Applying the same ideas discussed above to the full effective action \eqref{podd51}, we get the flux of bulk charge, stress, and spin currents into the edge 
\begin{subequations}\label{fullanomalies}
\begin{align}
\Delta Q &= \int_{\Sigma_4}\left(\sigma_H F\wedge F+\frac{\zeta_H}{2} dH+\frac{\kappa_H}{2}\mathrm{tr}\;R^{(-\torcplq)}\wedge R^{(-\torcplq)}+\frac{\lambda}{2}d*d*dH\right) \\
\Delta Q^a & = \int_{\Sigma_4} \left(\zeta_H F\wedge T^a+\kappa_H\;e^a\wedge d\mathcal{A}_2-\torcplq\kappa_H\; \mathcal{A}_2\wedge T^a+\lambda\; d*d*F\wedge T^a\right)\\
\Delta Q^{ab} & = -\int_{\Sigma_4}\left(\frac{\zeta_H}{2} F-\frac{\torcplq\kappa_H}{2}\;\mathcal{A}_2+\frac{\lambda}{2}\;d*d*F\right)\wedge e^a\wedge e^b
\end{align}
\end{subequations}
where we have defined
\beq
\mathcal{A}_2= \left(F\wedge R^{(-\torcplq)}_{ab}\right)(\ue^a,\ue^b)=\left(F^{ab}R^{(-\torcplq)}_{ab}+2F^a\wedge R^{(-\torcplq)}_a+R^{(-\torcplq)}F\right).
\eeq
These are the covariant \(U(1)\), diffeomorphism, and Lorentz anomalies of the edge theory in the presence of curvature. Note the appearance of the dimensionful viscosity term $\frac{\zeta_H}{2}\;dH$ in the chiral $U(1)$ anomaly. This might seem problematic given the topological character of the (integrated) chiral anomaly. However, note that $H$ is a globally well defined 3-form (unlike, for instance $A\wedge dA$), and $dH$ is truly a total derivative. On compact 4-manifolds then, this term drops out. On the physics side, we are interested in the \emph{local} anomaly densities -- which is why it is important for us to keep this term. In fact, this term is precisely the Nieh-Yan term discussed earlier, and it now has a clear meaning in the present context: its coefficient is the difference of magneto-Hall viscosities across a 3+1-d interface between two different topological phases.

Using the structure of the anomalous terms presented here, we will now go on to show the microscopic origin of a subset of the  anomalous currents using spectral-flow type arguments in the Hamiltonian formalism of the chiral boundary states. This will clarify the physical origin of the terms in which we are most interested, and will give a nice interpretation for some of the torsional contributions to the anomalous currents.

\section{Spectral flow}\label{sec:spectralflow}
In this section we will discuss the covariant anomalies of the boundary theory from the point of view of adiabatic spectral flow of the Hamiltonian spectrum of chiral boundary states. We will first review the well-known case of  the 4+1-d Hall conductivity and spectral flow induced by \(U(1)\) fluxes, and then move on to  magneto-Hall viscosity and the chiral anomaly due to torsion. 

\subsection{4+1-d Quantum Hall Effect}
First we will study the effects of the $U(1)$ second Chern-Simons term that enters the response action
\begin{equation}
S_{bulk}=\frac{\sigma_{H}}{3}\int_{M_5}A\wedge F\wedge F.
\end{equation}\noindent This term gives rise to the 4+1-d quantum Hall effect in which a charge current is carried through the bulk in a direction perpendicular to applied electric and magnetic fields. This is reminiscent of the 2+1-d effect where a current is generated perpendicular to an applied electric field. Here we have a non-linear topological response which requires  simultaneous electric and magnetic fields. The reason, of course, is well-known:  the bulk current is intertwined with the boundary chiral anomalies which require parallel electric and magnetic fields on the 3+1-d surface. In 2+1-d the bulk Hall current is also connected with the 1+1-d chiral anomaly on the edge, but in this case the anomalous current is generated in the presence of an electric field alone.

To simplify our discussion let us consider the spatial geometry to be \(\Sigma_3\times [0,L]\), where \(\Sigma_3 = \mathbb{R}\times S^1\times S^1\). We will label the bulk direction by \(w\in [0,L]\), while the coordinates on \(\Sigma_3\) will be labelled by \((x,y,z)\) with \(x\) being the non-compact direction. The edge states will be localized at \(w=0\) and \(w=L\). We turn on a magnetic field \(B\) perpendicular to the surface of the \((x,y)\)-cylinder, and an electric field \(E_z=\frac{2\pi}{qL_zT}\) (for some large and positive time scale \(T\) and with $\hbar=1$). This electric field can be generated by slowly threading magnetic flux through the hole of the \((z,w)\) cylinder. The corresponding gauge field configuration will be chosen to be
\beq
A=Bxdy+E_ztdz
\eeq
where the  \(U(1)\) flux is then given by
\beq
F=Bdx\wedge dy+E_zdt\wedge dz.
\eeq
From the  bulk Chern-Simons response we have the bulk Hall current
\beq
*J_{bulk}=\sigma_H F\wedge F = \frac{q^3}{8\pi^2}BE_zdt\wedge dx\wedge dy\wedge dz.
\eeq
This yields a constant current density through the bulk in the \(w\)-direction and leads to a charge transfer over a time period $T$ of
\beq
\Delta Q = \int_0^T\int_{\Sigma_3}*J_{bulk} = q^2\frac{BL_xL_y}{2\pi}\label{chargetrans}
\eeq
from one edge to the other. Given that the system is in the non-trivial topologically insulating phase, we have a left-handed chiral fermion localized at \(w=0\) and a right-handed chiral fermion localized at \(w=L\). From the boundary point of view, the above charge transfer is an anomalous process, which corresponds to the \(U(1)\) chiral anomaly in the boundary theory
\beq
d*J_{cov}=\sigma_H F\wedge F.
\eeq
Indeed, the anomalous charge created or destroyed on a boundary during the above process is precisely equal to the charge transferred across the bulk of the insulator by the Hall-current, as expected.

We can develop a more intuitive, microscopic picture of the anomaly from the Hamiltonian energy spectra of the chiral boundary states during the adiabatic flux threading process. In the presence of the above gauge field configuration, the low-energy spectrum on the boundary consists of two types of states (see Appendix \ref{app:Weylspectra}): (i) positive and negative energy towers of gapped states
\beq
E(\ell,p_z, \sigma)= \pm\left\{(p_z-qA_z)^2+2|qB|\left(\ell+\frac{1+\sigma}{2}\right)\right\}^{1/2},\;\;\;\ell=1,2,3\cdots,\;\;\sigma=\pm1
\eeq
and (ii) one gapless branch which depends on the chirality
\beq
E_L(p_z,t) = -\mathrm{sign}(qB)\;(p_z-qA_z(t)),\;\;\;E_R(p_z,t)= \mathrm{sign}(qB)\;(p_z-qA_z(t)) \label{gapless}
\eeq
all of which have a degeneracy of \(N = \frac{|q\Phi_B|}{2\pi}\) for every \(p_z\), where \(\Phi_B=BL_xL_y\) is the flux through the surface of the \((x,y)\)-cylinder. For the purpose of our discussion, it suffices to concentrate on the gapless states. Since the \(z\)-direction is compactified on a circle, we may take \(p_z = \frac{2\pi n}{L_z}\;(n\in\mathbb{Z})\) and re-write the gapless branches as
\beq
E_L(p_z,t) = -\mathrm{sign}(qB)\;\frac{2\pi}{L_z}\left(n-\frac{t}{T}\right),\;\;\;E_R(p_z,t)= \mathrm{sign}(qB)\;\frac{2\pi}{L_z}\left(n-\frac{t}{T}\right). \label{gapless2}
\eeq
Here \(T\) is taken to be large, and we assume that the spectrum flows \emph{adiabatically} as a function of time. We will put the chemical potential at \(E=0\) for convenience. 
\begin{figure}[!t]
\centering
\includegraphics[height=6.5cm]{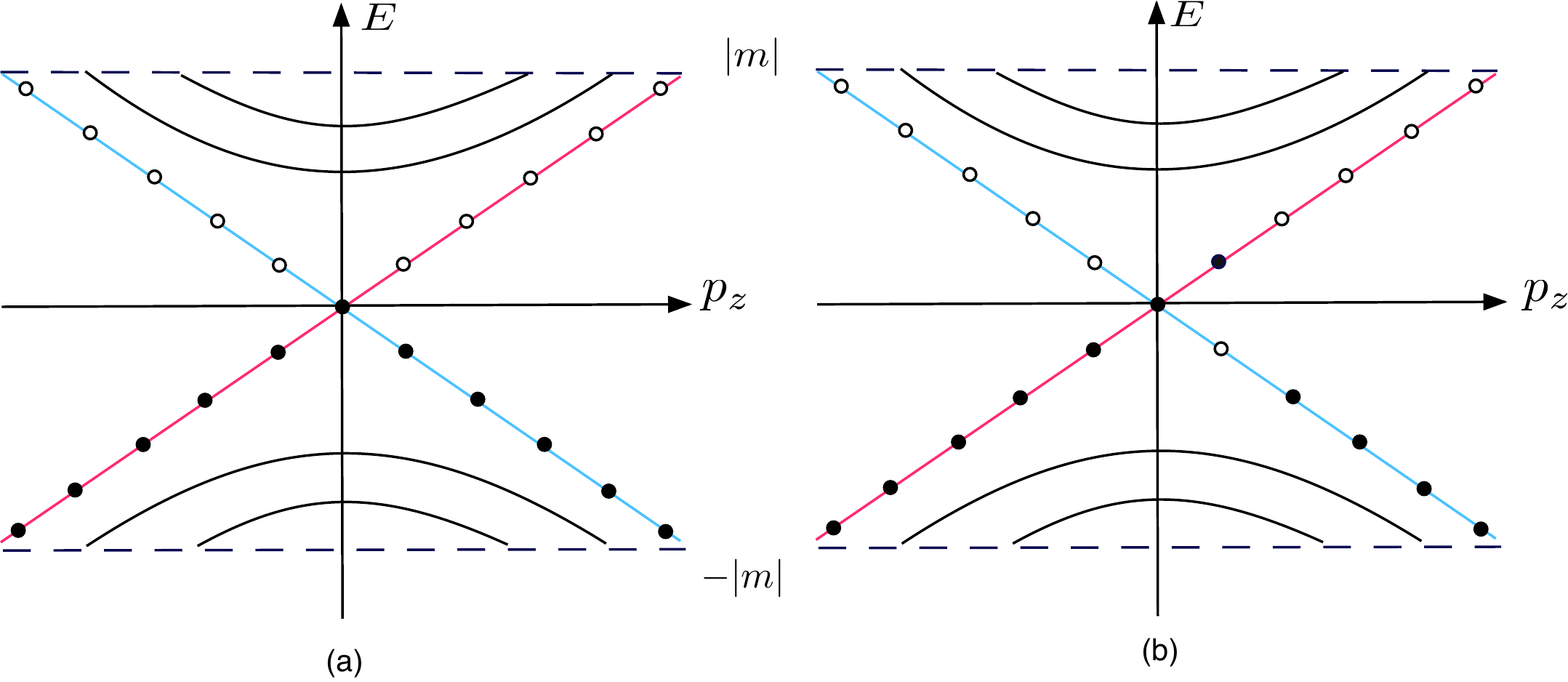}
\caption{The Hamiltonian energy spectrum for chiral fermions in the presence of a uniform background magnetic field in the $z$-direction. The (black) gapped states are higher Landau levels, while the linear gapless (blue, red) curves are the zeroth Landau levels for left and right handed fermions respectively. We can consider the left and right handed fermions to exist on opposite boundaries of a cylinder. Once the energies of the linearly dispersing modes reach $\pm\vert m\vert$ these states are no longer localized on the boundary and lose their sense of chirality. (a) Before an electric field is turned on the states are filled to $E=0$ on both boundaries. (b) After an electric field has acted and a single magnetic flux quantum has been threaded into the cylinder. Spectral flow has modified the level occupations such that one additional level of fermions appear in the right-handed branch and one level of fermions are missing from the left handed branch. }
\end{figure}
If \(\psi(\vec{x},t)\) is the boundary-fermion field operator (with \(\vec{x}=(x,y,z)\)) then the net charge may be defined as
\beq
Q(t)= q\int_{\Sigma_3}d^3\vec{x}\;\frac{1}{2}\;\langle vac|\left[\psi^{\dagger}(\vec{x},t),\psi(\vec{x},t)\right]|vac\rangle=\frac{q}{2}\sum_{\{|E_n| \leq |m|\}}\mathrm{sign}(E_n)
\eeq
where the summation is over all the Hamiltonian eigenstates with \(|E_n|\leq |m|\). The sum only includes these states because at energies beyond the mass gap of the bulk insulator there are no \emph{localized} chiral modes on the boundary. During the flux threading, we find that after a period of time \(t=rT\) for integral \(r\), the spectrum returns to itself, but after a translation by \(r\) units with respect to the chemical potential. In fact, \(r\) is the number of magnetic flux quanta which have been threaded through the hole of the \((w,z)\)-cylinder. For each flux quantum that is threaded, \(N=\frac{|q\Phi_B|}{2\pi}\) states cross the chemical potential, and the charge jumps by \(Nq\) - either increasing or decreasing depending on the chirality. Taking into account the factor of \(\mathrm{sign}(qB)\) in \eqref{gapless2}, we therefore reproduce precisely the charge transfer in Eq \eqref{chargetrans} due to the \(U(1)\) chiral anomaly. 

\subsection{Momentum and Charge Transport from Magneto-Hall Viscosity}\label{sec:torsionAnomaly}
In this section, we will consider the momentum and charge transport due to torsion flux. These transport processes both arise from the term 
\begin{equation}\label{SFviscosity}
S_{bulk}= \frac{\zeta_H}{2}\int_{M_5}F\wedge e^A\wedge T_A. 
\end{equation} 
To simplify the discussion of Hamiltonian spectral flow, we will set $\torcplq=1$ throughout this section. We can determine the momentum current by varying with respect to $e^A$ and the charge current by varying with respect to $A.$ We focus first on the momentum transport by turning on a \(U(1)\) magnetic flux and torsion electric field. To generate the necessary background fields we turn on a \(U(1)\) magnetic field through the \((x,y)\) cylinder using \(A=Bxdy\). We can thread torsion magnetic flux through the hole of the \((z,w)\) cylinder, represented by the co-frame
\beq
e^0=dt,\;\;e^1=dx,\;\;e^2=dy,\;\;e^3=(1+h(t))dz,\;\;e^5=dw
\eeq
where we take \(h(t)=\frac{bt}{L_zT}\), for some large and positive time-scale \(T\). The time-dependent torsion flux threading will generate a circulating torsion electric field in the z-direction. For simplicity, we will set the spin connection\footnote{In particular, we are supposing that the curvature $R^A{}_B$ vanishes. Consequently, $\omega^A{}_B$ is pure gauge, and we are choosing it to be zero here.} \(\omega_{AB}=0\). As a result, the above configuration is torsional with the torsion electric field given by \(T^3= \frac{b}{L_zT}dt\wedge dz\). 
%
%
The bulk stress current from the term \eqref{SFviscosity} in the action, in the presence of our set background fields, is
\beq
*J^3_{bulk} = \zeta_H\;F\wedge T^3 = q\frac{m^2Bb}{4\pi^2L_zT}dt\wedge dz\wedge dx\wedge dy.
\eeq 
In order to compute the momentum transferred due to this current over a time-period $t$, we introduce a \emph{covariant Killing vector field} \(\xi^A\ue_A=\partial_{z}\). Then the rate of momentum transfer from one edge to the other due to the constant stress-current density is
\beq
\frac{dP^3}{dt} = \int_{\Sigma_3}\xi_A*J_{cov}^A = \mathrm{sign}(qB)\frac{m^2 N}{2\pi }\left(1+\frac{bt}{L_zT}\right)\frac{b}{T}\label{momtrans1}
\eeq
where \(N = \frac{|q\Phi_B|}{2\pi}=\frac{|qB|L_xL_y}{2\pi}\). From the boundary point of view, this set of background fields gives rise  to the diffeomorphism anomaly
\beq
d*(\xi_AJ^A_{cov}) = \zeta_H\;F\wedge \xi_AT^A.
\eeq 
In order to understand this from the Hamiltonian point of view, it suffices once again to focus on the gapless boundary state branches for left- and right-handed chiral fermions in the presence of the uniform background magnetic field:
\beq
E_L(p_z,t)=-\mathrm{sign}(qB)\frac{p_z}{\left(1+\frac{bt}{L_zT}\right)},\;\;\; E_R(p_z,t)=\mathrm{sign}(qB)\frac{p_z}{\left(1+\frac{bt}{L_zT}\right)}
\eeq
with degeneracy of \(N=\frac{|q\Phi_B|}{2\pi}\) for every \(p_z\). 
\begin{figure}[!t]
\centering
\includegraphics[height=6.5cm]{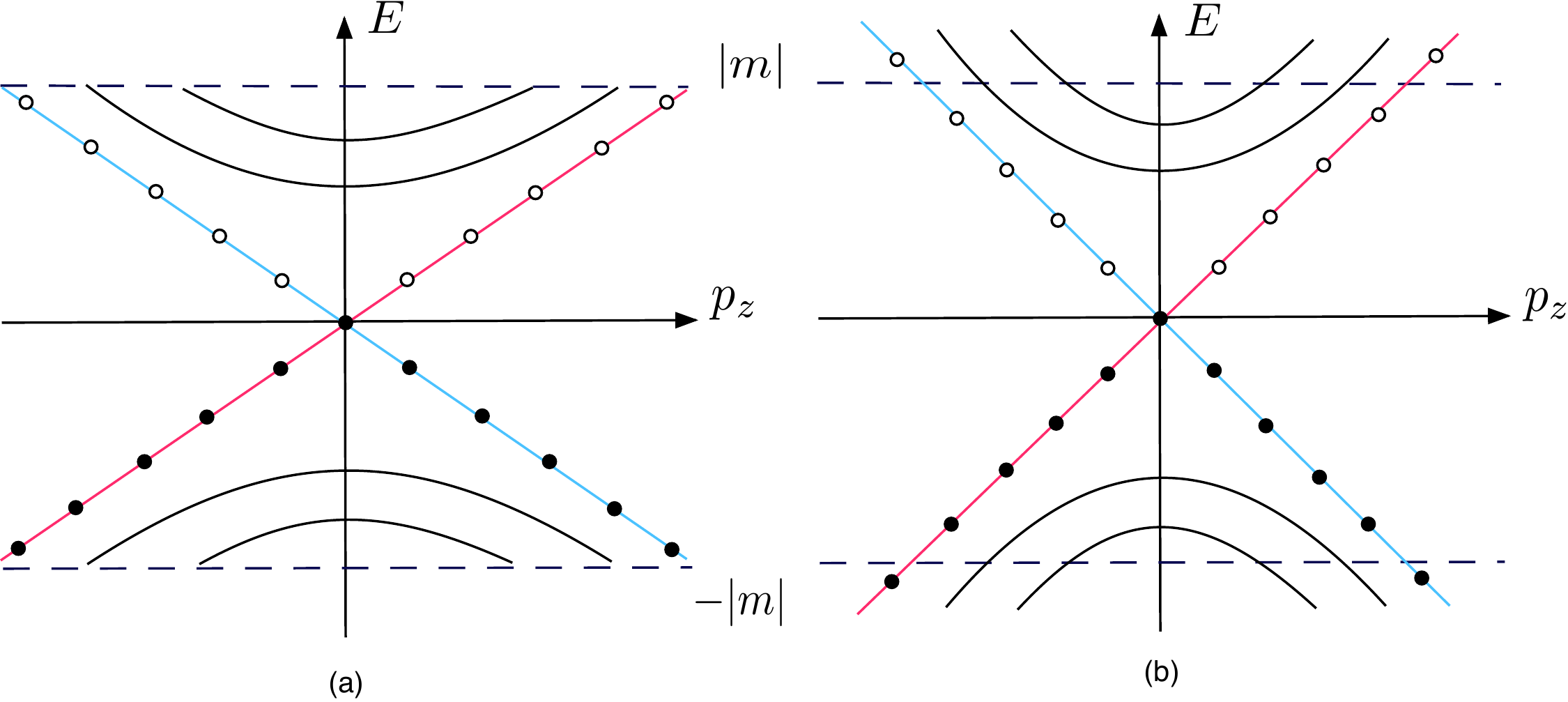}\label{fig4}
\caption{The Hamiltonian energy spectrum for chiral fermions in the presence of a uniform background magnetic field in the $z$-direction. The (black) gapped states are higher Landau levels, while the linear gapless (blue, red) curves are the zeroth Landau levels for left and right handed fermions respectively. We can consider the left and right handed fermions to exist on opposite boundaries of a cylinder. Once the energies of the linearly dispersing modes reach $\pm\vert m\vert$ these states are no longer localized on the boundary and lose their sense of chirality. (a) The initial state before the torsion electric field is applied. (b) A later state after some amount of torsional flux is threaded through the cylinder and the torsion electric field has had time to act on the system. The spectral rotation/stretching around $E=0$ pushes some occupied chiral modes outside of the topological insulator mass gap which causes them to be lost into the sea of gapped bulk states. The overall process changes the momentum localized on each edge since each chiral fermion state lost to the bulk carries momentum that originally was localized on the boundary.}
\end{figure}
Note that these Hamiltonian spectra differ from the usual spectra (for a trivial co-frame field) via a scaling of the momenta (or from another point of view a scaling of the velocity), on account of the torsional electric field.  In analogy with the boundary charge, we define the boundary momentum by
\beq
P^3(t)= \int_{\Sigma_3}d^3\vec{x}\;\frac{1}{2}\;\langle vac|\left[\psi^{\dagger}(\vec{x},t),\hat{P}_3\psi(\vec{x},t)\right]|vac\rangle=\frac{1}{2}\sum_{\{|E_n| \leq |m|\}}\mathrm{sign}(E_n)p_n^z
\eeq
where we recall that the summation is over all Hamiltonian eigenstates with \(|E_n|\leq |m|\). Using this, we can compute the net momentum along \(\xi\) on both the edges at a time \(t\)
\beq
P^3_L(t) = -\mathrm{sign}(qB)\frac{m^2NL_z}{4\pi}\left(1+\frac{bt}{L_zT}\right)^2,\;\;P_R^3(t)=\mathrm{sign}(qB) \frac{m^2NL_z}{4\pi}\left(1+\frac{bt}{L_zT}\right)^2
\eeq
where now we have taken \(L_z\) to be large. From here, we get the rate of momentum change
\beq
\frac{dP^3_L}{dt} = -\mathrm{sign}(qB)\frac{m^2N}{2\pi}\left(1+\frac{bt}{L_zT}\right)\frac{b}{T}
\eeq
\beq
\frac{dP_R^3}{dt}=\mathrm{sign}(qB) \frac{m^2N}{2\pi}\left(1+\frac{bt}{L_zT}\right)\frac{b}{T}.
\eeq
Comparing with Eq \eqref{momtrans1}, we find a precise agreement of the momentum transfer rates. Note that in contrast with the charge anomaly discussed in the previous section, the momentum anomaly in the present case is generated by a spectral rotation/stretching about $E=0$ which pushes some edge states to energies $|E|>|m|$, thus causing them to get lost into the sea of gapped bulk states (see figure \ref{fig4}).

We will now look at one final anomalous transport process. Interestingly, because of the mixed dependence of $S_{bulk}= \frac{\zeta_H}{2}\int_{M_5}F\wedge e^A\wedge T_A$ on $e^A, \omega^{AB}$ and $A,$ we can also generate a \emph{charge} current with a certain arrangement of background geometry fields. This is unusual as this type of transport does not occur in the 2+1-d effective action.
 Let us turn on a torsion magnetic field \(T^3=C dx\wedge dy\) on the \((x,y)\) cylinder, and thread torsion magnetic flux  (i.e., a dislocation) through the hole of the \((z,w)\) cylinder to generate the torsion electric field \(T^3=\frac{b}{L_zT}dt\wedge dz\). This can be achieved through the co-frame 
\beq
e^0=dt,\;\;e^1=dx,\;\;e^2=dy,\;\;e^3 = \left(1+\frac{bt}{L_zT}\right)dz+Cxdy,\;\;e^4=dw
\eeq
upon choosing \(\omega_{AB}=0\). From the bulk response action we get the bulk charge current
\beq
*J_{bulk}=\frac{\zeta_H}{2}d(e^A\wedge T_A)=\frac{qm^2}{8\pi^2}\frac{bC}{L_zT}\;dt\wedge dx\wedge dy\wedge dz.
\eeq
Just like in the case of the 4+1-d quantum Hall effect this gives a constant current density in the $w$-direction which transfers charge from one boundary to the other at a rate
\beq
\frac{dQ}{dt} = \frac{qm^2b\Phi_T}{8\pi^2 T}.\label{eq:torsionchiral}
\eeq
From the perspective of the boundary fermions, this current is due to another manifestation of the  \(U(1)\) chiral anomaly \(d*J_{cov}=\frac{\zeta_H}{2}T^a\wedge T_a\) for the chiral boundary states. This is of course the Nieh-Yan contribution to the (covariant) chiral anomaly, discussed previously.

Let us now explore how the anomaly can be understood microscopically from a Hamiltonian point of view. Once again, it suffices to focus on the lowest energy part of the spectrum of the chiral fermions in the background frame field (see Appendix \ref{app:Weylspectra} for a derivation):
\beq
E_L(t) = -\mathrm{sign}(Cp_z)\frac{p_z}{\left(1+\frac{bt}{L_zT}\right)},\;\;E_R(t) = \mathrm{sign}(Cp_z)\frac{p_z}{\left(1+\frac{bt}{L_zT}\right)}
\eeq
with degeneracy \(N(p_z,t)=\frac{|p_z\Phi_T|}{2\pi\left(1+\frac{bt}{L_zT}\right)}\). 
\begin{figure}[!t]
\centering
\includegraphics[height=6.5cm]{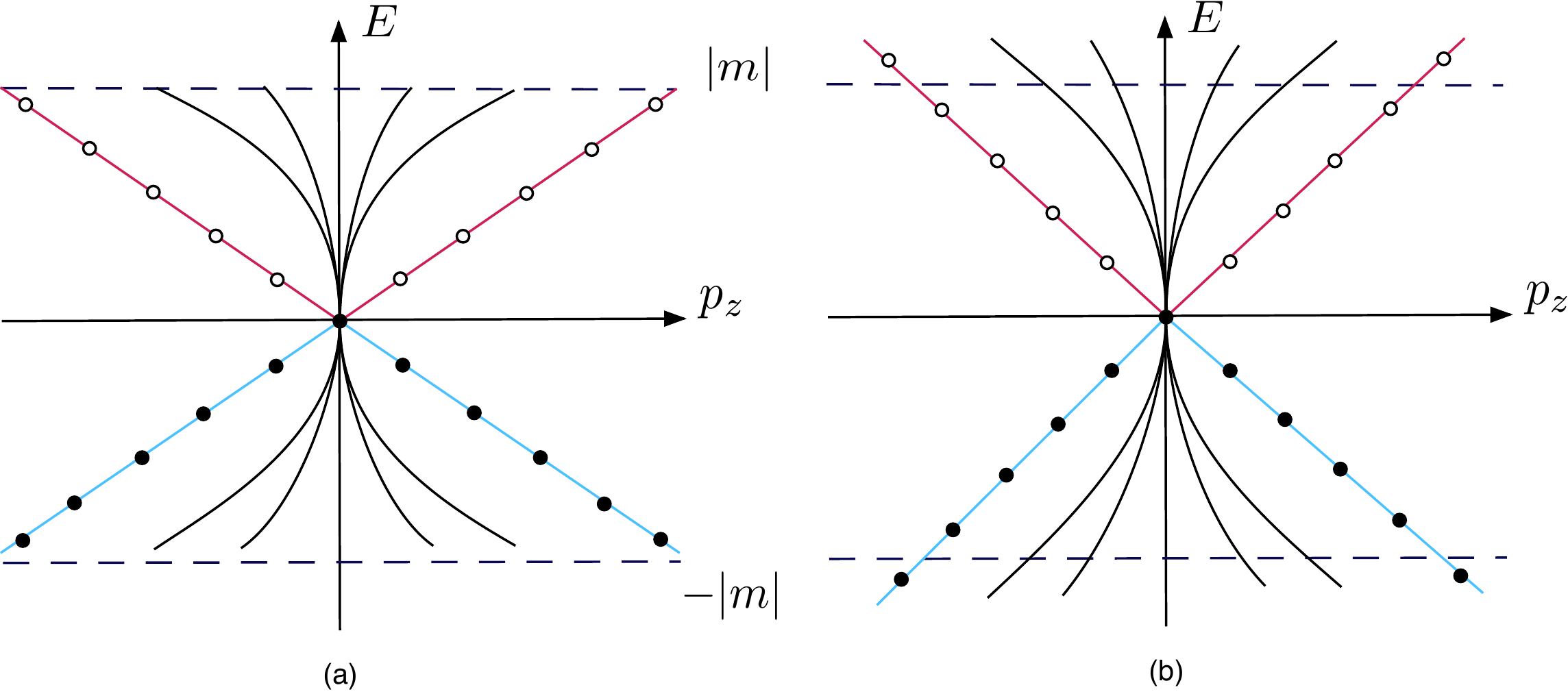}
\caption{The Hamiltonian energy spectrum for chiral fermions in the presence of a uniform background \emph{torsion} magnetic field in the $z$-direction. The (black) states are higher torsion Landau levels, while the linear gapless (blue, red) curves are the zeroth Landau levels for left and right handed fermions respectively. We can consider the left and right handed fermions to exist on opposite boundaries of a cylinder. Once the energies of the linearly dispersing modes reach $\pm\vert m\vert$ these states are no longer localized on the boundary and lose their sense of chirality. Note that something unusual happens here compared to the previous two figures. In a torsion magnetic field one chirality disperses upward while the other disperses downward. (a) The Hamiltonian spectrum before the application of a torsion electric field. (b) The spectral modification induced by an additional torsion electric field along the $z$ direction.}\label{fig:weyltorsion}
\end{figure}
From the definition
\beq
Q= q\int_{\Sigma_3}d^3\vec{x}\;\frac{1}{2}\;\langle vac|\left[\psi^{\dagger}(\vec{x}),\psi(\vec{x})\right]|vac\rangle=\frac{q}{2}\sum_{\{|E_n| \leq |m|\}}\mathrm{sign}(E_n)
\eeq
we see that the net left- and right-handed charges at a time \(t\) are given by (taking the large \(L_z\) limit)
\beq
Q_L =- \frac{qL_z}{2\pi}\int_0^{m\left(1+\frac{bt}{L_zT}\right)}dp_z\;\frac{\Phi_T}{2\pi}\frac{p_z}{\left(1+\frac{bt}{L_zT}\right)}=-\frac{qm^2\Phi_TL_z}{8\pi^2}\left(1+\frac{bt}{L_zT}\right)
\eeq
\beq
Q_R = \frac{qL_z}{2\pi}\int_0^{m\left(1+\frac{bt}{L_zT}\right)}dp_z\;\frac{\Phi_T}{2\pi}\frac{p_z}{\left(1+\frac{bt}{L_zT}\right)}=\frac{qm^2\Phi_TL_z}{8\pi^2}\left(1+\frac{bt}{L_zT}\right).
\eeq 
From here, we find the rates of change of net charge are given by
\beq
\frac{dQ_L}{dt}=-\frac{qm^2b\Phi_T}{8\pi^2 T},\;\;\frac{dQ_R}{dt}=\frac{qm^2b\Phi_T}{8\pi^2 T}\label{eq:torsionchiralfinal}
\eeq
which precisely agrees with the previous result in Eq. (\ref{eq:torsionchiral}).

We see here that the reason that the Nieh-Yan term can contribute to the covariant $U(1)$ anomaly is due to the structure of the low-energy chiral fermion branches in the presence of a uniform torsional magnetic field (see Appendix \ref{app:Weylspectra}). As a comparison, we know that in the case of a conventional $U(1)$ magnetic field the low energy states of a single Weyl node become quasi-1D branches that disperse chirally, i.e., the states coming from a left-handed (right-handed) Weyl node have a positive (negative) group velocity (if $qB<0$) $E=\pm vp_z.$ Heuristically, the magnetic field acts to convert a 3+1-d Weyl fermion into a highly degenerate quasi-1D Weyl fermion at low-energy which only disperses along the direction of the applied uniform magnetic field. The torsional magnetic field (which for instance can be thought of as a density of screw dislocations) acts differently. Instead it generates quasi-1D upward dispersing or downward dispersing branches depending on the chirality of the 3+1-d Weyl node $E=\pm v\vert p_z \vert.$ These branches contain both left- and right-movers but they have a fixed chirality. For example, for torsional field $C>0$ the downward dispersing branch of the low-energy modes are made up of left-handed modes alone, whereas the upward dispersing branch contains only right-handed modes. The degeneracy also depends on the value of the momentum $p_z$ as the torsional magnetic field is effectively stronger for larger $p_z$ charge. This seems a bit strange at first, but we can see that the microscopic calculation precisely matches the bulk anomaly calculation and thus it is a consistent interpretation. In the next section we will illustrate how this spectrum might be regularized if both chiralities are present, as must be the case, e.g., in 3+1-d Weyl semi-metals.

\section{Properties of Weyl Semi-metals with Torsion}\label{sec:wsm}
So far our work has focused on the general structure of the torsion anomalies associated to 3+1-d Weyl fermions. While such fermions can occur at the boundary of a 4+1-d topological insulator, they can also appear in a 3+1-d material, the so-called Weyl semi-metal. However, unlike the 4+1-d boundary modes, bulk Weyl fermions must always appear in pairs due to the Nielsen-Ninomiya no-go theorem\cite{Nielsen:1981hk}. Thus, our results do not immediately carry over to the discussion of the Weyl semi-metals. However, we can utilize the viewpoint taken by much of the recent work on the electromagnetic response properties of Weyl semi-metals, which casts the 3+1-d Weyl semi-metal as a 2+1-d family of Chern insulator Hamiltonians\cite{ran2011,wan2011,zyuzin2012}. Since we know the torsional response properties of the 2+1-d system, we can use those results to write down the correct response for the 3+1-d Weyl semi-metal in a manner analogous to what has already been done for the Hall conductance\cite{ran2011,wan2011,zyuzin2012}. We will first briefly review the electromagnetic case before proceeding to the geometric response.

\begin{figure}[!t]
\centering
\includegraphics[width=13cm]{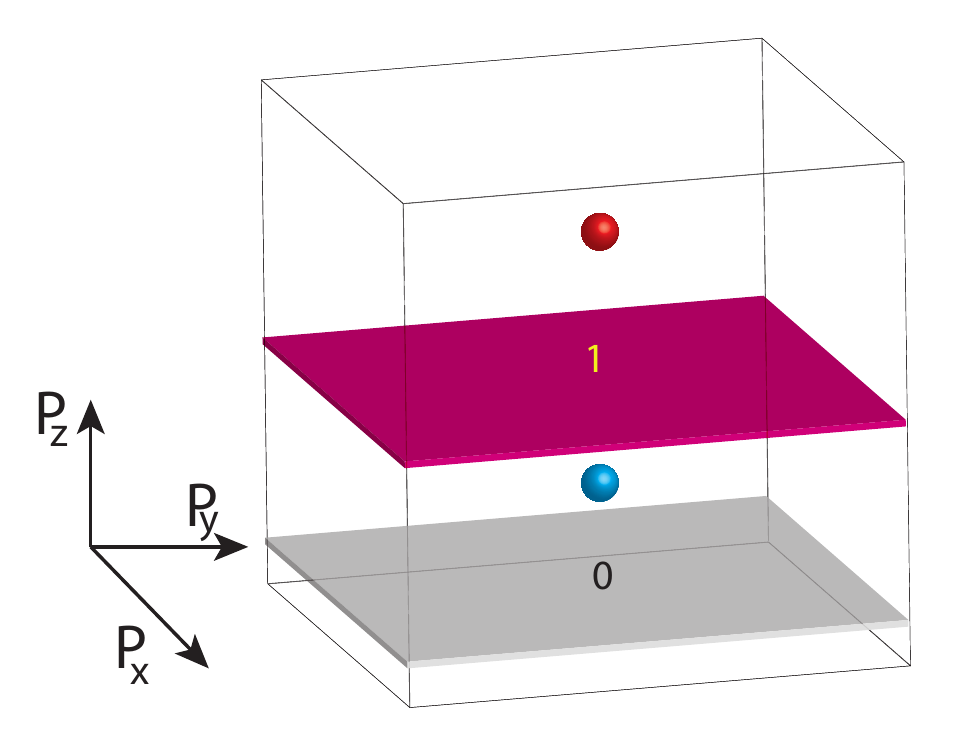}
\caption{Momentum space description of a simple Weyl semi-metal with two Weyl nodes of opposite chirality (red ands blue spheres) separated in the $p_z$ direction. The two planes represent two gapped 2+1-d insulator subspaces of the three dimensional Brillouin zone. The grey plane has a Hall conductance of $0$ and the magenta plane has a Hall conductance of $1$ in units of $\frac{q^2}{2\pi}.$ In fact the entire family of planes parameterized by $p_z$ that lies between the two Weyl nodes will each carry Hall conductance of $\frac{q^2}{2\pi}$ while the planes outside the nodes and inside the Brillouin zone boundaries carry no Hall conductance.}\label{fig:wsm}
\end{figure}

The properties of Weyl semi-metals (WSM) have been the focus of a large number of recent articles\cite{nielsen1981,wan2011,turner2013,haldane2014,matsuura2013,sid2013,grushin2013,jian2013,pavan2012,tewari2012,zyuzin2012,vazifeh2013,chen2013,chen2013weyl,vanderbilt2014comment,Landsteiner:2013sja, Chernodub:2013kya,rama2014}. As mentioned above, these materials are gapless in the bulk and have isolated point-like degeneracies between the valence and conduction bands. Each of these degeneracy points is a Weyl node, i.e., a bulk, 3+1-d Weyl fermion, and the total chirality of all the nodes in a single material must vanish. So, while the right and left-handed 3+1-d Weyl fermions are spatially separated on the surfaces of a 4+1-d topological insulator, there is no such spatial separation for the Weyl fermions in a WSM. To illustrate the basic physics, let us assume we have the simplest example of a WSM, i.e., one with two Weyl nodes that are separated in momentum space along the $p_z$ axis and located at $\vec{p}_{L,R}=(0,0,\pm p_{zc})$ (see Fig. \ref{fig:wsm}). Let us define the quantity $\vec{b}=\tfrac{1}{2}(\vec{p}_L-\vec{p}_R)=(0,0,p_{zc}).$ If the left and right-handed nodes are not degenerate in energy we can also define an energy separation $b_0=\tfrac{1}{2}(\epsilon_L-\epsilon_R)$ where $\epsilon_{L,R}$ are the energies of the nodes at $\vec{p}=\vec{p}_{L,R}$ respectively. We can combine these two quantities into a $1$-form $b_{\mu}dx^{\mu}.$ This definition is useful\footnote{We note that we have chosen the factors of $\tfrac{1}{2}$ in the definition of $b_\mu$ to match the convention in the literature.} because the quasi-topological electromagnetic response properties of WSM have been investigated, and it was found that the low-energy effective action takes the form\cite{nielsen1981,zyuzin2012}
\begin{equation}
\frac{q^2}{8\pi^2}\int (g^{-1}dg)\wedge A\wedge dA\label{eq:wsmEM}
\end{equation}\noindent for the space-time translation group element $g=\exp(i(2b_{\mu})x^{\mu}).$ This is usually written in terms of components as
\begin{equation}
\frac{q^2}{4\pi^2}\int d^4 x \epsilon^{\mu\nu\rho\sigma}b_{\mu}A_{\nu}\partial_\rho A_\sigma. \label{eq:wsmComp}
\end{equation}\noindent which is similar to the Lorentz-violating Chern-Simons terms discussed in Refs. \cite{carroll1990limits, jackiw2003}.

The origin of this response can be understood from the simple limit of two Weyl-nodes. Let us also assume that they are degenerate in energy. Then,  except for $p_z=-p_{zc}$ or $+p_{zc},$ the system is gapped, and thus every fixed-$p_z$ plane is a 2+1-d insulator apart from the two critical values of $p_z.$ Since at fixed $p_z$ the low-energy model near each Weyl node is that of a 2+1-d Dirac model with a mass given by the magnitude of $p_z$ away from the node, then every fixed-$p_z$ plane is either a trivial or topological 2+1-d Chern insulator. For the continuum models we have been considering, it only makes sense that the planes \emph{between} the critical values would be in the topological phase, i.e., for $-p_{zc}<p_z<p_{zc}.$  This implies that there is a finite contribution of Hall conductance given by $\sigma_{xy}=\tfrac{q^2}{2\pi}$ for each value of $p_z\in \left[-p_{zc},p_{zc}\right]$ which is exactly what Eq. (\ref{eq:wsmEM}) encodes when $b_0=0.$  If instead the region of $p_z$ outside of the range $-p_{zc}<p_z<p_{zc}$, but inside the Brillouin zone boundaries (assuming a lattice model), was topologically non-trivial, then the Hall conductance would differ by the addition of an amount $e^2/h$ per layer, i.e., the quantized amount due to fully occupied bands carrying a weak topological index\cite{halperin1987, FKM2007,MooreBalents07,haldane2014}. The WSM response action for a lattice system only uniquely determines the fractional piece of the response, i.e., only the piece corresponding to $2{\vec{b}}\mod \vec{G}$ where $\vec{G}$ is the set of reciprocal lattice vectors. 

For a generic set of Weyl-nodes located at a 3-momentum ${\textbf{P}}^{(\alpha)},$  with energy $\epsilon^{(\alpha)}$, and chirality $\chi_\alpha=\pm 1$ we can construct the 4 component 1-form $b_{\mu}=\tfrac{1}{2}\sum_{\alpha}\chi_{\alpha}P_{(\alpha),\mu}$ where $P^{(\alpha)}_{\mu}\equiv(\epsilon^{(\alpha)},{\textbf{P}}^{(\alpha)}).$ We can also represent this using the generic translation group element $g=\exp\left[i\sum_{\alpha}\left(\chi_{\alpha}P^{(\alpha)}_{\mu}\right)x^{\mu}\right].$ We note that for a lattice system the spatial translations can only take values in the real-space lattice which implies that the response only captures the fractional piece of the $\sum_{\alpha}\chi_{\alpha}{\textbf{P}}^{(\alpha)}$ which is less than a reciprocal lattice vector, i.e., it does not uniquely determine the response due to fully occupied bands. This is why it is not so important to specify which region of momentum space is topological and which is trivial (as in the simple example above), because they differ by an amount due to fully filled bands.

After having reviewed the electromagnetic response it is easy to see that this type of argument holds for more than just this case. Considering a family of Chern insulators parameterized by an additional momentum immediately leads us to the appropriate geometric responses. Terms with quantized coefficients (i.e., ones that only depend on the sign of the Dirac mass), such as the gravitational Chern-Simons term will yield
\begin{equation}
\frac{\kappa_H}{2\pi}\int  b\wedge \mtr\left(\lcw\wedge d\lcw+\frac{2}{3}\lcw\wedge\lcw\wedge\lcw\right).
\end{equation} However, for the torsional term, the Hall viscosity coefficient depends on the magnitude of the mass, i.e., $\zeta_H(p_z)$ depends on $p_z$ in a complicated fashion. In the context of the simple WSM discussed above this means that each 2+1-d Hamiltonian parameterized by $p_z$ yields a different contribution to the Hall viscosity, and thus the value of the Lorentz violating $1$-form that enters the response is not trivially determined from the energy-momentum locations of the Weyl-nodes as is the Hall conductance. In our simple example, since $\sigma_{xy}(p_z)$ is just a piecewise constant function which is quantized to be $q^2/2\pi$ for $-p_{zc}<p_z<p_{zc},$ and zero otherwise, we find that $2b_z=\tfrac{2\pi}{q^2}\int dp_z \sigma_{xy}(p_z)=p_{zc}-(-p_{zc})=2p_{zc}.$ However, we need to define a separate parameter for the torsion response $\lambda=\lambda_\mu dx^{\mu}$ such that $2\lambda_z=\int dp_z \zeta_{xy}(p_z)$ for our simple example. The $1$-form $\lambda$ will generically be a complicated function of the Weyl-node positions, and has units of $L^{-3}$ in natural units.  With this definition we see that $\lambda$ will contribute to the torsion response as
\begin{equation}
\frac{1}{2\pi}\int \lambda\wedge e^{a}\wedge T_a\label{eq:wsmTorsion}
\end{equation}\noindent where $a=0, 1, 2, 3.$  For the simple WSM we can use an almost identical argument as above to indicate that the collection of topological insulator planes will carry a total 3D Hall viscosity given by $\zeta_{xy}=\tfrac{\lambda_zL_z}{2\pi}.$  

An interesting phenomenon also occurs when the Weyl nodes are non-degenerate in energy. In this case, one finds the analog of the chiral magnetic effect (a non-zero electric current in the presence of a non-zero magnetic field but vanishing electric field), but for torsion. This would imply that with the insertion of a dislocation line, there should be a momentum current flowing in the direction of the Burgers' vector even without the application of a torsion electric field though there may be some subtleties\footnote{There has been some controversy in the literature about the existence of the chiral magnetic response in real systems and also the role played by the boundary states\cite{tewari2012,zyuzin2012,vazifeh2013,chen2013,haldane2014,chen2013weyl,vanderbilt2014comment}. These same complications may arise in the geometric response as well.}. For example, to generate a typical chiral magnetic effect one must violate the effective Lorentz invariance by either doping the system away from charge neutrality to induce a background density, or turning on a weakly time-dependent magnetic field and slowly taking the DC limit. These considerations will also enter the discussion of the chiral dislocation effect. We should also note that  Ref. \cite{kimura2012}  predicts a chiral heat effect at finite temperature which is related to the curvature response of a 3+1-d Weyl fermion, which is also contained in our bulk response calculation. The 3+1-d anomalous Hall viscosity and the chiral dislocation effect are two prominent geometric response features of the Weyl semi-metal. We will delay a more detailed discussion of the geometric response properties of Weyl semi-metals to future work. 

Before moving on to discuss 3+1-d TRI topological insulators  we want to illustrate one other interesting property of the Weyl semi-metal along the lines of the seminal Nielsen-Ninomiya paper that discussed the chiral anomaly in a crystal\cite{Nielsen:1983rb}. We know that because of the vanishing chirality in the semi-metal we cannot have an overall chiral anomaly. However, since the Weyl nodes are separated in momentum (and possibly in energy) we can have anomalous current flows in momentum space between the nodes. We will now illustrate this behavior for the anomaly due to the Nieh-Yan term, i.e., we will illustrate the anomalous chiral current due to parallel torsion electric and torsion magnetic fields arising from the anomalous Ward identity:
\beq
\partial_\mu j^{mu}_{5}=\int\frac{q}{32\pi^2\ell^2}\epsilon^{\mu\nu\rho\sigma}\left(\eta_{ab}T^a_{\mu\nu}T^b_{\rho\sigma}-2R_{ab;\mu\nu}e^a_{\rho}e^b_{\sigma}\right).
\eeq

\begin{figure}[!t]
\centering
\includegraphics[width=13cm]{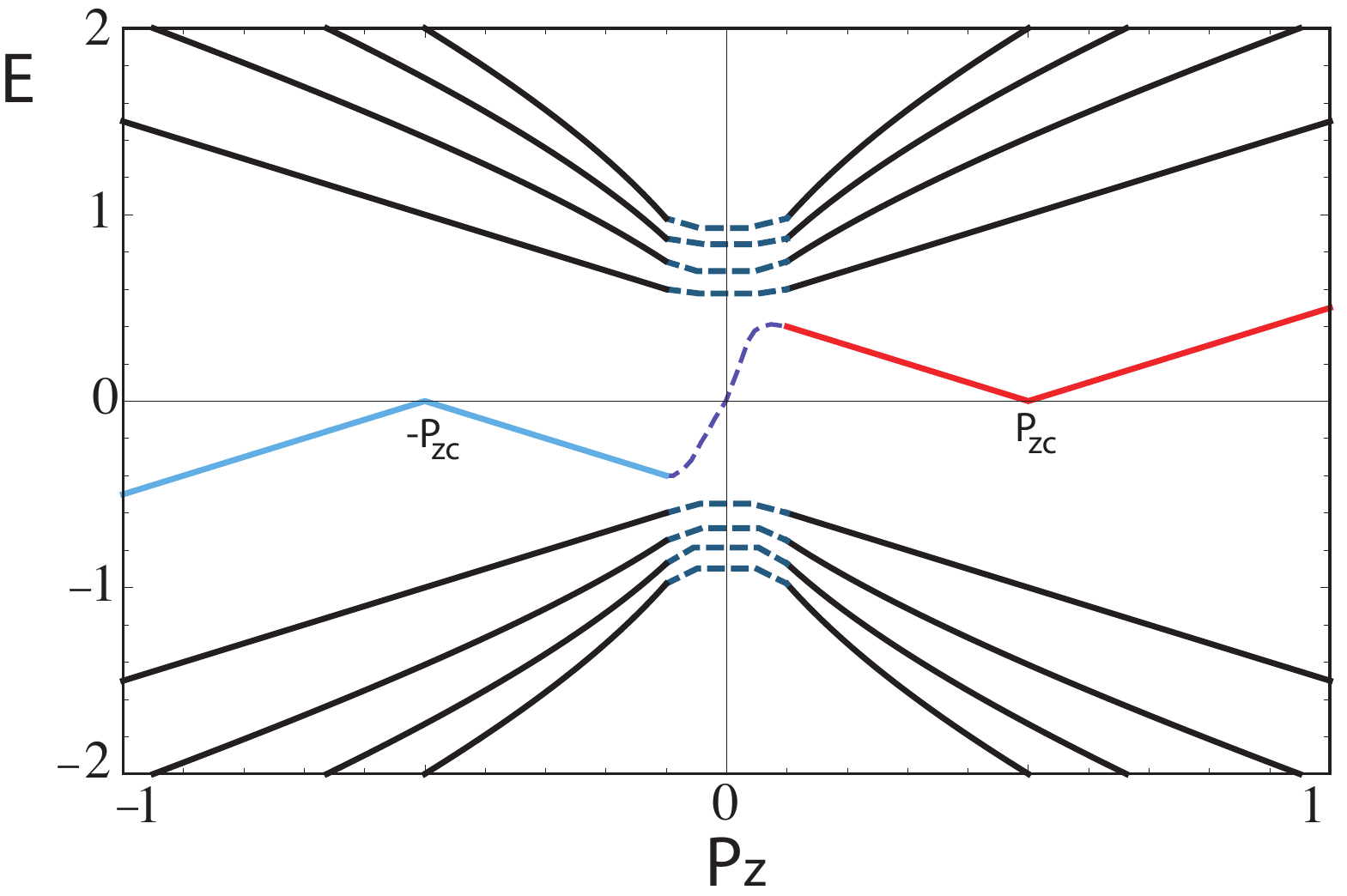}
\caption{The low-energy spectrum of a simple Weyl semi-metal in the presence of a uniform background torsion magnetic field. The Weyl nodes are located at $p_z=\pm p_{zc}$ in the absence of a field. The solid lines are from Eqs. \ref{eq:torsionzero}, \ref{eq:torsionNlevel} for the first few values of $n.$ The dotted lines are a conjectured continuation of the levels that show how they might be regularized in a lattice model. The red and blue colors represent left and right handed Weyl nodes. One can compare this to Fig. \ref{fig:weyltorsion} which shows the energy spectra of the Weyl nodes when they are both located at the same point in momentum space. }\label{fig:torsionLL}
\end{figure}

To calculate the anomalous current flow we need to understand the spectrum of a Weyl semi-metal in the presence of a uniform background torsion magnetic field. Suppose that the torsion magnetic field is applied using the co-frame
\begin{equation}
e^0 = d t,\qquad e^1 = d x,\qquad e^2 = d y,\qquad e^3 = dz + f(x)d y.
\end{equation}
The frame is torsional if we set the spin connection to zero (assuming zero curvature), with \(T^3 = d e^3 = f'(x)\;d x\wedge d y\), and hence $e^a\wedge T_a=f'(x) dx\wedge dy\wedge dz$. The Dirac operator is then given by
\begin{equation}
i\slashed{D} = i\gamma^a\ue_a^{\mu}\partial_{\mu} = i\left(\gamma^0\partial_t+\gamma^1\partial_x+\gamma^2(\partial_y-f(x)\partial_z)+\gamma^3\partial_z\right).
\end{equation}
We can project onto left-chiral modes, obtaining\footnote{The choice of representation for the Dirac matrices is \[\gamma^0=\left(\begin{array}{l r} 0 & 1\\1 & 0\end{array}\right),\;\gamma^i=\left(\begin{array}{l r} 0 & -\sigma^i\\\sigma^i & 0\end{array}\right),\; \gamma_5=\left(\begin{array}{l r} 1 & 0\\0 & -1\end{array}\right)\].}
\begin{equation}
i\gamma^0\slashed{D}P_L=\left(i\partial_t+i\sigma^1\partial_x-\sigma^2(p_y-f(x)p_z)-\sigma^3p_z\right)
\end{equation}
where, because \(p_y, p_z\) are good quantum numbers, we have Fourier transformed in the $y, z$ directions.

Since we want to represent a Weyl semi-metal whose nodes are shifted in the $p_z$ direction, we introduce a vector \(b\) in momentum space, such that the Dirac operator gets shifted to \(i\slashed{D}=i\gamma^a\ue_a^{\mu}(\partial_{\mu}+i\gamma_5b_{\mu})\), 
 and take \(b_a =\ue^{\mu}_ab_{\mu}= (0,0,b_3)\). In this case, 
\begin{equation}
i\gamma^0\slashed{D}P_L=\left(i\partial_t+i\sigma^1\partial_x-\sigma^2(p_y-f(x)p_z)-\sigma^3(p_z+b_3)\right).
\end{equation}
Upon solving the resulting Dirac equation, we find that the low-energy spectra of the left- and right-handed gapless modes shifts as
\beq
E_L = -sign(C)|(p_z+b_3)|,\;\; E_R= sign(C)|(p_z-b_3)|\label{eq:torsionzero}
\eeq
but the degeneracies remain \emph{unchanged}
\beq
N_L  =\left|\frac{p_z\Phi_T}{2\pi}\right|,\;\; N_R  =\left|\frac{p_z\Phi_T}{2\pi}\right|.
\eeq
The higher energy modes now do not completely shift, they simply get gapped and distorted (see Fig. \ref{fig:torsionLL})
\beq
E_{n,\pm} = \pm \left((p_z \pm b_3)^2+2n|Cp_z|\right)^{1/2},\qquad n=1,2,\ldots\label{eq:torsionNlevel}
\eeq

If we now add a torsion electric field then we will see that chiral charge is transferred between the two low-energy branches of the Weyl-nodes in the uniform torsion magnetic field. The calculation is identical to that presented at the end of Section \ref{sec:torsionAnomaly} which culminates with Eq. \ref{eq:torsionchiralfinal} so we will not reproduce it here. 

\section{3+1-d Topological Insulator via Dimensional reduction}\label{sec:3dti}
Given our derivation of the 4+1-d response action we can now discuss the properties of the 3+1-d time-reversal invariant strong topological insulator\cite{FKM2007,MooreBalents07,Roy07}. 
As shown in Ref. \cite{Qi:2008ew}, if one knows the anomaly structure in odd space-time dimensions, one can dimensionally reduce the relevant effective actions to study the properties of topological phases in one or two dimensions lower. There is a cost for this, namely one expects to have to make symmetry constraints on the lower-dimensional system in order to have a robust topological phase, and the integer topological invariants of the higher-dimensional systems get reduced to $\bZ_2$ invariants in the lower-dimensional systems. 

As an example, let us briefly review the theory for electromagnetic response of the $4+1$-d topological insulator reduced to $3+1$-d. We will be a little imprecise here, but the overall picture is correct (for more detail see Ref. \cite{Qi:2008ew}). The action for the $4+1$-d topological insulator is
\begin{equation}
S_{eff}[A]=\frac{q^3C_2}{24\pi^2}\int d^5 x \epsilon^{abcde}A_a\partial_b A_c\partial_d A_e
\end{equation}\noindent which is the second Chern Simons term, where $a, b, c, d, e =1,2 \ldots 5,$ and $C_2$ is the second Chern number, the value of which is an integer which depends on the phase of the underlying massive fermions of the topological insulator. To dimensionally reduce this system we can assume that the fields do not depend on the $4$-th spatial coordinate $w$ (which we have compactified to a circle with circumference $L$). Then we can take the limit as $L_w\to 0$ from which we find the action
\begin{eqnarray}
S_{eff}[A]&=&3\frac{q^2}{24\pi^2}\int d^4 x \left[\int dw\ qC_2 A_w\right]\epsilon^{\mu\nu\rho\sigma}\partial_\mu A_\nu\partial_\rho A_\sigma\nonumber\\
&=&\frac{q^2}{8\pi^2}\int d^4 x\;\theta\epsilon^{\mu\nu\rho\sigma}\partial_\mu A_\nu\partial_\rho A_\sigma
\end{eqnarray}\noindent where $\theta\equiv \int dw\ qC_2 A_w$ which gives us the amount of flux threaded through the $w$ circle. For example, for one flux quantum $\theta = 2\pi.$ 

Since we want to consider time-reversal invariant insulators in $3+1$-d there is a constraint on $\theta.$ Under time-reversal $\theta \to -\theta$. However $\theta$ is only well-defined mod $2\pi$: $\theta\equiv \theta+2\pi n$ for some integer $n.$ Thus, if we require time-reversal then $\theta =0$ or $\theta =\pi$ are the only two allowed values. So while the $4+1$-d insulator was classified by an integer $C_2$ and there were no required symmetries, it turns out that the time-reversal invariant 3+1-d case is classified by the $Z_2$ invariant $\theta.$ The physical consequence of this term is as follows. If $\theta$ is a constant in space-time then our dimensionally reduced action is a total derivative and thus there is no measurable response. There are two exceptions to this: (i) when magnetic monopoles exist then this term attaches an electric charge $q_{monopole}=\tfrac{q\theta}{2\pi}$ to the monopole via the Witten effect\cite{Witten:1979ey} (ii) if we have a boundary to the vacuum  or trivial insulator then $\theta$ necessarily changes from $\pi$ to $0$ and the action will have a non-zero contribution. For case (ii) the effect of this action is to endow the boundary with a quantum Hall effect localized at the boundary with a half Hall conductance $\sigma=\tfrac{q^2}{4\pi}.$ Generically at such a 2+1-d boundary,  a $\theta$-term will attach its corresponding Chern-Simons action to that localized region. For topological insulators the coefficient of the Chern-Simons term can be half of that required for a properly regularized intrinsically 2+1-d system. These are the general features of the dimensional reduction.

Given the general anomaly structure in 4+1-d we now want to dimensionally reduce the effective response action to find the relevant action for 3+1-d time-reversal invariant topological insulators in the presence of curvature and torsion. The calculation here is more complicated since our fields are intimately related to the geometry. Thus, to perform this reduction we need to split the fields up into appropriate pieces. We take the following co-frame, frame, and connections
\beqn\label{bg}
e^a &=&b^a dw +\ble^a_idx^i\nonumber\\
e^4 &=& N dw\nonumber\\
\ue_a &=& \blue_a\nonumber\\
\ue_4 &=& N^{-1}\left(\partial_w-b^a\blue_a^i\partial_i\right)\\
A &=& \Theta dw  + \blA_idx^i\nonumber\\
{\omega^a}_b &=&  {\theta^a}_b dw +{\blw^a}_{i;b}dx^i\nonumber\\
{\omega^a}_4 &=& 0\nonumber
\eeqn
where $a=0, 1,2 ,3,$ and the intrinsic $3+1$-d co-frame, frame, and connections are now labelled by a tilde. As usual for a dimensional reduction, all the fields are only allowed to depend on the intrinsic $3+1$-d coordinates, but not on the fourth spatial direction $w$. Also note that we have set ${\omega^a}_4=0$ because this is related to extrinsic geometric effects, which are not of interest to us here.

We want to compute our fermion effective action with this choice, which we will now do term by term. Let us begin with $A\w F\w F$ which we already calculated above in component notation. Using
\beq
F = d\Theta\w dw+ \blF
\eeq
we obtain
\beq\label{DRHcond}
\frac{1}{L}\oint A\w F\w F = 3\Theta\; \blF\w \blF - 2d\left(\Theta\;\blA\w\blF\right)
\eeq
where the integral above is over the $w$-direction. Next, for the $F\wedge e_A\wedge T^A$ term, we need to use
\beq
T^a = \blD b^a\w dw +\blT^a+{\theta^a}_bdw \w \ble^b
\eeq 
\beq
T^4= dN\w dw
\eeq
and we find
\beqn\label{DRvisc}
\frac{1}{L}\oint F\wedge e^A\w T_A &=& \Theta\; d(\ble^a\w\blT^a) + 2b_a\;\blF\w \blT^a-\blF \w \ble^a\w \ble^b\theta_{ab}
\nonumber\\
&-& d(\Theta\;\ble^a\w \blT^a)- d(b_a \ble^a\w \blF). 
\eeqn
Notice above, that terms linear in $\Theta,\;b^a$ and $\theta_{ab}$ seem to be related to $3+1-d$ covariant anomalies. Of course, this is no coincidence, and we will return to this point shortly in Section \ref{sec:lincov}.

Next, we need to deal with the curvature terms. These are quite complicated in general, and involve many terms which are not easy to interpret physically. In order to avoid cluttering our discussion, we will defer some of these calculations to Appendix \ref{app:curvature}. Nevertheless, there is a straightforward way to extract the dimensionally reduced action to linear order in $b^a$ and $\theta_{ab}$. Fortunately, these are also the most interesting terms from the point of view of our discussion so far.

\subsection{Linear terms and Covariant anomalies}\label{sec:lincov}
The choice of frame and connections in Eq. \eqref{bg} can be broken up into the separable background
\beqn
e^a &=& \ble^a\nonumber\\
e^4 &=& Ndw\nonumber\\
A &=& \blA\label{bg1} \\
\omega_{ab} &=& \blw_{ab}\nonumber\\
\omega_{a4} &=& 0\nonumber
\eeqn
and the perturbations about this background
\beq
\delta e^a = b^adw,\;\;\delta A=\Theta dw,\;\; \delta\omega_{ab} = \theta_{ab}dw.
\eeq 
Note that all of these are proportional to $dw$. Since we are interested in computing the intrinsic $d=3+1$ effective action, we need terms in the $4+1$-d Lagrangian density of the form $dw\w (\cdots)$. If further, we decide to focus on terms linear in $\Theta,\;b^a,$ and $\theta_{ab}$, then the terms of interest are precisely
\beq
\mathcal{L}_{4+1} = \delta A\wedge *J_{bulk}+\delta e_a\wedge *J^a_{bulk}+\delta\omega_{ab}\wedge *J^{ab}_{bulk}+O(b^2,\theta^2).
\eeq
Performing the integration over $w$, we then arrive at the intrinsic $3+1$-d Lagrangian density 
\beq
\mathcal{L}_{3+1} = \Theta\;\left. *J_{bulk}\right|_{bg}+b^a\;\left. *J^a_{bulk}\right|_{bg}+\theta_{ab}\; \left.*J^{ab}_{bulk}\right|_{bg}+O(b^2,\theta^2)
\eeq
where the subscript $bg$ means that these currents are to be evaluated on the separable background \eqref{bg1}. Indeed, the currents above are precisely the covariant $U(1)$, diffeomorphism, and Lorentz anomalies in $3+1$-d, as calculated from the Callan-Harvey argument. Having computed these anomalies previously (see Eq. \eqref{fullanomalies}), we merely state the result\footnote{In this language, boundary terms such as those present in Eqs.  \eqref{DRHcond} and \eqref{DRvisc} are the same as the induced boundary currents (or Bardeen-Zumino terms) from the Callan-Harvey discussion. These are however not important in what follows.} 
\beqn\label{dimred}
\mathcal{L}_{3+1}&=&\frac{q^2}{8\pi^2}\Theta\;\blF\w\blF+\frac{q_Tm^2}{8\pi^2}\Theta\; d(\ble^a\w\blT^a)+\frac{1}{192\pi^2}\Theta\;\mathrm{tr}\;\blR^{(-q_T)}\w\blR^{(-q_T)}\nonumber\\
&+&\frac{q_T}{96\pi^2}\Theta\;d*d*d(\ble^a\w \blT^a)+\frac{qq_Tm^2}{4\pi^2}b_a\;\blF\w\blT^a+\frac{q}{96\pi^2}b^a\;\ble_a\w d\mathcal{A}_2\nonumber\\
&-& \frac{q_Tq}{96\pi^2}b_a\;\mathcal{A}_2\w \blT^a+\frac{qq_T}{48\pi^2}b_a\;d*d*\blF\w \blT^a-\frac{m^2qq_T}{8\pi^2}\theta_{ab}\;\blF\w \ble^a\w \ble^b\nonumber\\
&+&\frac{qq_T}{96\pi^2}\theta_{ab}\;\mathcal{A}_2\w \ble^a\w \ble^b-\frac{qq_T}{96\pi^2}\theta_{ab}\;d*d*\blF\w \ble^a\w \ble^b+O(b^2,\theta^2)
\eeqn
where recall the definition
\beq
\mathcal{A}_2 = (\blF\w\blR^{(-q_T)}_{ab})(\blue^a,\blue^b).\nonumber
\eeq

Unfortunately there are still a lot of terms to understand, though some of them are simpler than others. The first three terms are variations on the electromagnetic $\Theta$ term action found in 3+1-d time-reversal invariant topological insulators. As explained above, all three terms can be interpreted as giving rise to 2+1-d response coefficients on the surface (domain wall of $\Theta$) of the topological insulator. Explicitly, the terms 
\begin{equation}
\frac{q^2}{8\pi^2}\Theta\;\blF\w\blF+\frac{q_Tm^2}{8\pi^2}\Theta\; d(\ble^a\w\blT^a)+\frac{1}{192\pi^2}\Theta\;\mathrm{tr}\;\blR^{(-q_T)}\w\blR^{(-q_T)}\
\end{equation}\noindent give rise to a surface Hall conductivity, a surface Hall viscosity, and a surface gravitational 
Chern-Simons 
term respectively. Perhaps one can view the third term as a response of angular momentum to intrinsic curvature deformations of the surface. That is. at locations on the surface where there is a non-zero curvature, the gravitational Chern-Simons term may bind spin/angular momentum to that location similar to the charge Chern-Simons term binding electric charge on locations with non-zero magnetic flux ($U(1)$ curvature). In addition to this interpretation, Ref. \cite{kimura2012} shows that at finite temperature the surface gravitational Chern-Simons term is related to a thermal response. 

Although there are a large number of terms, they can be organized in a way which is easier to interpret. Besides the $U(1)$ anomaly term, the other types of terms can each be grouped into (i) a leading order piece which goes as $m^2$, (ii) a curvature dependent universal piece, and (iii) a higher-order derivative piece. There are three separate groupings I, II, and III which depend on the parameters $\Theta, b_a,$ and $\theta_{ab}$ respectively:
\begin{equation}\label{group1}
I\colon \frac{q_T}{8\pi^2}\Theta\left[ m^2 d(\ble^a\w\blT^a)+\frac{1}{24q_T}\mathrm{tr}\;\blR^{(-q_T)}\w\blR^{(-q_T)}
+\frac{1}{12}d*d*d(\ble^a\w \blT^a)\right]
\end{equation}
\begin{equation}\label{group2}
II\colon\frac{q_Tqb_a}{4\pi^2}\left[ m^2 \blF\w\blT^a+\frac{1}{24q_T}\left(\ble_a\w d\mathcal{A}_2
- q_T\mathcal{A}_2\w \blT^a\right)+\frac{1}{12}d*d*\blF\w \blT^a\right]
\end{equation}

\begin{equation}\label{group3}
III\colon-\frac{qq_T\theta_{ab}}{8\pi^2}\left[m^2\blF\w \ble^a\w \ble^b
-\frac{1}{12}\mathcal{A}_2\w \ble^a\w \ble^b+\frac{1}{12}d*d*\blF\w \ble^a\w \ble^b\right].
\end{equation}

Grouping I shows the response terms which all depend on the parameter $\Theta.$ In the bulk of a non-trivial $Z_2$ 3+1-d topological insulator $\Theta$ is quantized to be an odd multiple of $\pi$, while outside the material $\Theta=0.$ Thus, as has been mentioned above, these terms imply that on the surface of a topological insulator (if time-reversal symmetry is weakly broken by a magnetic layer) there will be a surface quantum Hall viscosity and its associated curvature correction. If we assume that at a given surface $\Theta$ varies like a step function from $\pi$ inside to zero outside then the effective surface action becomes
\begin{equation}
S_{surf}=\int_{bdry}-\frac{q_T}{8\pi}\left[ m^2 \ble^a\w\blT^a+\frac{1}{24q_T}CS[\omega^{(-q_T)}]
+\frac{1}{12}*d*d(\ble^a\w \blT^a)\right]
\end{equation}\noindent where $CS$ is the Chern-Simons 3-form. This means the surface of a 3+1 d topological insulator has a viscosity coefficient which is exactly half that found in 2+1-d. This is similar to the surface Hall conductance which also carries exactly half the value of the bulk Hall conductance of a 2+1-d Chern insulator. Note that the gravitational Chern-Simons term can be expanded in powers of torsion, to obtain the Levi-Civita Chern Simons term plus the curvature correction to surface Hall-viscosity, etc. While we have written the higher derivative term as well, this term (a) depends on the metric (through the bulk Hodge star operator) and thus is not a topological response term, and (b) captures effects which are extrinsic to the surface. 

The second grouping is a response when the parameter $b_a$ is non-vanishing. We know that 3+1-d time-reversal invariant topological insulators have a non-vanishing $\Theta$, however it is not known what materials would have a non-vanishing $b_a,$ though it seems they must somehow be anisotropic. For now let us assume we have a material in which $b_a\neq 0$ inside, and we will calculate the consequences (assuming the vacuum has $b_a=0$). From the first term in this grouping we see that in such a material we will find a localized charge density at places where dislocation lines intersect the surface, but only if the Burgers vector of the dislocation is not orthogonal to $b_a.$ We can see this for the simple case where we set the spin connection to zero, i.e., in flat space. If we assume $b_a$ changes as a step function at a surface we find that the leading order term in the surface action contains the mixed Chern-Simons term
\begin{eqnarray}
S_{surf}&=&\frac{q_Tqm^2\Delta b_a}{4\pi^2}\int_{bdry}\tilde{A}\wedge d\ble^a\\
\ast j &=& \frac{q_Tqm^2}{4\pi^2}\Delta b_a d\ble^a.
\end{eqnarray}\noindent 
Thus for a dislocation line with Burgers vector $B^a$ that intersects the surface, there will be a bound charge density $\rho = \tfrac{qm^2}{4\pi^2}\Delta b_a B^a.$ Conversely, magnetic flux lines will carry momentum density along the direction $\Delta b^a$ at points where they intersect the surface. As usual, the second term in equation \eqref{group2} can then be thought of as a universal curvature correction to this mixed Chern-Simons response.

The sensitivity to dislocations reminds one of the properties of \emph{weak} topological insulators which have been shown to trap low-energy modes on dislocations\cite{ran2009}. In fact, naively, an action of the form $S\sim b_a \blF\wedge \blT^a$ looks like the action for a massive 1+1-d Dirac fermion bound to dislocation lines with Burgers vectors parallel to $b_a.$  However, despite the similarity, we must resist, for now, the temptation to identify $b_a$ with a weak topological index (e.g., by letting $b_a$ be proportional to half a reciprocal lattice vector) until we more carefully consider the properties of $b^a.$ The weak invariant arises purely from the Lorentz-violating lattice structure which is not taken into account  so far. Additionally, $b_a$ has units of length, not inverse length as would be required for a weak invariant.  We could consider the quantity $m^2 b_a$ instead which does have the correct units. If one chose to ``quantize" the inverse area scale $m^2$ to be proportional to a lattice plaquette area, and have $b^a$ proportional to the lattice constant in the $a$-th direction, as would be appropriate for a spatial lattice vector, then this combined number would have the correct units and structure. Thus, it could be that for lattice models with discrete translation symmetry we would find a quantized $b_a,$ but in our continuum calculations this is not yet obvious. In fact, since the spatial components of $b_a$ are odd under time-reversal it should vanish identically in the dimensionally reduced time-reversal invariant insulator. We will discuss this further in the next subsection where we show that imposing a lattice structure induces a modular/periodic structure in $b_a$ that allows it to be non-vanishing even in a system with time-reversal symmetry. We will also see in the next subsection that in 3+1-d the parameter $b^a$ intrinsically arises from chiral translations in space(time) and for translationally invariant systems it gives rise to a momentum dependence of the chiral mass angle $\Theta.$

The third grouping of terms is harder to physically interpret. These terms arise in a material where $\theta_{ab}$ is non-zero, but we know of no such material. Just as the parameter $b_a$ is related to translations, $\theta_{ab}$ is related to rotations, and so similar terms to those in grouping III might appear in materials with topological phases determined by discrete rotation symmetries. It is possible that topological crystalline insulators/superconductors\cite{FuKaneWeak2007,teo2008,fu2011topological,hughes2011inversion,turner2012, fang2012bulk,teohughes1,slager2012, Benalcazar2013,Morimoto2013,RyuReflection2013,fang2013entanglement,HughesYaoQi13,Sato2013,Kane2013Mirror,jadaun2013} might generate such a response, or even secondary weak topological systems which have a non-trivial antisymmetric tensor as a topological invariant\cite{ran2010weak,HughesYaoQi13}. For these cases dislocation (torsion) and disclination (curvature) defects may have bound charges, e.g., electric charge, momentum, or spin. The spatial components of $\theta_{ab}$ are also odd under time-reversal ($a,b$ both spatial) and thus must vanish unless an additional symmetry structure is added such that $\theta_{ab}$ is only well-defined modulo some quantized amount. We will leave further discussion of this to future work.

\subsection{Intrinsic point of view}
In addition to understanding how these terms arise from dimensional reduction, it is also important to understand how they appear intrinsically in 3+1-d without reference to a 4+1-d parent system. We will carry out this calculation now. 
The Dirac operator in $d=4+1$ is given by
\beq
\slashed{\nabla}_{(5)} = \gamma^A\ue_A^{\mu}\left(\pa_{\mu}+\frac{1}{4}\omega_{\mu;AB}\gamma^{AB}+A_{\mu}+B_{\mu}\right)
\eeq\noindent where we remind the reader that $B\equiv \frac{1}{2}T^B(\underline{e}_A,\underline{e}_B)\;e^A.$
For the choice of frame in \eqref{bg}, we find
\beqn
\slashed{\nabla}_{(5)} &=& \gamma^a\blue_a^i\left(\pa_i+\frac{1}{4}\blw_{i;ab}\gamma^{ab}+\blA_i+\widetilde{B}_i\right)+\frac{1}{N}\gamma^4\pa_t-\frac{1}{N}\gamma^4b^i\left(\pa_i+\frac{1}{4}\blw_{i;ab}\gamma^{ab}+\blA_i\right)\nonumber\\
&+&\frac{1}{N}\gamma^4\Theta+\frac{1}{4N}\gamma^4\gamma^{ab}\theta_{ab}+\frac{1}{2N}\gamma^a\pa_aN-\frac{1}{2N}\gamma^4(\blD_ab^a+2\widetilde{B}_ab^a)
\eeqn
where we have used 
\beqn
B_a &=& \widetilde{B}_a + \frac{1}{2N}\pa_aN\nonumber\\
B_4 &=& \frac{1}{2N}\blD_ab^a-\frac{1}{N}b^c\widetilde{B}_c.
\eeqn
The  $w$-independent modes of the parent fermions $\Psi$ can be written in terms of $d=3+1$ fermions $\psi$ as $\Psi = \frac{1}{\sqrt{L N}}\psi$, where $L$ is some length scale. The intrinsic Dirac action becomes
\beq
S_{3+1}[\psi]=\int_{M_4}vol_4\left\{i\bar\psi\widetilde{\slashed{\nabla}}_{(4)}\psi-m\bar\psi\psi +\frac{i}{NL}\bar\psi\gamma^5\left(\Theta-b^i\widetilde{\nabla}_i+\frac{1}{4}\theta_{ab}\gamma^{ab}-\frac{1}{2\sqrt{g_5}}\pa_i\left(\sqrt{g_5}\;b^i\right) \right)\psi\right\}
\eeq
where $\sqrt{g_5} = N\;\mathrm{det}(\tilde{e})$, and we have relabeled $\gamma^4$ as $\gamma^5$. From the 3+1-d point of view, the first two terms look like the action of a Dirac fermion. The remaining $\gamma^5$ terms can be gauged away by performing a chiral gauge transformation, a chiral diffeomorphism, and a chiral Lorentz transformation with parameters $\Theta$, $b^a,$ and $\theta_{ab}$ respectively. However, these chiral transformations are anomalous in $d=3+1$, and the removal of the $\gamma^5$ terms from the above action can be done at the cost of accounting for the corresponding anomaly contributions in the effective action. These are precisely the terms which appear in the action \eqref{dimred} which we derived previously from dimensional reduction.
If we consider a trivial flat space geometry then the action reduces to 
\beq
S_{3+1}[\psi]=\int_{M_4}vol_4\left\{i\bar\psi\slashed{\partial}_{(4)}\psi-m\bar\psi\psi +im\bar\psi\gamma^5\left(\Theta-b^i\widetilde{\partial}_i+\frac{1}{4}\theta_{ab}\gamma^{ab} \right)\psi\right\}
\eeq where we have used the convention that $NL=1/m.$ If we chose a different convention then we would have to rescale $\Theta, b^i,$ and $\theta_{ab}$ so that their periodicity relations take simple forms, e.g., $\Theta \equiv \Theta+2\pi.$

Let us now try to understand the intrinsic meaning of the $\Theta, b^{i},$ and $\theta_{ab}$ parameters in a 3+1-d time-reversal invariant topological insulator, which is represented by this action. Under time-reversal  it is well-known that $\Theta$ transforms to $-\Theta.$ Thus, if time-reversal is a required symmetry, we must have the constraint that $\Theta=-\Theta$ or $2\Theta=0.$ If $\Theta$ is defined uniquely there is only one solution, i.e., $\Theta=0.$ However, there is a physical ambiguity such that  $\Theta$ is only well-defined up to a multiple of $2\pi$ and thus the symmetry condition becomes $2\Theta=0\mod\; 2\pi.$ This equation has two solutions: $\Theta=0 ,\pi$ which represent the trivial and topological time-reversal invariant insulator classes respectively. At the surface of the topological insulator phase $\Theta$ changes from $\pi$ to $0$ which has the effect of binding a half-quantum Hall effect to the region where $\Theta$ is varying. The ambiguity in $\Theta$ can be understood from the boundary perspective where we can add extra 2D layers to the surface that can change the quantized Hall conductance by an integer amount. If we add a layer with Hall conductance $\sigma=n\tfrac{q^2}{2\pi}$ then effectively $\Theta\to\Theta+2\pi n.$ The physical property determined by the time-reversal invariant bulk is the parity of $\Theta\mod\; \pi$ which is not changed by adding extra integer layers onto the surface. Thus, the parity of $(\tfrac{\Theta}{\pi}\mod 2)$ determines a $Z_2$ topological invariant.

Since we will need this type of argument soon, let us recount the periodicity argument for $\Theta.$ Once we have integrated out the fermions we recall that we produce the term in the effective action 
\begin{eqnarray}
S_{eff}&=&\frac{q^2}{8\pi^2\hbar}\int d^4x \Theta \epsilon^{\mu\nu\rho\tau}\partial_\mu A_\nu \partial_\rho A_\tau\nonumber\\
&=&\frac{q^2\Theta}{2\pi h}\int d^4 x  {\vec{E}}\cdot{\vec{B}}\nonumber\\
&=&\hbar N_{\phi_E}N_{\phi_B}\Theta.
\end{eqnarray}\noindent where $N_{\phi_{E/B}}$ are the integer numbers of electric and magnetic fluxes (where we have assumed all of the space-time directions are compact and only $E_z$ and $B_z$ are non-zero for simplicity). This means that the phase picked up by this term in a path-integral is
\begin{equation}
\exp \left[ \frac{i}{\hbar}S_{eff}\right]=\exp \left[iN_{\phi_E}N_{\phi_B}\Theta\right]
\end{equation}\noindent from which we clearly see that $\Theta$ is only defined mod $2\pi.$

Now, we want to consider the other intrinsic quantities $b_i$ and $\theta_{ab}.$ We also find that the spatial components of $b_i$ and the components of $\theta_{ab}$ where $a,b$ are both spatial indices are odd under time-reversal. If these intrinsic quantities are uniquely defined it implies that they must vanish identically in a time-reversal invariant insulator. However, if we require discrete spatial symmetries it is possible to induce periodicity relations such that we can find non-trivial values even in a time-reversal symmetric system. As an example, let us impose a discrete translation symmetry with spatial lattice vectors $\vec{a}_1,\vec{a_2},$ and $\vec{a}_3$ such that the system is symmetric under the discrete translations by $\vec{R}_{map}=m\vec{a}_1+n\vec{a}_2+p\vec{a}_3$ for any $m,n,p \in \mathbb{Z}.$ For every spatial lattice there is a corresponding reciprocal lattice spanned by ${\vec{G}}_1, {\vec{G}}_2,$ and ${\vec{G}}_3$ which satisfy 
${\vec{a}}_{i}\cdot{\vec{G}}_j = 2\pi\delta_{ij}.$

For $b_a$ we will focus on one piece of the effective action:
\begin{eqnarray}
S_{eff}&=&\frac{q_Tqm^2b_a}{4\pi^2}\int d^4 x\epsilon^{\mu\nu\rho\tau} \partial_{\mu}A_{\nu}\partial_{\rho} e_{\tau}^{a}\\
&=&\hbar \frac{q_T m^2 b_i}{2\pi}N_{\phi_E}{\cal{B}}^i.
\end{eqnarray}\noindent where $N_{\phi_{E}}$ is the integer number of electric flux quanta and ${\cal{B}}_i$ is the total Burgers' vector coming from the torsion magnetic flux (where again we have assumed all of the space-time directions are compact  and only $E_z$ and $T_{xy}^i$ were non-zero for simplicity). This means that in a path-integral formalism the phase picked up due to this term is
\begin{equation}
\exp \left[i\tfrac{q_Tm^2}{2\pi} N_{\phi_E}b_i{\cal{B}}^i\right]
\end{equation}\noindent
 from which we see that $\frac{q_T m^2 b_i{\cal{B}}^i}{2\pi}$ is only defined mod $2\pi.$ To clearly see the implications of this condition let us rewrite the phase as ${\cal{G}}_i{\cal{B}}^i$ (which is defined mod $2\pi$) where we have defined ${\cal{G}}_i=\frac{q_T m^2}{2\pi} b_i.$ 
 
 Now, under time-reversal ${\cal{G}}_i\to -{\cal{G}}_i,$ and thus we must have $ {\cal{G}}_i{\cal{B}}^i=-{\cal{G}}_i{\cal{B}}^i$ for a time-reversal invariant insulator. Because of the periodicity we can have ${\cal{G}}_i{\cal{B}}^i=n\pi$ for some integer $n.$ Since the total Burgers' vector ${\cal{B}}^i$ is itself a real-space lattice vector this constraint implies  that ${\cal{G}}_i$ is either a reciprocal lattice vector (for $n$ even) or a half-reciprocal lattice vector (for $n$ odd). The latter is the non-trivial case, and is the familiar result of a weak topological invariant. 
 
 One consequence of this result can be determined from this effective action. Let us assume that ${\cal{G}}_i$ is non-vanishing such that the term in the effective action above becomes
\begin{equation}
 S_{eff}=\frac{q}{2\pi}\int d^4 x{\cal{G}}_i\epsilon^{\mu\nu\rho\tau} \partial_{\mu}A_{\nu}\partial_{\rho} e_{\tau}^{i}.
   \end{equation}\noindent For a straight dislocation-line localized at the origin in the $xy$-plane, and extended in the $z$-direction with Burgers' vector ${\cal{B}}^i,$ we can evaluate the action to find
   \begin{equation}
  \frac{q}{2\pi} \int dz dt {\cal{G}}_i{\cal{B}}^i \epsilon^{\mu\nu}\partial_{\mu}A_{\nu}
   \end{equation}\noindent where now $\mu,\nu=t, z.$ This is exactly $N_D=\tfrac{1}{\pi} {\cal{G}}_i{\cal{B}}^i$ copies of the response action for a 1D Dirac fermion localized on the dislocation coupled to a scalar/axion field. This result matches what was found using more conventional methods in Ref. \cite{ran2009}. Thus, for a lattice system with discrete translation symmetry we can interpret the vector $b_i$ as being connected to a weak topological invariant. This hints that $\theta_{ab}$ might be non-zero in systems with time-reversal symmetry and discrete rotation symmetries. We will leave the treatment of these systems to future work.

\section{Discussion and Conclusion}
In this article we set out to understand the response of several classes of condensed matter systems to geometric perturbations. By utilizing the anomaly polynomial technology in a high space-time dimension, we were able to cleanly derive the response coefficients of the charge, stress, and spin currents in the presence of the full range of geometric and electromagnetic perturbations including curvature and torsion contributions. Our results include both universal quantized responses, e.g., the Hall conductance, gravitational Chern-Simons response, and curvature corrections to the Hall viscosity, and seemingly less universal quantities, e.g., the leading-order Hall viscosity term, the magneto-Hall viscosity, and the torsion contribution to the chiral anomaly. These latter response coefficients all share a dependence on a (possibly non-universal) intrinsic length scale of the system and are not generically quantized since they are attached to terms in the effective action which are completely gauge, diffeomorphism, and Lorentz invariant. This invariance does not allow for the enforcement of a quantizing constraint in contrast to what is found, for example, for a non-Abelian Chern-Simons term under gauge transformations.

In addition to providing the bulk response coefficients, we presented a spectral-flow/Callan-Harvey analysis for many of the different types of responses. One of the most interesting examples is the explanation of how 3+1-d Weyl fermions are anomalous in the presence of torsion. This anomaly is encoded in the Nieh-Yan term and can be explained by considering the low-energy physics of a Weyl fermion in a uniform background torsion magnetic field. Such a field generates torsional Landau level type states and there is a special zeroth Landau level. For the more conventional configuration of Weyl fermion in a uniform $U(1)$ magnetic field, this zeroth Landau level has a 1+1-d chiral dispersion along the direction of the magnetic field. The resulting low-energy theory has many degenerate copies of a 1+1-d chiral fermion, which are anomalous in the presence of an electric field due to the 1+1-d chiral anomaly. For the torsional case, the dispersion is not linear. In fact, for a Weyl node with a fixed chirality, the low-energy theory in the presence of a torsion magnetic field has 1+1-d modes with group velocities parallel \emph{and} anti-parallel to the field. It is exactly this difference which allows for the anomaly when a torsion electric field is applied as we discussed earlier. The torsion electric field deforms the velocities of the low-energy modes and transfers states past the high-energy cutoff which, in total, results in an anomalous process.  

Finally we provided two possible applications of our calculations in the visco-elastic response of Weyl semi-metals and 3+1-d time-reversal invariant topological insulators. For the Weyl semi-metals we showed that there is both a 3D anomalous Hall viscosity and an analog the chiral magnetic effect in which momentum current flows along dislocation lines in the absence of an applied torsional electric field.  For the 3+1-d topological insulator we showed that the surface, in the presence of a time-reversal breaking perturbation, will exhibit a half Hall viscosity (though the half just means that the regularized coefficient is half of the coefficient for a regularized bulk 2+1-d Dirac fermion, and not that it is quantized), and in fact all of the 2+1-d geometric responses, but with half of the coefficient of the intrinsic, bulk 2+1-d Dirac fermion.  We also found anisotropic response terms that have not previously been discussed. We argued that these anisotropic responses are connected to topological phases protected by translation and rotation symmetries.

\section*{Acknowledgements}
We would like to acknowledge G. Y. Cho, C. Hoyos, K. Landsteiner, S. Ramamurthy, and M.A.H. Vozmediano for useful discussions. TLH thanks the NSF CAREER DMR-1351895 for support and the ICMT at UIUC. Some support (OP,RGL) for this research has been provided by the U.S. DOE contract DE-FG02-13ER42001. 

\appendix
\section{Asymptotic expansions from supersymmetric quantum mechanics}\label{app:asymptotic}
In section \ref{sec:anomalies} we encountered traces of the form
\beq\label{traceeg}
\mathrm{Tr}_{2n}\;\Gamma^{2n+1}e^{s\slashed{\msD}_{2n}^2}
\eeq
and in particular, their asymptotic expansions (in powers of \(s\)) in the limit \(s\rightarrow 0\). We can use \(\mathcal{N}=1\) supersymmetric quantum mechanics to evaluate these expressions. We will not provide details, but rather only sketch the essential ideas involved; see \cite{AlvarezGaume:1983ig, BVN, DeBoer:1995hv, deBoer:1995cb} for details. We also note that the use of \(\mathcal{N}=1\) Supersymmetric quantum mechanics (SQM) in computing Chiral anomalies or Atiyah Singer index densities on torsional backgrounds has been discussed before in \cite{Peeters:1999ks} (see also \cite{Chandia:1998nu}), and in the special case of vanishing Nieh-Yan four form in \cite{Mavromatos:1987ru, Kimura:2007xa, Bismut:1989} (see also the older works like \cite{Obukhov:1983mm,Yajima:1985} etc.).

Let \(\Sigma\) be a manifold with metric \(g_{ij}\), a torsional connection \(\omega_{i;ab}\), and a $U(1)$ gauge field $A$. The action for \(\mathcal{N}=1\) SQM in the presence of torsion is given by
\beqn
S_{SQM} &=&\int ds\;\left(\frac{1}{2}g_{ij}\dot x^i\dot x^j+\frac{i}{2}\chi^a(\delta_{ab}\dot \chi^b+\dot x^k\lcw_{k;ab}\chi^b)-i\frac{\torcplg}{4}\dot x^k\chi^a\chi^b H_{kab}-\frac{\torcplg}{2}\frac{1}{4!}N_{abcd}\chi^a\chi^b\chi^c\chi^d\right.\nonumber\\
&+&\left. i\bar c(\dot c+i\dot x^kA_kc) + \frac{i}{2}\bar c F_{ab}\chi^a\chi^bc\right)
\eeqn
where \(x^i\) are local coordinates on \(\Sigma\), \(\chi^a\) are one-component real fermions, while \(c\) and \(\bar c\) are one-component complex fermions, and the notation $\dot{x}_j\equiv \partial_s x^j.$ We have also introduced the notation $N=dH$, and $F=dA$. The theory is invariant under the supersymmetry transformations \(\delta x^i = i\epsilon \chi^i,\;\delta \chi^i = -\epsilon \dot{x}^i\), with the supercharge
\beq 
Q = i\chi^a\underline{e}^i_a(p_i-\frac{i}{2}\lcw_{i,bc}\chi^b\chi^c+\bar cA_ic)-\frac{\torcplg}{2}\frac{1}{3!}H_{a;bc}\chi^a\chi^b\chi^c
\eeq
(\(p_i\) being the momentum conjugate to \(x^i\)), and the Hamiltonian \(\mathcal{H}=-Q^2\).
Upon quantization, we must replace \(p_i \rightarrow -i\partial_i\) and \(\chi^a \to \frac{1}{\sqrt{2}}\gamma^a\). The supercharge becomes \(Q = \frac{1}{\sqrt{2}}\slashed{\msD}+\cdots\), while the Hamiltonian is \(\mathcal{H} = -\frac{1}{2}\slashed{\msD}^2+\cdots\), up to operator ordering ambiguities indicated by \(\cdots\). Further, the fermion number operator in SQM, \((-1)^F\), is proportional to the chirality matrix \(\Gamma^{2n+1}\). 

This is what allows us to compute traces of the type \eqref{traceeg} - the Hilbert space of \(\mathcal{N}=1\) SQM essentially furnishes a representation of Dirac fermions on \(\Sigma\). In fact, the trace \eqref{traceeg} is proportional to the \emph{Witten index} of supersymmetric quantum mechanics 
\beq
\mTr\;(-1)^Fe^{-\beta\hat {\mathcal{H}}}
\eeq
with \(s=\frac{1}{2}\beta\). Such a trace over the Hilbert space is easiest to compute using the path integral representation. To handle the operator ordering ambiguities, we follow the time-slicing prescription for the path integral \cite{BVN}, at the expense of the counter-terms 
\beq
L_{ct}=\frac{1}{8}g^{ij}{{\mathring{\Gamma}}^k}_{\;\;il}{{\mathring{\Gamma}}^l}_{\;jk}+\frac{1}{16}{\omega^{(\torcplg)}}_{i;ab}{\omega^{(\torcplg)}}^{i;ab}-\frac{\torcplg^2}{16}\frac{1}{3!}H_{a;bc}H^{a;bc}.
\eeq
The path integral corresponding to \(\mTr\;(-1)^Fe^{-\beta\hat{\mathcal{H}}}\) is then given by
\beq
\mTr\;(-1)^Fe^{-\beta\hat{\mathcal{H}}} = \int_{PBC} [dx^id\chi^ada^idb^idc^i]e^{-\int_{-\beta}^0ds\; L_E}
\eeq
where \(a_i\) are commuting ghosts, \(b_i\) and \(c_i\) are anti-commuting ghosts,\footnote{The ghosts are introduced to exponentiate factors of \(\mathrm{det}(e)\) which arise due to insertion of complete set of position eigenstates in the discretized path integral.} and \(L_E\) is the Euclidean time Lagrangian given by
\beqn
L_E  &=&\frac{1}{2}g_{ij}\dot x^i\dot x^j+\frac{1}{2}\delta_{ab}\chi^a\dot \chi^b+\frac{1}{2}\dot x^k\omega^{(\torcplg)}_{k;bc}\chi^b\chi^c+\frac{\torcplg}{2}N_{abcd}\chi^a\chi^b\chi^c\chi^d\nonumber\\
&+& \bar c(\dot c+\dot x^kA_k c) - \frac{i}{2}\bar c F_{ab}\chi^a\chi^b c+ \frac{1}{2}g_{ij}(a^ia^j+b^ic^j)+L_{ct}.
\eeqn
Here \(x^i\) and \(a^i\) have periodic boundary conditions, \(\chi^a\) have periodic boundary conditions because of the \((-1)^F\) in the trace (which is what the subscript \(PBC\) indicates), and \(b_i,\; c_j,\;c\) and \(\bar c\) all have anti-periodic boundary conditions. In the absence of \((-1)^F\), \(\chi^a\) acquire anti-periodic boundary conditions \((APBC)\). Finally, the \(\beta \rightarrow 0\) limit is just the weak coupling limit in SQM, where we can do perturbation theory. In this way, \(\mathcal{N}=1\) SQM allows us to compute the asymptotic expansions in \eqref{traceeg} using standard techniques of field theory. For instance, using the method described above, we find the asymptotic expansion for \(\mTr_4\;\Gamma^5e^{s\slashed{\msD}^2}\) in four dimensions is given by
\beq
\mTr_4\;\gamma^5e^{s\slashed{\msD}_4^2}\simeq\int_{\Sigma_4}\left(\frac{\torcplg}{16\pi^2s} dH+\frac{1}{8\pi^2}F\wedge F+\frac{1}{192\pi^2}\mathrm{tr}\;{R^{(-\torcplg)}}\wedge {R^{(-\torcplg)}}+\frac{\torcplg}{96\pi^2}d*d*dH+ O(s)\right).
\eeq
The same procedure can be applied for computing such asymptotic expansions in higher dimensions. For instance, in six dimensions we get
\beqn
\mTr_6\;\Gamma^7e^{s\slashed{\msD}_6^2} &\simeq& \int_{\Sigma_6}\left(-\frac{\torcplq}{32\pi^3 s}F\w dH-\frac{1}{384\pi^3}F\w \mtr\;R^{(-\torcplq)}\wedge R^{(-\torcplq)}-\frac{1}{48\pi^3}F\w F\w F\right.\nonumber\\
&-&\left. \frac{\torcplq}{192\pi^3}d\left(F\wedge *d*dH\right)+\frac{\torcplq}{384\pi^3 }d*d*(F\w dH)+O(s)\right)
\eeqn

\section{Divergences in higher dimensions}\label{app:expansion}
 In this section, we discuss the torsional divergences in anomaly polynomials in arbitrary dimensions, and their Pauli-Villar's regularization. As we noted in Section \ref{sec:anomalies}, divergences of the anomaly polynomials in \(d=4n\) and \(d=4n+2\) are the same. Therefore, to study the cancellation of divergences, it suffices to focus on the anomaly polynomials in \(d=4n\). We have dealt with the case of \(n=1\) explicitly in section \ref{sec:anomalies}, so we now take \(n>1\). Now in \(d=4n\), we have the asymptotic expansion
\beq
\mathrm{Tr}_{4n}\Gamma^{4n+1}e^{s\slashed{\msD}_{4n}^2}\simeq \frac{1}{s^{n}}\sum_{k=0}^{\infty}b_ks^k=\frac{1}{s^{n}}\sum_{k=0}^{n}b_ks^k+O(s)
\eeq
where the \(b_k\) are \(4n\)-form polynomials made out of curvature, torsion, and their covariant derivatives (see Eqs. \eqref{ind4} and \eqref{ind6}). For instance, in \(d=4n\) we have \(b_0 \propto \int_{M_{4n}}(dH)^n\), while in \(d=4n+2\) we have \(b_0\propto \int_{M_{4n+2}}F\wedge (dH)^n\).\footnote{The explicit form of \(b_k\) is difficult to compute in arbitrary dimension in the presence of torsion.}  As before, we will not consider \(O(s)\) terms because these lead to \(1/m\) corrections in the anomaly polynomial. The un-regulated anomaly polynomial thus takes the form
\beq
\mathcal{P}^{(0)}(m) = \lim_{\epsilon \to 0}i\sqrt{\pi}m\sum_{k=0}^{n}\Gamma_{\epsilon}(-n+\frac{1}{2}+k,m^2)\;b_k
\eeq
where 
\beq
\Gamma_{\epsilon}(\alpha,m^2) = \int_{\epsilon}^{\infty} s^{\alpha-1}e^{-sm^2}
\eeq
with \(\epsilon = \frac{1}{\Lambda^2}\). Therefore, the UV divergences of the anomaly polynomial in \(d=4n\) are contained in
\beq
\left\{ m\Gamma_{\epsilon}(-n+\frac{1}{2}+k,m^2)\right\},\;\; 0\leq k < n
\eeq
where \(\epsilon = \frac{1}{\Lambda^2}\). Let us examine these integrals schematically:
\beq
m\Gamma_{\epsilon}(-n+\frac{1}{2}+k,m^2)= a^{(k)}_{0}m\Lambda^{2n-2k-1}+a^{(k)}_{1}m^3\Lambda^{2n-3-2k}+\cdots a^{(k)}_{n-k-1}m^{2n-1-2k}\Lambda+a^{(k)}_{n-k}\mathrm{sign}(m) m^{2n-2k} \label{schmdivs}
\eeq 
where the \(a^{(k)}_{\ell}\) are finite numerical coefficients. As before, we introduce Pauli-Villar's regulator fermions with masses $M_I$ and parities $C_I$, where $I=1,2\cdots N$. For convenience, we label the original low-energy fermion as $I=0$ with $M_0 = m$ and $C_0 = 1$. From equation \eqref{schmdivs}, it is amply clear that to cancel all the UV divergences, we must require
\beq
\sum_{I=0}^{N}C_IM_I=0,\;\sum_{I=0}^NC_IM_I^3=0,\cdots, \sum_{I=0}^NC_IM_I^{2n-1}=0. \label{constraints}
\eeq
Additionally, we must also check the finiteness of the remaining \(\Lambda\)-independent coefficients 
\beq
\alpha_0=\sum_{I=0}^{N}a^{(0)}_nC_I\mathrm{sign}(M_I)M_I^{2n},\;\alpha_1=\sum_{I=0}^{N}a^{(1)}_{n-1}C_I\mathrm{sign}(M_I)M_I^{2n-2},\cdots,\; \alpha_n=\sum_{I=0}^{N}a^{(n)}_0C_I\mathrm{sign}(M_I) \label{coeffs}
\eeq 
in both the topological and trivial phases, where we note that \(a^{(k)}_{n-k} = \widetilde\Gamma(-n+k+\frac{1}{2})\), where \(\widetilde{\Gamma}\) stands for analytic continuation of the Gamma function. Having done so, the regulated anomaly polynomial is
\beq
\mathcal{P}(m)=\sum_{k=0}^n\alpha_k(m)\;b_k.
\eeq
In order to see that the constraints in \eqref{constraints} can be satisfied, and that the coefficients \(\{\alpha_k\}\) are finite, we go back to the lattice Dirac model in \(d=4n-1\). We will work with the lattice Hamiltonian
\beq
H = \sum_{\vec{k}}c^{\dagger}_{\vec{k}}\left\{m+\mu_{bw}\left(4n-2-\sum_{\mu=1}^{4n-2}\cos(k_\mu)\right)\gamma^{4n-1}+v_F\sum_{\mu=1}^{4n-2}\sin(k_\mu)\gamma^\mu\right\}c_{\vec{k}}.
\eeq
The Hamiltonian has \(2^{4n-2}\) Dirac points - the one at \(\vec{k}= (0,0,\cdots, 0)\) will be labelled by \(I=0\) and interpreted as the low-energy Dirac fermion, while the other fermions will be labelled by \(I\) from 1 to \(4n-2\) and interpreted as Pauli-Villar's regulator fermions. The fermions have a degenracy of \(N_I=\dbinom{4n-2}{I}\), parities \(C_I = (-1)^I\), and masses \(M_I = (m+2I\mu_{bw})\). Now in this model, all of the UV constraints \eqref{constraints} translate to
\beq
\sum_{I=0}^{4n-2}C_IN_I=0,\;\sum_{I=0}^{4n-2}C_IN_II=0,\;\sum_{I=0}^{4n-2}C_IN_II^2=0\cdots, \sum_{I=0}^{4n-2}C_IN_II^{2n-1}=0.
\eeq
These constraints are obviously satisfied on account of the following identity
\beq
\sum_{I=0}^{4n-2}\dbinom{4n-2}{I}(-1)^II^{k}=\left.\left(x\frac{\partial}{\partial x}\right)^k(1-x)^{4n-2}\right|_{x=1}=0,\;\;\;\;\;\; \forall\;0\leq k \leq 2n-1. \label{ident}
\eeq
Moving on to the finiteness of the coefficients \eqref{coeffs}, we have to deal with these separately for \(m<0\) and \(m>0\). For \(m>0\), these are all zero (for \(n>1\)) as a result of identity \eqref{ident}. On the other hand for \(m<0\), we get
\beq
\alpha_k = -2m^{2n-2k}\widetilde\Gamma\left(-n+k+\frac{1}{2}\right).
\eeq
This proves that the parity-odd fermion effective action for the lattice Dirac model is finite in arbitrary dimension even in presence of torsion, provided we take into account the contributions from spectator fermions. 

\section{Energy Spectra for 3+1-d Weyl Fermions}\label{app:Weylspectra}
\subsection{$U(1)$ Magnetic Field}
Let us  consider the energy spectra of isolated Weyl fermions in the presence of a uniform $U(1)$ magnetic field. This result is well-known but we recount it here to compare it with the case of the torsional magnetic field. We take the spatial geometry to be $\Sigma_3 = \re\times S^1\times S^1$, parametrized by $x^i=(x^1,x^2,x^3)$ respectively. The $U(1)$ gauge field is taken to be \(A = f(x) dy\). We chose the Weyl basis for gamma matrices
\begin{equation}
\gamma^0= \left(\begin{matrix}0 & 1\\ 1 & 0\end{matrix}\right),\;\gamma^i= \left(\begin{matrix}0 & -\sigma^i\\ \sigma^i & 0\end{matrix}\right),\;\gamma^5= \left(\begin{matrix}-1 & 0\\ 0 & 1\end{matrix}\right).
\end{equation}
With this, the Dirac equation for the left and right modes \(\psi_L = \frac{1-\gamma^5}{2}\psi_L\), \(\psi_R = \frac{1+\gamma^5}{2}\psi_R\) becomes
\begin{equation}
i\left(\partial_0 -\sigma^i(\partial_i+iqA_i)\right)\psi_R = 0,\;i\left(\partial_0 +\sigma^i(\partial_i+iqA_i)\right)\psi_L = 0.
\end{equation}
Let us now concentrate on the left handed modes, and we will drop the \(L\) subscript from here on. If \(\psi\) is a zero mode of \(\partial_0 +\sigma^i(\partial_i+iqA_i)\), then so is \((\partial_0 -\sigma^i(\partial_i+iqA_i))\psi\) (because the \(A_i\) are time independent), and hence we try to solve the second order equation\footnote{Eventually, we should be careful to discard solutions of \((\partial_0 -\sigma^i(\partial_i+iqA_i))\psi = 0\)}
\begin{eqnarray}
\left(\partial_0^2 -\sigma^i(\partial_i+iqA_i)\sigma^j(\partial_j+iqA_j)\right)\psi &=& 0.
\end{eqnarray}
Using \(\sigma_i\sigma_j = \delta_{ij}+i\epsilon_{ijk}\sigma_k\) 
and the fact that \(p_2, p_3\) are good quantum numbers, we find that energy eigenfunctions must satisfy
\begin{equation}
\left(-\partial_1^2+(p_2+qA_2)^2+p_3^2+\frac{q}{2}\epsilon_{ijk}F_{ij}\sigma^k\right)\psi = E^2\psi.\label{mastereq}
\end{equation}

Now let us consider the special case of a uniform magnetic field. Choose \(A = Bx^1dx^2\) corresponding to a uniform magnetic field \(B\) parallel to \(x^3\). Substituting into Eq. \ref{mastereq} we find
\begin{equation}
\left(-\partial_1^2+(qB)^2\left(x^1+\frac{p_2}{qB}\right)^2+p_3^2+qB\sigma^3\right)\psi = E^2\psi
\end{equation}
which is the simple harmonic oscillator equation with frequency \(|qB|\). The dispersion relations are
\begin{equation}
E(\ell, p_3, \sigma_3) = \pm \left(p_3^2+2|qB|(\ell+\frac{1}{2})+qB\sigma_3\right)^{1/2},\hspace{0.5cm}\ell=0,1,2,\cdots, \sigma_3 =\pm 1
\end{equation}
and the wavefunctions are
\begin{equation}
\psi(\ell,p_3, \sigma_3) = A_{\ell}e^{ip_3x^3+ip_2x^2}e^{-|qB|x_1^2/2}H_{\ell}\left(\sqrt{|qB|}(x^1+\frac{p_2}{qB})\right)|\sigma_3\rangle
\end{equation}
with \(A_{\ell} = \frac{1}{2^{\ell} \ell!}(|qB|)^{1/4}\) being the normalization. 

The solutions corresponding to \(\ell=0, \sigma_3 = -\mathrm{sign}(qB)\) are the gapless modes \(E(p_3) = \pm p_3\). But note that we still need to eliminate the spurious solutions which satisfy \((i\partial_0 -i\sigma^i(\partial_i+iqA_i))\psi = 0\), i.e.
\begin{equation}
\left(\begin{matrix} E+p_3 & (p_1-ieB(x^1+p_2/qB))\\p_1+iqB(x^1+p_2/qB) & E - p_3\end{matrix}\right)\psi(\ell,p_3,\sigma) = 0.
\end{equation}
\begin{figure}[!t]
\centering
\includegraphics[height=5.5cm]{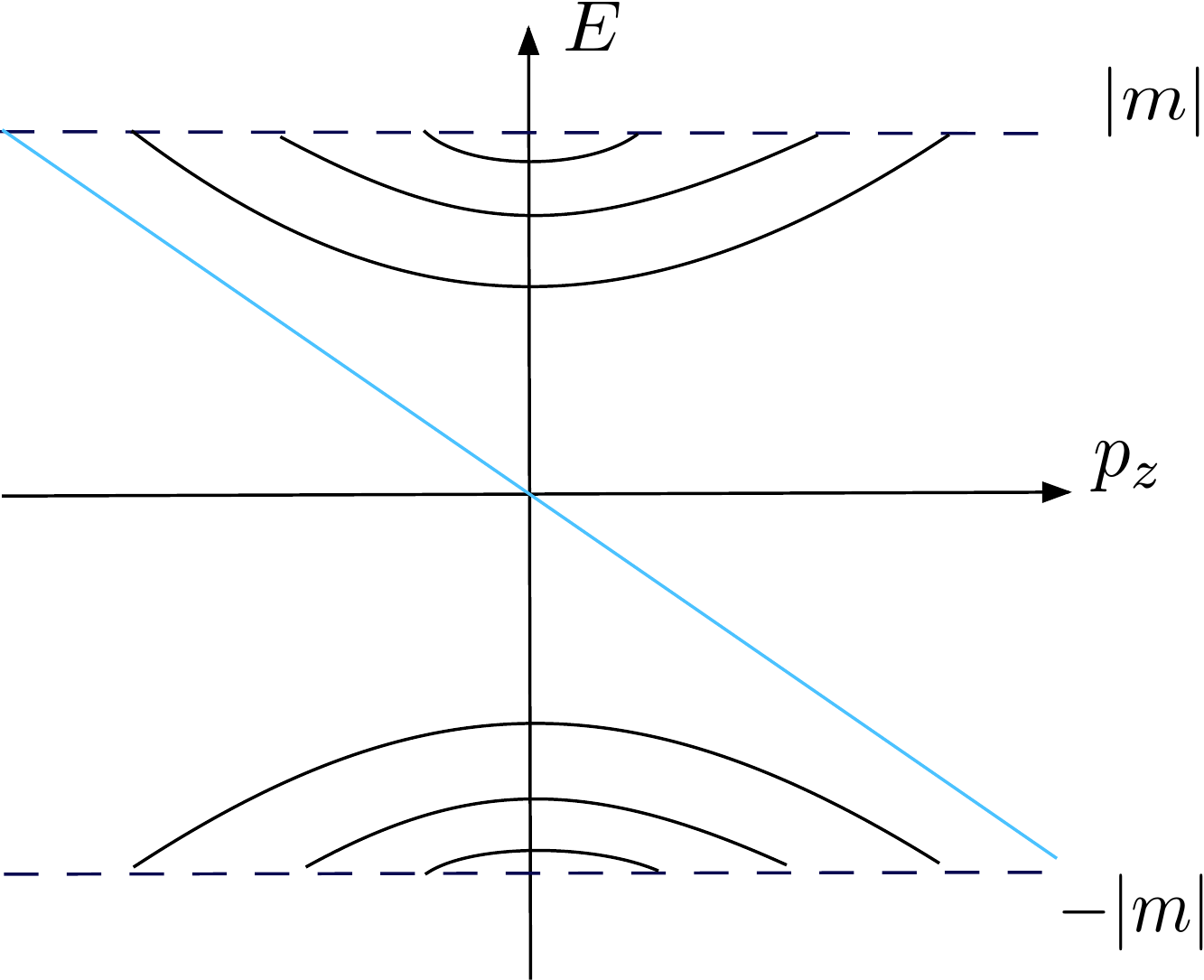}
\caption{An illustration of the energy spectrum for a left-handed Weyl fermion in the presence of a uniform background $U(1)$ magnetic field. The linear dispersing mode is the zeroth Landau level and the gapped modes are higher Landau levels (or bulk states). We have drawn a mass cut-off $\pm\vert m\vert$ to represent the energy at which the low-energy chiral modes begin to couple with the bulk modes in the gapped topological insulator and lose their chirality and boundary localization properties.}\label{fig:chiralDisp}
\end{figure}
Thus, the \(E= \mathrm{sign}(qB)p_3\) mode gets eliminated, and we are left with only one gapless branch
\begin{equation}
E = -\mathrm{sign}(qB) p_3.
\end{equation}
The number of states for each \(p_3\) is given by \(\frac{|q\Phi_B|}{2\pi}\), which comes from demanding \(-\frac{L_1}{2}<\frac{p_2}{qB}<\frac{L_1}{2}\); here \(\Phi_B\) is the magnetic flux. If we had chosen to study the right-handed chirality then $-\mathrm{sign}(qB) p_3$ would have been eliminated and the remaining mode would be $E=+\mathrm{sign}(qB) p_3.$

\subsection{Torsion Magnetic Field}
Now set the $U(1)$ magnetic field to zero, and consider the following co-frame and its dual frame
\begin{equation}
e^0 = dt,\; e^1 = dx^1,\; e^2 = dx^2,\; e^3 =dx^3 + f(x^1)dx^2,
\end{equation}
\begin{equation}
\ue_0 = \partial_0,\; \ue_1 = \partial_1,\; \ue_2 = \partial_2- f(x^1)\partial_3,\; \ue_3 = \partial_3.\nonumber
\end{equation}
We will set the spin connection to zero for simplicity. In this case, the above co-frame is torsional with \(T^3 =  de^3 = \partial_1f(x^1)dx_1\wedge dx_2\). The Dirac operator becomes
\begin{equation}
i\slashed{D} =  i\left(\gamma^0\partial_0+\gamma^1\partial_1+\gamma^2(\partial_2-f(x^1)\partial_3)+\gamma^3\partial_3\right).
\end{equation}
For the left-handed Weyl fermions, the Dirac equation reduces to
\begin{equation}
i\left(\partial_0+\sigma^1\partial_1+\sigma^2(\partial_2-f(x^1)\partial_3)+\sigma^3\partial_3\right)\psi_L = 0,
\end{equation}
and since \(p_2, p_3\) are good quantum numbers, we can write the above as
\begin{equation}
\left(i\partial_0+i\sigma^1\partial_1-\sigma^2(p_2-f(x^1)p_3)-\sigma^3p_3\right)\psi_L = 0.
\end{equation}

We notice that this looks exactly like the Dirac equation with a $U(1)$ gauge field \(A = -\frac{p_3}{q} f(x^1) dx_2= -\frac{p_3}{q}\delta e^3\) and field strength \(F = -  \frac{p_3}{q} T^3\). Thus \eqref{mastereq} becomes
\begin{equation}
\left(-\partial_1^2+(p_2-p_3\delta e^3_2)^2+p_3^2-\frac{p_3}{2}\epsilon_{ijk}T^3_{ij}\sigma^k\right)\psi = E^2\psi.
\end{equation}
To understand the spectrum, we first notice that for \(p_3 = 0\), the spectrum is just \(E(p_1,p_2,p_3 = 0) = \pm (p_1^2+p_2^2)^{1/2}\). This must be the case because the $p_3=0$ mode is not sensitive to translations/torsion. In order to proceed, we choose \(f(x^1) = Cx^1\), this leads to a uniform torsion magnetic field \(T^3 = C dx^1\wedge dx^2\). The spectrum for \(p_3 \neq 0\) is similar to the case of the uniform magnetic field
\begin{equation}
E(\ell,p_3,\sigma_3) = \pm \left(p_3^2+2|Cp_3|(\ell+\frac{1}{2})-Cp_3\sigma^3\right)^{1/2}\hspace{0.5cm}\ell=0,1,2\cdots, \sigma_3=\pm 1.\label{eq:TLL}
\end{equation}
\begin{figure}[!t]
\centering
\includegraphics[height=5.5cm]{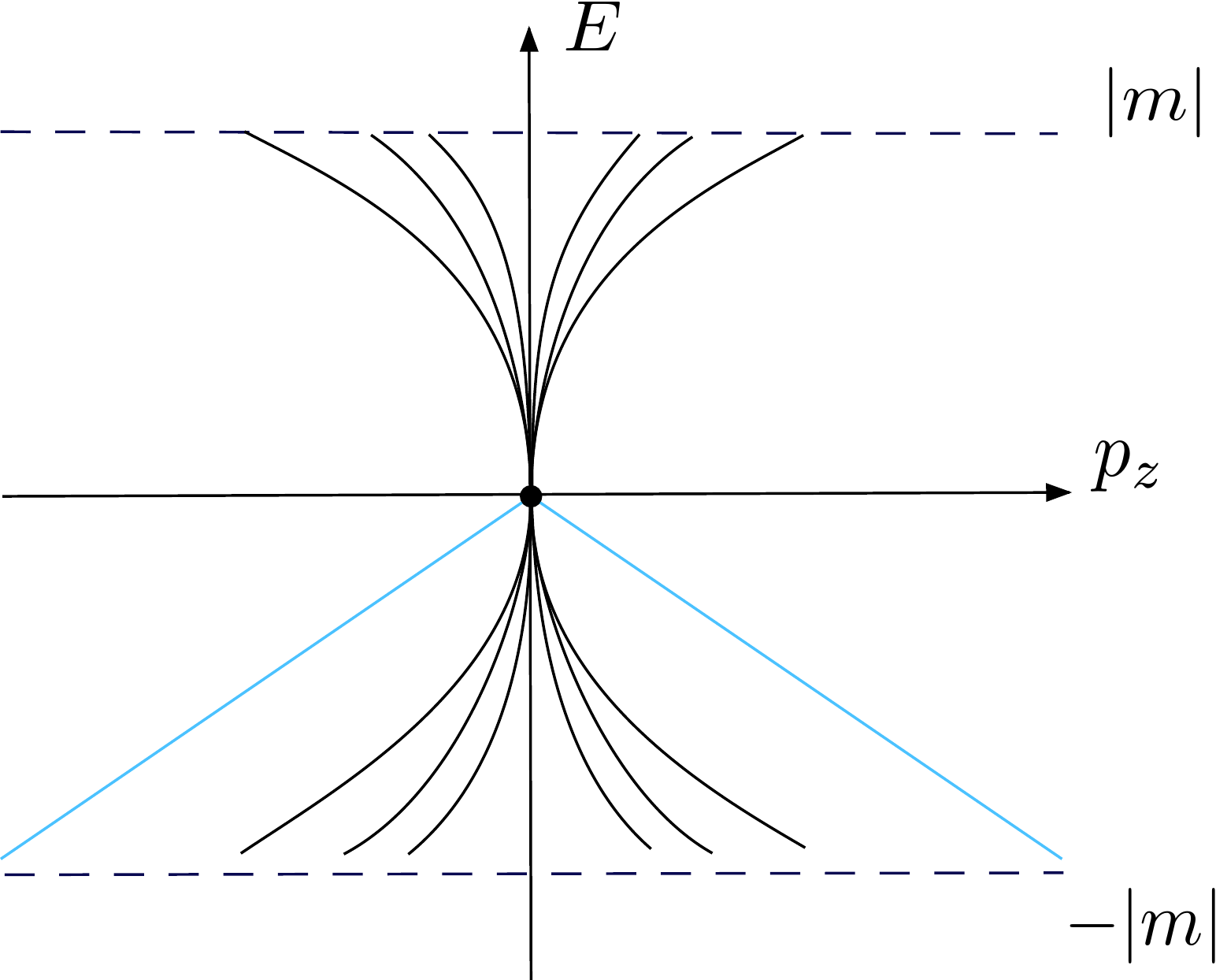}
\caption{An illustration of the energy spectrum for a 3+1-d left-handed Weyl fermion in the presence of a uniform background torsion magnetic field. The downward dispersing (blue) curve represents the zeroth Landau level while the non-linear (black) curves represent higher Landau levels as given in Eq. \ref{eq:TLL}. This should be compared with the result for a $U(1)$ magnetic field shown in Fig. \ref{fig:chiralDisp}.}
\end{figure}
Notice that for \(\ell=0, \sigma_3=\mathrm{sign}(Cp_3)\), the spectrum is simply given by \(E= \pm p_3\). But once again we have to be careful to eliminate the spurious zero mode. This is delicate, so let us work this out explicitly; the spurious mode satisfies 
\begin{equation}
\left(\begin{matrix} E+p_3 & p_1-i(p_2-Cp_3x^1)\\ p_1+i(p_2-Cp_3x^1) & E - p_3\end{matrix}\right) \psi = 0.
\end{equation}
We find that \(E=-\mathrm{sign}(Cp_3)p_3\) should be eliminated. Thus the remaining gapless (\(p_3 \neq 0\)) mode is
\begin{equation}
E  = \mathrm{sign}(Cp_3)p_3,\;\sigma_3 = \mathrm{sign}(Cp_3).
\end{equation}
The opposite chirality mode will have $E  = -\mathrm{sign}(C p_3)p_3,\;\sigma_3 = -\mathrm{sign}(Cp_3)$. 
This is different from the case of the $U(1)$ magnetic field in two important ways. First, the number of states for each \(p_3\neq 0\) is now given by \(\frac{|p_3\Phi_T|}{2\pi}\), where \(\Phi_T = CL_1L_2\) is the torsion magnetic flux. Second the right-handed and left-handed fermions do not give rise to 1+1-d fermion branches with a constant group velocity. In fact, one chirality disperses upward and the other chirality disperses downward. The fact that the association between the different 1+1-d branches and the chirality is modified is exactly what gives rise to the torsional contribution to the chiral anomaly.

\section{Dimensional reduction of Curvature terms}\label{app:curvature}
In Section \ref{sec:3dti}, we performed the dimensional reduction from the 4+1-d topological insulator to the 3+1-d topological insulator. Here, we wish to demonstrate the additional terms which arise due to curvature. We recall the form of the geometry fields we employ
\beqn
e^a &=&b^a dw +\ble^a_idx^i\nonumber\\
e^4 &=& N dw\nonumber\\
\ue_a &=& \blue_a\nonumber\\
\ue_4 &=& N^{-1}\left(\partial_w-b^a\blue_a^i\partial_i\right)\\
A &=& \Theta dw  + \blA_idx^i\nonumber\\
{\omega^a}_b &=&  {\theta^a}_b dw +{\blw^a}_{i;b}dx^i\nonumber\\
{\omega^a}_4 &=& 0\nonumber
\eeqn
where $a=0, 1,2 ,3$ and the intrinsic $3+1$-d co-frame, frame and connections are now labelled by a tilde. For simplicity, we will take $N$ to be constant. 

Now we wish to compute the dimensional reduction of the Levi-Civita Chern Simons term, but this can be straightforwardly done for the full torsional case as well.
We start by computing the dimensionally reduced Levi-Civita connection. Using
\beq
{\lcw}_{AB}= \frac{1}{2}\left\{de_A(\ue_B,\ue_C)-de_B(\ue_A,\ue_C)-de_C(\ue_A,\ue_B)\right\}e^C
\eeq
we find
\beq
\lcw_{ab}=\bllcw_{ab}+K_{[ab]}dw 
\eeq
\beq
\lcw_{a4}=-\frac{1}{N}K_{(ac)} \left(\ble^c+b^cdw\right)
\eeq
and we have defined $K_{ab}$ 
\beq
K_{ac}=(\bllcD b_a)(\blue_c)=(db_a+\bllcw_{ad}b^d)(\blue_c).
\eeq
The Levi-Civita curvature two-form is given by
\beq
\lcR_{ab}= \bllcR_{ab}+\bllcD K_{[ab]} \w dw-\frac{1}{N^2}K_{(ac)}K_{(bd)}(\ble^c+b^cdw)\wedge (\ble^d+b^ddw)
\eeq
\beq
\lcR_{a4}= -\frac{1}{N}\left(\bllcD K_{(ac)}\w\ble^c+\bllcD\left(K_{(ac)} b^c\right)\w dw+K_{[ac]} dw\w K_{(cd)}\ble^d\right)
\eeq
where note that $\bllcD K_{ab}=dK_{ab}+[\bllcw,K]_{ab}$. 

Let us now proceed to computing the LC Chern Simons term in the effective action. Up to unimportant boundary terms, we find
\beqn
\oint F\w CS[\lcw] &=& \oint F\w \left(CS[\lcw_{ab}]+2\lcw_{a4}\w \lcR_{4a}\right)\\
&=&-d\Theta\w CS[\bllcw]+2\blF\w K_{[ab]} \bllcR_{ba}-\frac{2}{N^2}d\Theta\w K_{a}\w\bllcD K_{a} \nonumber\\
&-&\frac{2}{N^2}K_{(ac)}b^c\;\blF\w\bllcD K_{a}+\frac{1}{N^2}K_{[ab]}\;\blF\w K_{a}\w K_{b}\nonumber
\eeqn
where we have introduced the 1-form $K_a = K_{(ab)}\ble^b$. Note that up to terms of $O(b^2)$, we find
\beq
\oint F\w CS[\lcw] = -d\Theta\w CS[\bllcw]+b_a\ble^a\w d\mathring{\mathcal{A}}_2+O(b^2)
\eeq
where $\mathring{\mathcal{A}}_2 = (\blF \w \bllcR_{ab})(\blue^a,\blue^b)$, which is the result we arrived at previously, albeit in the presence of torsion. 

\providecommand{\href}[2]{#2}\begingroup\raggedright\endgroup

\end{document}